\documentclass[a4paper,twocolumn,11pt]{article}
\pdfoutput=1

\usepackage[T1]{fontenc}
\usepackage[utf8]{inputenc}
\usepackage{lmodern}
\usepackage[english]{babel}
\usepackage[numbers,sort&compress]{natbib}
\usepackage{microtype}
\usepackage{geometry}

\usepackage{amsmath,amssymb,amsthm,mathtools,bm}
\usepackage{braket}
\usepackage{mathrsfs}

\usepackage{booktabs}
\usepackage{enumitem}
\usepackage{xcolor}
\usepackage{graphicx}

\graphicspath{{ProjectorRaychaudhuriDraft/Draft/}{./}}
\makeatletter
\def\input@path{{ProjectorRaychaudhuriDraft/Draft/}{./}}
\makeatother
\usepackage{tikz}
\usetikzlibrary{arrows.meta,positioning,calc,decorations.markings}
\usepackage{caption}
\usepackage{subcaption}
\usepackage{hyperref}
\usepackage[nameinlink,noabbrev]{cleveref}

\crefname{appendix}{Appendix}{Appendices}
\Crefname{appendix}{Appendix}{Appendices}
\allowdisplaybreaks

\makeatletter
\renewcommand\section{\@startsection{section}{1}{\z@}%
  {-3.5ex \@plus -1ex \@minus -.2ex}%
  {2.3ex \@plus .2ex}%
  {\normalfont\Large\bfseries\raggedright}}
\renewcommand\subsection{\@startsection{subsection}{2}{\z@}%
  {-3.25ex \@plus -1ex \@minus -.2ex}%
  {1.5ex \@plus .2ex}%
  {\normalfont\large\bfseries\raggedright}}
\renewcommand\subsubsection{\@startsection{subsubsection}{3}{\z@}%
  {-3.25ex \@plus -1ex \@minus -.2ex}%
  {1.5ex \@plus .2ex}%
  {\normalfont\normalsize\bfseries\raggedright}}
\makeatother

\newtheorem{theorem}{Theorem}[section]
\newtheorem{proposition}[theorem]{Proposition}

\theoremstyle{definition}

\newcommand{\Gr}{\operatorname{Gr}}

\newcommand{\Tr}{\operatorname{Tr}}
\newcommand{\Ric}{\operatorname{Ric}}
\newcommand{\TF}{\operatorname{TF}}
\newcommand{\FS}{\mathrm{FS}}

\newcommand{\CP}{\mathbb{CP}}
\newcommand{\dd}{\mathrm d}
\newcommand{\ii}{\mathrm i}
\newcommand{\ee}{\mathrm e}
\newcommand{\cQ}{\mathcal Q}
\newcommand{\cF}{\mathcal F}
\newcommand{\cG}{\mathcal G}

\newcommand{\cB}{\mathcal B}
\newcommand{\cV}{\mathcal V}
\newcommand{\cE}{\mathcal E}

\newcommand{\cI}{\mathcal I}
\newcommand{\proj}{\Pi}
\newcommand{\norm}[1]{\left\lVert #1\right\rVert}

\begin{document}

\title{Raychaudhuri-Ehlers Kinematics on Quantum State-Space:  Caustics and Extrinsic Quantum Geometry}
\author{%
Neellohit Karmakar%
\thanks{Neellohit Karmakar and Suraj S. Hegde contributed
equally to this work and share first authorship.},
\quad
Suraj S. Hegde\footnotemark[1]
\thanks{Email: \href{mailto:surajhegde@iisertvm.ac.in}
{\texttt{surajhegde@iisertvm.ac.in}}},
\quad
Aryan Bhatia\\[0.5em]
{\small School of Physics, Indian Institute of Science
Education and Research}\\
{\small Thiruvananthapuram, Vithura, Kerala, India}
}
\date{}
\maketitle

\begin{abstract}
Quantum mechanics is inherently geometric in that the pure states form a curved complex manifold and the Schrödinger evolution defines a Hamiltonian flow on this space. Geometry identifies constraints on flows on quantum state-space that hold independently of their microscopic realization. We uncover such universal behavior in families of geodesic paths on the quantum state-space  using the Raychaudhuri-Ehlers equations. We show that the complex structure leads to tidal anisotropy in the curvature, and for Hilbert space dimension $N\geq3$, complete point-emitted geodesic bundles  necessarily develop shear and undergo direction-selective focusing at finite distance. We give these geometric results an information-theoretic interpretation in terms of fidelity susceptibility and quantum Fisher information. For ground-state manifolds constrained by physical parameters, we resolve the tidal curvature into Berry-curvature-dependent ambient tides and contributions from extrinsic bending within the full ambient state space. We further relate extrinsic bending to variations in the unitary generator of motion along intrinsic geodesics. We present exact many-body geometric results for the transverse-field XY spin chain. The product structure of the ground state leads to an exact cancellation of the ambient tide, including its Berry-curvature enhancement, by extrinsic curvature contributions. At the zero-field Ising point, we obtain exact results at finite size and show that, for every direction of motion, the shear source (for anisotropy-directed motion) is entirely extrinsic relative to the full many-body state-space. For the metric in the thermodynamic limit, the geodesic families exhibit repeated $A_2$ fold catastrophes inside the ferromagnetic phase. Our framework thus separates universal state-space effects from those of physical restrictions and exposes their consequences in a paradigmatic many-body system.
\end{abstract}

\section{Introduction}

Quantum mechanics admits a completely geometric description apart from the familiar 
wavefunction and path integral formalisms \cite{Kibble1979,ProvostVallee1980,AnandanAharonov1990,
MukundaSimon1993I,AshtekarSchilling1999,BrodyHughston2001,Manko2005}. The geometric formulation begins with the physical meaning of a state: normalized vectors that differ only by an overall phase represent the same pure state. Physical pure states therefore form a space of rays. The Hermitian inner product and multiplication by $\mathrm{i}$ in Hilbert space endow this state-space with three structures: \emph{a Riemannian metric, symplectic form, and complex structure}. Intuitively, the metric quantifies state distinguishability, infinitesimal state-mixing, and quantum fluctuations. The symplectic form gives the antisymmetric structure associated with Poisson brackets and corresponding Hamiltonian
evolution. The complex structure brings out the complex
character of quantum amplitudes. Together, these compatible
structures allow Born probabilities, observable expectation
values, and Schr\"odinger evolution to be expressed geometrically,
with the same physical content as the wavefunction formulation
\cite{Kibble1979,AshtekarSchilling1999,BrodyHughston2001}.
 For parameterized families of states, these structures appear as the well-known \emph{quantum metric and Berry curvature}. 

Nevertheless one might inquire: \emph{Why do geometry at all instead of working with wavefunctions?}
Geometry helps us to identify what is universal. It organizes information already
contained in wavefunctions into physical statements that survive
changes of basis, phase convention, and parameterization.
Geometric relations can therefore constrain entire families of
quantum transformations without a separate calculation for every
trajectory. They also distinguish consequences of quantum state
space itself from features of a particular Hamiltonian or
preparation protocol, and clarify how restriction to ground states or other parametrised family of states modify those consequences.

We pursue this perspective through the question:
\emph{How can we probe the curvature structure of quantum
state space? What universal physical consequences does
this structure imply?}
We address these questions by studying the collective
behaviour of geodesics on quantum state space.
Geodesics are particularly relevant because they locally minimize
the accumulated magnitude of mixing between a state and its
orthogonal complement. Such mixing underlies changes in the
physical state during quantum evolution and in response to
variations of fields and couplings. A single geodesic does not reveal the full curvature structure of the underlying space. Curvature governs how neighboring geodesics separate and approach one another, causing a bundle to spread, change shape, or focus. These collective effects expose geometric constraints on relative evolution of states that are not apparent from an individual trajectory. The separations of neighboring geodesics obey the Jacobi equation, while the Raychaudhuri-Ehlers equations organize their collective
evolution into expansion, shear, and rotation
\cite{Raychaudhuri1955,Ehlers1993}. In general relativity, congruence kinematics has played a central role in understanding the global structure of
spacetime. In particular, geodesic focusing is a key
ingredient in the Hawking-Penrose singularity theorems,
which establish geodesic incompleteness under suitable
curvature, causality, and initial or boundary conditions
\cite{Penrose1965,Hawking1966,HawkingPenrose1970,HawkingEllis1973}. In this work, we apply these equations to flows on the curved quantum state-space. 

Our question also has a direct historical precedent. Discussing
the quantum metric on parameter space, M. Berry asked in his 1989 article `The Quantum Phase, Five Years After':
``Do families of geodesics (for example those issuing in different directions from the same point) exhibit the generic
caustic singularities classified by catastrophe theory? Do any such caustics have physical meaning?''
\cite[Sec.~4, p.~15]{Berry1989FiveYears}. Berry's questions motivate us to explore caustics, catastrophes and their interpretation for a broader class of geodesic families in quantum state space and on parameterized state manifolds, including families that originate from a common point.

We address these questions by connecting the geometry of quantum-state paths to the quantum information theoretic concepts. Existing work provides several ingredients for this connection. Geodesics play a foundational role in the kinematic theory of geometric phases and Bargmann invariants, while constrained geodesics and null-phase curves extend this connection to submanifolds of ray space \cite{MukundaSimon1993I,RabeiArvindMukundaSimon1999,
MukundaEtAl2003}. For parameterized ground-state families,
geodesics have also been studied in connection with
adiabatic state preparation and paths through phase
diagrams
\cite{Kolodrubetz2013,Tomka2016,Bukov2019,LiskaGritsev2021,GollJonsson2026,HuangProjectors2026}. Geometric approaches relate geodesic
deviation to quantum uncertainty
\cite{Benvegnu2007,Spera2011}, while classical curvature
spectra and conjugate loci characterize geodesic
separation and the limits of length minimization in
projective and Grassmannian spaces
\cite{Helgason2001,Berceanu1997}.

The physical interpretation of geodesic separation comes
from state distinguishability. For a parameterized
pure-state family, the quantum metric determines fidelity
susceptibility and is one quarter of the quantum Fisher
information \cite{BraunsteinCaves1994,Facchi_2010}. This relation connects geodesic separation to the precision
with which preparation parameters can be estimated.
When the Fisher matrix becomes singular, some parameter
combinations cease to produce distinguishable state
changes to first order \cite{LiYeh2012,Namkung2025}.
Related perspectives include geodesic spread on
Fisher-Rao manifolds \cite{CafaroMancini2011}, sensitivity
and extrinsic geometry in models with widely differing
parameter sensitivities \cite{Transtrum2011}, and Bures
geodesics in quantum metrology \cite{Spehner2025}. Bohmian approaches have also explored quantum corrections
to spacetime congruences
\cite{Das2014,AliDas2015,CheriyodathillathuEtAl2026}.
In a different setting of quantum-circuit complexity,
conjugate points diagnose loss of local length
minimization for penalized metrics on the unitary group
\cite{DowlingNielsen2008,Auzzi2021,
Balasubramanian2020,Balasubramanian2021}.

To apply the congruence viewpoint to a physical system,
one must also specify which state variations are allowed by the
parameters. Parameters such as applied fields, couplings, or crystal momenta
usually select a restricted family of states, such as ground states or occupied
Bloch subspaces, within the full
quantum state space. Quantum metrics and Berry curvatures
of these families already play important roles in
adiabatic transport, phase characterization, and
electronic and optical responses
\cite{KolodrubetzReview2017,TormaEssay2023,
YuReview2025,JiangReview2025}. Their intrinsic geometry
describes distances and curvature measured within the
accessible family. Their extrinsic geometry describes
how that family bends within the surrounding state
space. A locally shortest path constrained to the
family can therefore bend when viewed in the full
space.

Extrinsic state geometry and gauge-invariant projector
calculus provide tools for describing this distinction
\cite{AvdoshkinPopov2023,Mitscherling2025}. The Gauss
equation relates the intrinsic curvature to the
ambient curvature and the contribution from bending
\cite{DajczerTojeiro2019}. Related questions arise in
classical embedded-worldsheet deformations
\cite{CapovillaGuven1995} and in distance-based geometric
descriptions of correlated many-body states
\cite{HassanShankarChakrabarti2018,
ChakrabartiHassanShankar2019}. The shape operator of an
occupied sub-bundle has also been studied
\cite{Oancea2026}. These developments motivate resolving
the curvature governing geodesic separation into
contributions from the surrounding state space and
from the bending of the parametrised family.

The transverse-field XY chain provides an explicit
setting in which these intrinsic and extrinsic
descriptions can be compared. Its ground-state quantum
metric and critical scaling have been extensively
analyzed
\cite{ZanardiGiordaCozzini2007,VenutiZanardi2007,
Kolodrubetz2013,MeraPaunkovicAminVieira2022}, while
studies of its geodesics and hidden isometries provide
a foundation for examining neighboring paths
\cite{KumarMahapatraPhukonSarkar2012,LiskaGritsev2021}.
More directly related to the present question,
geodesic expansion and the Raychaudhuri equation have
been studied on quantum parameter manifolds, including
the XY chain. These studies examined geometric critical exponents and
also discussed divergent expansion without divergent
curvature \cite{KumarSarkar2014}. A subsequent renormalization-group
treatment examined curvature and expansion scaling
\cite{MaityMahapatraSarkar2015}. The XY model has also
been used to compare Fubini-Study and Nielsen complexity
as diagnostics of its critical lines
\cite{JaiswalGautamSarkar2021}.

Building on these results, we examine the full transverse
deformation of quantum-state geodesic families, resolve
its ambient and extrinsic curvature contributions, and
identify its fidelity and quantum Fisher signatures.

\paragraph{Main results.} 
\emph{Our results combine a synthesis of geometric results on quantum-state
congruence kinematics with its information-theoretic
interpretation in terms of fidelity susceptibility and quantum Fisher
information. We highlight the importance of understanding the extrinsic geometry  of embedding of restricted manifold of states in the full state-space.  We complement this general framework with
exact many-body curvature calculations and fold catastrophes
in the XY ground-state manifold.}
We present a summary of our results below.

\begin{enumerate}[leftmargin=2.2em]

\item \textbf{Universal shear, selective focusing, and quantum-information signatures.}
Using the established positive and anisotropic curvature of Fubini-Study geometry, we determine the expansion and shear of complete geodesic families emitted from a common pure state. For Hilbert-space dimension $N\ge3$, the enhanced tidal curvature along the complex partner of the velocity necessarily generates shear. At the first
conjugate point, this direction collapses while the remaining transverse directions stay regular. We express this selective focusing through fidelity susceptibility
and quantum Fisher information: sensitivity to one initial-direction parameter vanishes, while independent transverse variations remain distinguishable. We also explain their consistency with family of Schr\"odinger evolutions under a common Hamiltonian, whose normalized flow has zero transverse expansion and shear because acceleration and twist balance the curvature contribution.

\item \textbf{Berry curvature and extrinsic bending in congruence dynamics.}
Using the Gauss equation, we resolve the tidal tensor of
a parameterized state manifold into an isotropic state-space
contribution, a Berry-curvature-dependent enhancement,
and contractions of its second fundamental form. This
decomposition identifies how the ambient state-space complex structure and the bending of the parametrised state family
enter the expansion and shear equations. Extrinsic bending can reinforce or counteract the ambient
tide, so the conditions that ensure focusing in full projective space need not hold on a ground-state manifold.
We further relate the second fundamental form to the covariant variation of the minimal generator of intrinsic geodesic motion, giving the bending tensor a direct transport interpretation.

\item \textbf{Product-state cancellation and exact finite-size XY tides.}
The ground state structure of the XY spin chain is expressed as a product of momentum-pair states. Using the product structure of the XY ground state, we evaluate the extrinsic curvature contributions to the Gauss equation and show that they exactly cancel the collective ambient tide, including its Berry-curvature enhancement. The surviving terms determine the intrinsic tidal tensor governing geodesic focusing and shear.  At the zero-field Ising point, we evaluate the surviving tidal tensor exactly at finite size for every geodesic direction. For anisotropy-directed motion, the ambient tide is isotropic, and the shear is therefore sourced entirely by extrinsic curvature relative to the full many-body projective space. Along this direction, we also calculate the normal acceleration and the corresponding variation of the intrinsic geodesic generator.

\item \textbf{$A_2$-Fold catastrophes within a quantum phase and compensated fidelity.}
Using the conserved momenta associated with the known isometries of the XY spin chain quantum metric \cite{Kolodrubetz2013,LiskaGritsev2021}, we analytically obtain a geodesic family with two fixed nonzero momenta. We show that its simple turning points are $A_2$ folds of the geodesic-family map within the ferromagnetic phase. Between folds, the congruence stretches in one transverse direction and contracts in the other, while its expansion scalar decreases. At each fold, one combination of path labels produces no first-order transverse separation, although the QFI evaluated at equal arclength remains nonsingular. To isolate the transverse response, we compare neighboring states on a common Hamilton-Jacobi wavefront, removing their displacement along the geodesic. For the specified label variation in the collapsing direction, the logarithmic fidelity loss per site scales as the fourth power of the label change. We derive its exact coefficient.

\end{enumerate}

We illustrate shear and selective focusing in a simple qutrit preparation scheme through their distinct signatures in the quantum Fisher information.

The paper is organized as follows.
Section~\ref{sec:projectors} gives a pedagogical review of the quantum state-space geometry in terms of projectors and the interpretation of geodesics as paths of minimum accumulated mixing. Section~\ref{sec:qfi-congruence} develops congruence kinematics and the Raychaudhuri-Ehlers equations. Section~\ref{sec:cp-exact} resolves the curvature-induced deformation and caustics for pure states. Section~\ref{sec:schrodinger-rigidity} examines the contrasting rigidity of evolution under a common Hamiltonian. Section~\ref{sec:extrinsic} introduces embedded state manifolds and separates ambient and extrinsic contributions to their curvature. Section~\ref{sec:xy} applies this framework to the XY chain, presenting exact results for extrinsic shear and geodesic congruences. Section~\ref{sec:operational} discusses how some of the results could be seen in an example of a qutrit, followed by the discussion and
conclusions in Section~\ref{sec:discussion}.

\section{Geometry of quantum states}
\label{sec:projectors}

The geometrical formulation of quantum mechanics has been known for many decades but several of its aspects remain underappreciated. For the reader’s convenience, we review the geometric formulation in the language of projectors on pure states. Such a description is
gauge invariant and is widely used in condensed matter systems \cite{Kato1950,AvronSeilerSimon1983,MeraOzawa2021,Mitscherling2025}.
We review the geometry of the space of projectors following
Refs.~\cite{Kibble1979,AshtekarSchilling1999,ProvostVallee1980}. Until
Section~\ref{sec:extrinsic}, we deal with an abstract state-space and do not assume any Hamiltonian eigenspace or a parametrised family of states.

\subsection{Three geometric structures on the state-space}

A normalized wavefunction $\ket{\psi}\in\mathbb C^N$ represents
a pure quantum state, but its overall phase has no physical
significance. The projector
\begin{equation}
P=\ket{\psi}\bra{\psi}
 \label{eq:pure-state-projector}
\end{equation}
removes this ambiguity as $\ket{\psi}$ and
$e^{i\chi}\ket{\psi}$ give the same $P$.
It contains the full physical information about the pure state,
with expectation values given by
$\langle A\rangle=\Tr(PA)$.
As an operator, $P$ selects the component of any vector along
$\ket{\psi}$, while $Q=I-P$ selects its orthogonal complement. The defining properties of the projector are
\begin{equation}
 P^\dagger=P,\qquad P^2=P,\qquad \Tr P=1.
 \label{eq:grassmannian}
\end{equation}
Varying $P$ over all such projectors gives the space of physical
pure states, $\mathbb{CP}^{N-1}$. The collection of all rank-\(r\) orthogonal projectors on \(\mathbb C^N\) forms the Grassmannian \(\Gr(r,N)\).\footnote{Equivalently,
$\Gr(r,N)\simeq U(N)/[U(r)\times U(N-r)]$:
unitary transformations generate all rank-$r$ subspaces,
while rotations within the selected subspace or its
complement leave the projector unchanged.} For $r=1$, the selected subspace is a single ray. The Grassmannian then
reduces to the projective state space
$\Gr(1,N)=\mathbb{CP}^{N-1}$. For a single spin-$\tfrac12$,
this is the Bloch sphere. For a lattice of $L$ spin-$\tfrac12$
degrees of freedom, $N=2^L$, and the full pure-state space is
$\mathbb{CP}^{2^L-1}$.

We now introduce geometric structures on the projector manifold, beginning with the tangent space at each point. Differentiating the constraint $P^2=P$ yields the conditions satisfied by infinitesimal tangential displacements $dP$: $ P\,dP\,P=0, Q\,dP\,Q=0$ and $dP = P\,dP\,Q + Q\,dP\,P$

Thus, tangent vectors $X\in T_P\Gr(r,N)$ are  Hermitian matrices with nonzero matrix elements only between the subspace selected by $P$ and its orthogonal complement selected by $Q$. They therefore describe infinitesimal mixing between these two subspaces:
\begin{equation}
X=PXQ+QXP=[P,[P,X]].
\label{eq:tangent-projector}
\end{equation}
The orthogonal projection of an arbitrary Hermitian matrix $X$ onto the tangent space is given by
$\proj_{\Gr,P}(X)=[P,[P,X]]$

The projector manifold carries three compatible geometric structures: a Riemannian metric \(g\), a symplectic form \(\omega\), and a complex structure \(J\). For rank-one projectors, these are the structures underlying the geometric formulation of pure-state quantum mechanics~\cite{Kibble1979,AshtekarSchilling1999,BrodyHughston2001}.  We introduce each structure and its physical interpretation below.

The {\it Riemannian metric} assigns a real number \(g(X,Y)\) to two tangent vectors $X,Y$ at \(P\). It is bilinear, symmetric, and positive definite, and therefore defines lengths and angles between infinitesimal displacements. On the projector manifold, we take
\begin{equation}
g(X,Y)=\frac12\operatorname{Tr}(XY).
\label{eq:RiemannMetric}
\end{equation}
Its physical meaning follows directly from the off-diagonal structure of tangent vectors. Writing $ X=B+B^\dagger, B=QXP$, we obtain
\begin{equation}
g(X,X)=\operatorname{Tr}(B^\dagger B).
\end{equation}
Thus, the squared length of a tangent displacement measures the total squared magnitude of the infinitesimal mixing between the subspace selected by \(P\) and its orthogonal complement. 
For rank-one projectors, this is the Fubini-Study metric~\cite{AshtekarSchilling1999,BrodyHughston2001}:
\begin{equation}
ds^2=\langle d\psi|d\psi\rangle
-|\langle\psi|d\psi\rangle|^2.
\end{equation}
It measures changes of the physical state after removing the contribution from an overall phase.
To see its physical content, choose an orthonormal basis
$\{\ket{\beta}\}$ for the subspace orthogonal to $\ket{\psi}$.
Then
\begin{equation}
ds^2=\sum_\beta|\langle\beta|d\psi\rangle|^2.
\end{equation}
The metric therefore measures the total squared amplitude
admixed into orthogonal states by an infinitesimal change of
$\ket{\psi}$. When $\ket{\psi}$ is a Hamiltonian eigenstate,
these are admixtures of other eigenstates.
The same quantity determines the leading reduction in overlap between neighboring states, connecting this mixing picture
directly to state distinguishability.

The {\it symplectic form} is an antisymmetric, nondegenerate bilinear form on tangent vectors that is also closed, \(d\omega=0\). On the projector manifold, it is given by
\begin{equation}
\omega_P(X,Y)=-\frac{i}{2}\operatorname{Tr}\left(P[X,Y]\right).
\label{eq:SymplecticForm}
\end{equation}
Unlike the metric, which measures lengths, the symplectic form measures oriented areas in tangent planes and supplies the geometric structure needed for Hamiltonian evolution.

The symplectic form allows unitary quantum evolution to be expressed as a Hamiltonian flow on the state manifold ~\cite{Kibble1979,AshtekarSchilling1999}. In particular, let \(H\) be a Hamiltonian operator and we define the function

$$
\mathcal E(P)=\operatorname{Tr}(PH).
$$

For a rank-one projector, \(\mathcal E\) is the energy expectation value.
With the normalization above, the Hamiltonian flow is defined by 
\begin{equation}
2\hbar\omega_P(\dot P,Y)
=d\mathcal E_P(Y)
=\operatorname{Tr}(HY)
\end{equation}
for every tangent vector \(Y\). With $\dot{P}=dP/ds$
\begin{equation}
\dot P=-\frac{i}{\hbar}[H,P],
\end{equation}
which is Schrödinger evolution expressed in terms of projectors. The factor \(2\hbar\) converts our geometrically normalized symplectic form into the symplectic form associated with the physical energy function.

The {\it complex structure} is a linear map from each tangent space to itself, defined by
\begin{equation}
J_PX=-i[P,X],\qquad J_P^2X=-X.
\end{equation}
It provides the tangent-space counterpart of multiplication by \(i\). Indeed, for \(X=B+B^\dagger\), $J_PX=iB-iB^\dagger.$ Thus, \(J\) multiplies the mixing amplitude \(B\) by \(i\), while preserving the Hermiticity of the tangent vector. For a pure state, a tangent displacement can be represented by a state-vector variation satisfying \(\langle\psi|\delta\psi\rangle=0\). Then \(J\) acts as \(|\delta\psi\rangle\mapsto i|\delta\psi\rangle\).

These three structures are compatible:
\begin{align}
\omega_P(X,Y)=g(J_PX,Y), \\
g_P(J_PX,J_PY)=g(X,Y).
\end{align}
Consequently, \(J\) rotates a tangent displacement into an orthogonal displacement of equal length, and the symplectic pairing is its metric overlap with another displacement. Together, these compatible structures define the Grassmannian's K\"ahler geometry. Their role in curvature and the relative motion of quantum states will become explicit in the congruence analysis.

The full
pure-state space forms an \emph{ambient space} within which the parametrised state manifolds are embedded.
The ambient geometric structures are  the origin of the familiar quantum metric and Berry curvature of a parameterized state family. If $P=P(\lambda)$, a parameter change produces the tangent displacement $dP=(\partial_\mu P)d\lambda^\mu$. Evaluating the ambient tensors on these tangent vectors gives
\[
 \begin{aligned}
 g_{\mu\nu}
 &=g(\partial_\mu P,\partial_\nu P),\\
 \cF_{\mu\nu}
 &=-2\omega(\partial_\mu P,\partial_\nu P).
 \end{aligned}
\]
This construction is called a pullback: it expresses the geometry of the parametrised family in its physical parameters. The ambient tensors exist before a particular family is selected. In \cref{sec:extrinsic}, we examine how the intrinsic curvature of such a family also depends on its bending within the surrounding state space.

\subsection{Geodesics on the state-space}

A smooth curve \(P(s)\) on the projector manifold describes a continuously varying quantum state or subspace. Its tangent velocity is
\begin{equation}
U(s)=\dot P(s)=\frac{dP}{ds}.
\end{equation}

As we move along the curve, a tangent vector can change both within the projector manifold and in directions normal to it. The covariant derivative retains only the tangential part of this change. In projector language, this means keeping only the matrix elements that mix the subspaces selected by \(P\) and \(Q\):

\begin{equation}
\begin{aligned}
\nabla_s U
&=\proj_{\Gr,P}(\dot U)\\
&=P\dot UQ+Q\dot UP.
\end{aligned}
\label{eq:covariant-derivative-curve}
\end{equation}

A geodesic is a curve whose tangent velocity is parallel transported along itself,
\begin{equation}
\nabla_s U=0.
\label{eq:geodesic-intrinsic}
\end{equation}
 Thus, a geodesic has no tangential acceleration. Its ordinary matrix acceleration \(\ddot P\) need not vanish: a normal acceleration is generally required to keep the curve on the manifold, just as a great circle on a sphere bends in the surrounding Euclidean space. Geodesics locally minimize length for the metric \(g(X,Y)=\frac12\Tr(XY)\).

For \(U=\dot P\), the geodesic condition therefore becomes
\begin{equation}
\proj_{\Gr,P}(\ddot P)
=[P,[P,\ddot P]]=0.
\label{eq:geodesic-tangent-projection}
\end{equation}
In other words, the acceleration has no matrix elements connecting the subspaces selected by \(P\) and \(Q\).

The remaining, diagonal blocks of the acceleration are fixed by the projector constraint. Differentiating \(P^2=P\) twice gives
$\ddot PP+2\dot P^2+P\ddot P=\ddot P$,
and hence $P\ddot PP=-2P\dot P^2P,
Q\ddot PQ=2Q\dot P^2Q$. Combining these identities with the vanishing off-diagonal blocks yields the \emph{geodesic equation} entirely in projector form \cite{HelmkeHuperTrumpf2007,HuangProjectors2026}:
\begin{equation}
\ddot P=[\dot P,[P,\dot P]].
\label{eq:projector-geodesic-equation}
\end{equation}
The parameter \(s\) is an affine parameter, with respect to which the speed is constant. It equals arclength when the speed is normalized to unity.

\subsection{Subspace mixing and the physical meaning of geodesics}
\label{sec:mixing}

The quantum metric measures infinitesimal mixing between eigenstates, a relation expressed through adiabatic gauge potentials \cite{KolodrubetzReview2017}. Describing the selected states by a projector gives the quantum metric and Berry curvature a common geometric setting on the Grassmannian \cite{MeraOzawa2021}. Geodesics on this manifold describe finite changes of the selected subspace \cite{HelmkeHuperTrumpf2007}, with their interpretation in terms of interband mixing developed in Ref.~\cite{HuangProjectors2026}. We review it here to establish the physical interpretation of the paths whose collective deformation we study.  Consider a smooth path
\begin{equation}
P(s)=\sum_\alpha\ket{\alpha(s)}\bra{\alpha(s)},
\end{equation}
where the states $\ket{\alpha(s)}$ form an orthonormal basis
of the selected subspace, and let $\ket{\beta(s)}$ span its
complement. The operator
\begin{equation}
Z=Q\dot P P
=\sum_{\beta,\alpha}
\ket{\beta}\langle\beta|\dot\alpha\rangle\bra{\alpha}
\label{eq:mixing-block}
\end{equation}
contains the amplitudes that change the subspace itself. Since $\dot P=Z+Z^\dagger$,
\begin{equation}
\begin{aligned}
g(\dot P,\dot P)
&=\frac12\Tr(\dot P^2)
=\Tr(Z^\dagger Z)\\
&=\sum_{\beta,\alpha}
|\langle\beta|\dot\alpha\rangle|^2.
\end{aligned}
\label{eq:mixing-speed}
\end{equation}
Thus the squared speed measures the sum of squared magnitudes
of mixing per unit path parameter. This quantity is independent
of the bases chosen within the two subspaces and requires
no Hamiltonian or assignment of energies.

For a pure state, geodesic motion mixes the initial state with
an orthogonal state selected by the initial tangent.
Choose an orthonormal basis whose first vector represents
$P_0$. In this basis,
\begin{equation}
P_0=
\begin{pmatrix}1&0\\0&0_{N-1}\end{pmatrix},
\qquad
U_0=
\begin{pmatrix}
0&Z_0^\dagger\\
Z_0&0_{N-1}
\end{pmatrix}.
\label{eq:U-block}
\end{equation}
The $(N-1)$-component column $Z_0$ contains the amplitudes
for mixing the initial state with its orthogonal complement.
The anti-Hermitian operator
\begin{equation}
K=[U_0,P_0]
=\begin{pmatrix}
0&-Z_0^\dagger\\
Z_0&0_{N-1}
\end{pmatrix}
\end{equation}
satisfies $[K,P_0]=U_0$ and generates the geodesic
\begin{equation}
P(s)=\ee^{sK}P_0\ee^{-sK}.
\label{eq:projector-geodesic}
\end{equation}
Differentiating this expression verifies
\cref{eq:projector-geodesic-equation} and gives
$\dot P(s)=\ee^{sK}U_0\ee^{-sK}$. Its speed is therefore
constant:
\begin{equation}
g(\dot P,\dot P)
=\frac12\Tr(U_0^2)
=Z_0^\dagger Z_0.
\end{equation}
For unit speed, $Z_0^\dagger Z_0=1$, so $Z_0$ specifies
a normalized state orthogonal to the initial state.

The motion is confined to the two-dimensional Hilbert subspace spanned by these two states. At unit speed,
\begin{equation}
\ket{u(s)}
=\cos s\,\ket{u_0}+\sin s\,\ket v,
\qquad \braket{u_0|v}=0,
\label{eq:cp-geodesic-state}
\end{equation}
where both states are normalized. This representative satisfies
$\braket{u(s)|\dot u(s)}=0$, excluding an additional overall
phase rotation. More generally, the same mixing speed determines
the leading overlap loss along any smooth pure-state path:
\begin{equation}
1-|\langle u(s)|u(s+ds)\rangle|^2
=g(\dot P,\dot P)\,ds^2+O(ds^3).
\end{equation}
Mixing and state distinguishability therefore describe the
same infinitesimal change.

The corresponding distance between pure-state projectors is
\begin{equation}
d_{\FS}(P_0,P)
=\arccos|\braket{u_0|u}|
\in[0,\pi/2].
\label{eq:FS-distance}
\end{equation}
Along \cref{eq:cp-geodesic-state}, this distance equals $s$
for $0\le s\le\pi/2$, when the state first becomes orthogonal
to $\ket{u_0}$. The geodesic continues smoothly beyond that
point, but the continued segment no longer gives the shortest
path from $P_0$.

The accumulated magnitude of mixing along the path is its geometric length,
\begin{equation}
\begin{aligned}
\mathcal L_{\rm mix}[P]
&=\int_0^T\sqrt{\Tr(Z^\dagger Z)}ds\\
&=\int_0^T
\sqrt{\sum_{\beta,\alpha}
|\langle\beta|\dot\alpha\rangle|^2}ds.
\end{aligned}
\label{eq:mixing-length}
\end{equation}
This length is independent of the traversal rate and reduces to the Fubini-Study length for rank one. A shortest geodesic therefore minimizes the accumulated mixing between the fixed endpoints.

For a fixed duration \(T\), the quadratic mixing cost satisfies
\begin{equation}
\begin{aligned}
\mathcal A_{\rm mix}[P]
&=\frac12\int_0^T\Tr(Z^\dagger Z)ds
\ge\frac{\mathcal L_{\rm mix}^{2}}{2T},
\end{aligned}
\label{eq:mixing-action}
\end{equation}
where the equality holds at constant metric speed. A shortest geodesic traversed at constant speed consequently minimizes this quadratic cost as well.

 If the path is restricted to a smooth parametrised family \(P(\lambda)\), the cost becomes
\begin{equation}
\mathcal A_{\rm mix}
=\frac12\int_0^T
g_{\mu\nu}(\lambda)\dot\lambda^\mu\dot\lambda^\nu ds.
\end{equation}
On a patch where the induced metric is nondegenerate, its stationary paths satisfy
\begin{equation}
\ddot\lambda^\mu
+\Gamma^\mu_{\nu\rho}
\dot\lambda^\nu\dot\lambda^\rho=0.
\end{equation}
These are geodesics corresponding to the pulled back metric on the parameter manifold.

This mixing also has a direct dynamical interpretation. For a prescribed path traversed as $P(s(t))$, the
parallel-transport Hamiltonian is
\begin{equation}
 H_{\rm mix}(t)
 =i\hbar[\partial_tP,P]
 =i\hbar\frac{ds}{dt}[\partial_sP,P].
 \label{eq:mixing-CD}
\end{equation}
It satisfies $\partial_tP=-i[H_{\rm mix},P]/\hbar$ and $\Tr(H_{\rm mix}^2)/(2\hbar^2) =g(\partial_tP,\partial_tP)$, giving the metric speed a Hamiltonian interpretation. When $P(t)$ is an instantaneous eigenprojector of a reference Hamiltonian, $H_{\rm mix}$ acts as a counterdiabatic term that ensures exact following of the selected eigenspace \cite{DemirplakRice2003,Berry2009}.

\section{Raychaudhuri-Ehlers kinematics of geodesic congruences}
\label{sec:qfi-congruence}

Now we proceed to consider a collection of geodesics. The relative motion of neighboring geodesics connects curvature to
changes in the size, shape, and rotation of a bundle, and hence to
the quantum metric on labels of the geodesics.

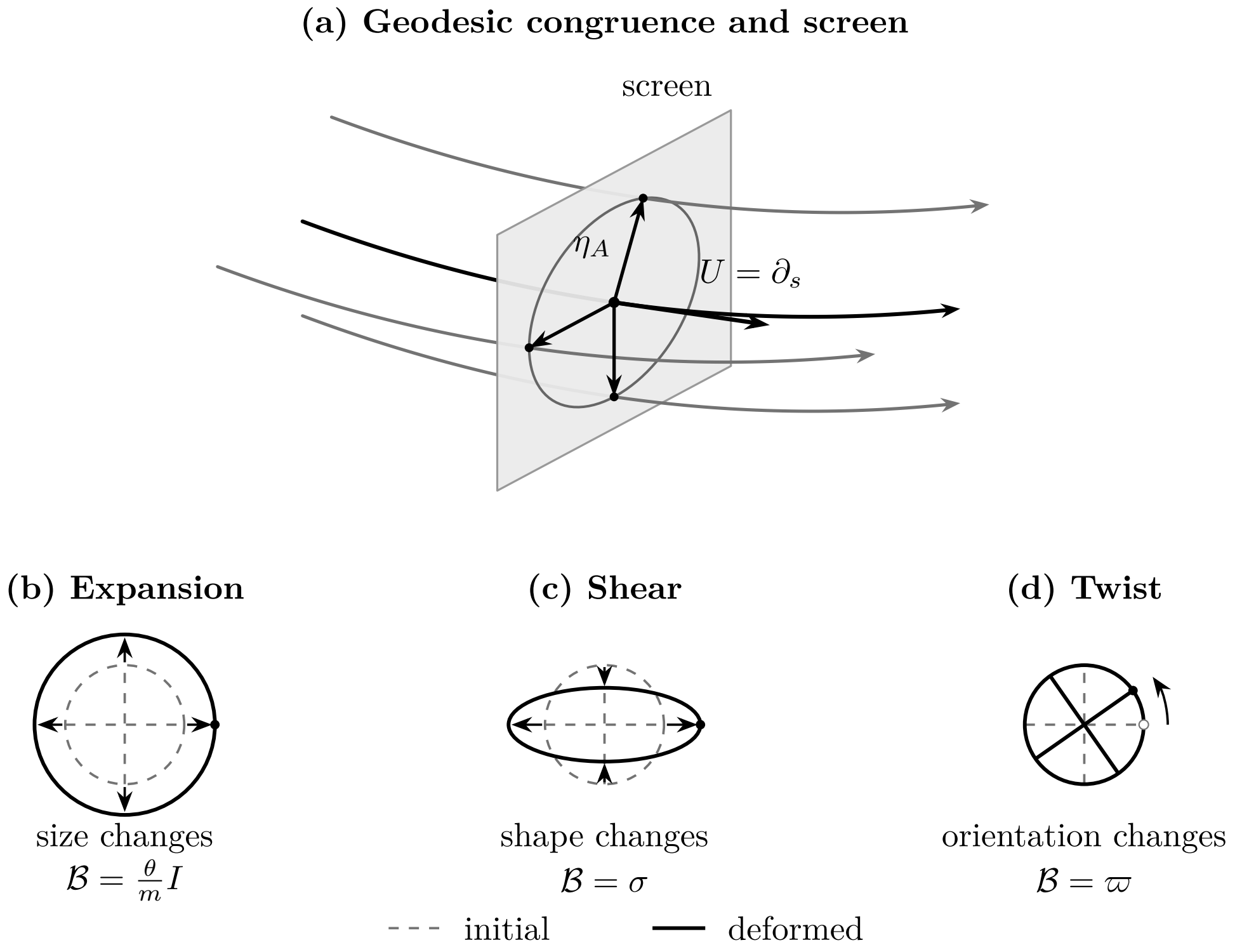
\begin{figure*}[t]
 \centering
\begin{tikzpicture}[x=1cm,y=1cm,>=Stealth,
  initial/.style={black!55,dashed,line width=.7pt},
  final/.style={black,line width=1.1pt},
  velocity/.style={->,black,line width=.7pt},
  every node/.style={font=\small}]
  \path[use as bounding box] (-2.3,-2.3) rectangle (12.3,7.65);
  \node[font=\small\bfseries] at (5,7.30) {(a) Geodesic congruence and screen};
  \begin{scope}[shift={(5.1,4.4)}]
\begin{scope}[scale=.82,
  x={(1.1cm,-.16cm)},y={(-.9cm,-.48cm)},z={(0cm,1cm)},
  line cap=round,line join=round,every node/.style={font=\small},
  geodesic/.style={black!55,line width=.95pt},
  deviation/.style={black,line width=1pt,-{Stealth[length=2.3mm,width=1.6mm]}}]
% Reference geodesic and three neighboring geodesics, behind the screen.
\draw[black,line width=1.15pt]
  plot[domain=-3.6:0,samples=40,variable=\t] (\t,0,{.035*\t*\t});
\foreach \a in {0,110,270}{
  \draw[geodesic]
    plot[domain=-3.6:0,samples=40,variable=\t]
      (\t,{1.2*cos(\a)},{1.2*sin(\a)+.035*\t*\t});
}
% Orthogonal screen and a circular cross-section (an ellipse in perspective).
\path[fill=black!8,fill opacity=.94,draw=black!40,line width=.6pt]
  (0,-1.65,-1.6)--(0,1.65,-1.6)--(0,1.65,1.65)--(0,-1.65,1.65)--cycle;
\draw[black!60,line width=.75pt]
  plot[domain=0:360,samples=100,variable=\a]
    (0,{1.2*cos(\a)},{1.2*sin(\a)});
% Continuations in front of the screen.
\foreach \a in {0,110,270}{
  \draw[geodesic,-{Stealth[length=2mm]}]
    plot[domain=0:4,samples=40,variable=\t]
      (\t,{1.2*cos(\a)},{1.2*sin(\a)+.035*\t*\t});
}
\draw[black,line width=1.15pt,-{Stealth[length=2mm]}]
  plot[domain=0:4,samples=40,variable=\t] (\t,0,{.035*\t*\t});
% Tangential deviation vectors within the screen.
\foreach \a in {0,110,270}{
  \draw[deviation] (0,0,0)--(0,{1.2*cos(\a)},{1.2*sin(\a)});
  \fill (0,{1.2*cos(\a)},{1.2*sin(\a)}) circle (1.6pt);
}
\node[anchor=east,inner sep=5pt] at (0,-.21,.60) {$\eta_A$};
% Unit tangent at the central geodesic's intersection.
\draw[black,line width=1.25pt,-{Stealth[length=2.6mm,width=1.7mm]}]
  (0,0,0)--(1.8,0,0) node[pos=.88,above=6pt] {$U=\partial_s$};
\fill (0,0,0) circle (2pt);
\node[anchor=south,inner sep=6pt] at (0,-.75,2.0) {screen};
\end{scope}

  \end{scope}
  \foreach \x/\title/\formula in {0/{(b) Expansion}/{$\mathcal{B}=\tfrac{\theta}{m}I$},5/{(c) Shear}/{$\mathcal{B}=\sigma$},10/{(d) Twist}/{$\mathcal{B}=\varpi$}}{
    \node[font=\small\bfseries] at (\x,1.4) {\title};
    \node at (\x,-1.62) {\formula};
  }
  \begin{scope}
    \draw[initial] (0,0) circle (.62);
    \draw[initial] (-.62,0)--(.62,0) (0,-.62)--(0,.62);
    \draw[final] (0,0) circle (.94);
    \foreach \a in {0,90,180,270}{
      \draw[velocity] (\a:.65)--(\a:.91);
    }
    \fill (.94,0) circle (1.4pt);
    \node at (0,-1.18) {size changes};
  \end{scope}
  \begin{scope}[shift={(5,0)}]
    \draw[initial] (0,0) circle (.62);
    \draw[initial] (-.62,0)--(.62,0) (0,-.62)--(0,.62);
    \draw[final] (0,0) ellipse (1.00 and .3844);
    \draw[velocity] (.65,0)--(.97,0);
    \draw[velocity] (-.65,0)--(-.97,0);
    \draw[velocity] (0,.60)--(0,.40);
    \draw[velocity] (0,-.60)--(0,-.40);
    \fill (1,0) circle (1.4pt);
    \node at (0,-1.18) {shape changes};
  \end{scope}
  \begin{scope}[shift={(10,0)}]
    \draw[initial] (0,0) circle (.62);
    \draw[initial] (-.62,0)--(.62,0) (0,-.62)--(0,.62);
    \draw[final] (0,0) circle (.62);
    \draw[final] (35:.62)--(215:.62) (125:.62)--(305:.62);
    \draw[velocity] (0:.87) arc[start angle=0,end angle=35,radius=.87];
    \fill (35:.62) circle (1.4pt);
    \draw[black!55,fill=white] (.62,0) circle (1.4pt);
    \node at (0,-1.18) {orientation changes};
  \end{scope}
  \draw[initial] (2.75,-2.12)--(3.3,-2.12);
  \node[anchor=west] at (3.4,-2.12) {initial};
  \draw[final] (5.5,-2.12)--(6.05,-2.12);
  \node[anchor=west] at (6.15,-2.12) {deformed};
\end{tikzpicture}

 \caption{A geodesic bundle and the local action of its deformation
 tensor. Dashed curves and spokes mark the initial circular cross-section.
Solid curves and spokes show the separate effects of expansion,
shear, and rotation. Expansion changes the size uniformly in all
directions. Shear changes the shape, with the illustrated finite map preserving
area. Twist rotates the marked directions while preserving their
inner products.}
 \label{fig:congruence-deformation}
\end{figure*}

\subsection{Jacobi fields, transverse deformation, and quantum distinguishability}

\paragraph{Jacobi fields and geodesic labels.}
A Jacobi field describes the infinitesimal separation between
neighboring geodesics, obtained by slightly varying their
initial positions or velocities (\cref{fig:congruence-deformation}(a)). To follow these separations,
we consider a smooth family of geodesics with velocity field $U$ and the Jacobi field $\eta_A$,
\begin{equation}
 P=P(s,\lambda^A),\qquad U=\partial_sP,\qquad
 \eta_A=\partial_AP,
 \label{eq:geodesic-family}
\end{equation}
where $s$ is arclength and the parameters/labels $A=1,\ldots,q$ distinguish the paths. At fixed $s$, a label change produces $\delta P=\eta_A\delta\lambda^A$.

The physical meaning of this separation is supplied by state distinguishability. At fixed arclength, the Jacobi fields induce the label-space metric
\begin{equation}
 \cG_{AB}=g(\eta_A,\eta_B)
=\tfrac12\Tr(\partial_AP\,\partial_BP).
 \label{eq:label-metric}
\end{equation}
For rank-one projectors, the squared fidelity between neighboring members of the family satisfies
\begin{equation}
 \begin{aligned}
 \mathscr F(s;\lambda,\lambda+\delta\lambda)
 &:=\Tr[P(s,\lambda)P(s,\lambda+\delta\lambda)]\\
 &=1-\cG_{AB}\delta\lambda^A\delta\lambda^B
   +O(|\delta\lambda|^3).
 \end{aligned}
 \label{eq:label-fidelity}
\end{equation}
Thus, for a label direction $c^A$, the squared norm of $\eta_c=c^A\eta_A$ is the fidelity susceptibility $\chi_F(c)=\cG_{AB}c^Ac^B$. The corresponding pure-state quantum Fisher information (QFI) is
\begin{equation}
 \cI_{AB}=4\cG_{AB}.
 \label{eq:label-qfi}
\end{equation}
These relations connect the separation of neighboring geodesics to sensitivity to their preparation labels \cite{BraunsteinCaves1994,YouLiGu2007}.

Our question is therefore how this sensitivity changes as the paths advance. The Jacobi equation determines the evolution of their separation vectors. To describe the spreading and deformation of the bundle, we will also distinguish separation along the reference path from separation transverse to it. These need not give the same sensitivity: two outputs at equal arclength can differ partly by a displacement along the path.

We consider the case where the path and label derivatives commute
$[U,\eta_A]_{\mathrm{Lie}}=\nabla_U\eta_A-\nabla_{\eta_A}U=0$.  For a tangent field $X(s)$, its covariant derivative
along the path retains the part of its change tangent to state space:
\begin{equation}
 \nabla_UX=P\frac{dX}{ds}Q+Q\frac{dX}{ds}P.
 \label{eq:covariant-derivative-projector}
\end{equation}
Thus $\nabla_U\eta_A$ is the relative velocity and
$\nabla_U^2\eta_A$ the relative covariant acceleration. The geodesic
condition $\nabla_UU=0$ gives the \emph{Jacobi equation},
also called the geodesic-deviation equation:
\begin{equation}
 \nabla_U^2\eta_A+R(\eta_A,U)U=0.
 \label{eq:jacobi}
\end{equation}

Here $R$ is the Riemann curvature tensor associated with
the Fubini-Study metric  of the quantum state-space.
The metric determines the Levi-Civita connection, which defines
parallel transport and the covariant derivative $\nabla$.
Curvature measures how this transport depends on the path:
transporting a tangent vector around an infinitesimal closed
loop generally changes it. We use the convention
\begin{equation}
 R(X,Y)Z=\nabla_X\nabla_YZ-\nabla_Y\nabla_XZ
                    -\nabla_{[X,Y]}Z.
 \label{eq:curvature-convention}
\end{equation}
The vectors $X$ and $Y$ specify the directions of the loop,
while $Z$ is the vector being transported. Here $[X,Y]$ is
the Lie bracket.

Although $g(X,Y)=\tfrac12\Tr(XY)$ has a simple form, the
projector manifold is curved because the constraint $P^2=P$ makes
its tangent space vary with $P$. The resulting curvature is therefore intrinsic to the geometry of quantum states.

In \cref{eq:jacobi}, this geometric property has a direct
meaning: $-R(\eta_A,U)U$ is the relative covariant acceleration
of neighboring geodesics. Curvature thus governs how their
separation evolves from a specified initial separation and
relative velocity.

\paragraph{The screen and transverse separation.}
Now we consider a bundle of neighbouring geodesics. To follow a bundle around a reference path, we use the tangent
directions perpendicular to its unit velocity. They form its
\emph{screen space},
\begin{equation}
 \begin{gathered}
 S_P=U^\perp=\{V\in T_P\Gr(r,N):g(V,U)=0\},\\
 m=2r(N-r)-1.
 \end{gathered}
 \label{eq:screen-space}
\end{equation}
We denote its dimension by $m$. Since the tangent space
has $2r(N-r)$ real dimensions, excluding the direction along
$U$ leaves $m:=\dim_{\mathbb R}S_P=2r(N-r)-1$.
For pure states ($r=1$), this gives $m=2N-3$.
The separation fields project onto this space as
\begin{equation}
 \eta_A^\perp=\eta_A-a_AU,
 \qquad a_A=g(U,\eta_A).
 \label{eq:screen-jacobi-fields}
\end{equation}
The family spans the full screen at $P$ if its transverse separation vectors $\{\eta_A^\perp\}$ have rank $m$. Otherwise, they span only a lower-dimensional subspace of $S_P$. The screen space is a subspace of the tangent
space.

For a unit-speed geodesic family, the longitudinal components are
constant. Commuting derivatives and metric compatibility give
\begin{equation}
 \dot a_A=g(U,\nabla_{\eta_A}U)
         =\tfrac12\partial_Ag(U,U)=0.
 \label{eq:longitudinal-constant}
\end{equation}
An initially transverse separation therefore remains transverse.
In a \emph{radial}, or point-emitted, family, $P(0,\lambda)=P_0$
and the labels vary the unit initial direction. Then $\eta_A(0)=0$
and $a_A=0$ throughout. Equal-arclength angular variations already
lie in the screen.

\paragraph{The deformation tensor.}
Where the family defines a smooth unit tangent field, the
deformation tensor measures how its velocity varies across
the screen:
\begin{equation}
 \cB(V)=\nabla_VU,\qquad V\in S_P.
 \label{eq:deformation-operator}
\end{equation}
Its output is transverse because
$g(U,\nabla_VU)=\tfrac12 V[g(U,U)]=0$.

Write $V_A=\eta_A^\perp$ for the transverse separation
associated with the label $\lambda^A$. To express this evolution in components, choose an orthonormal
screen basis $E_a$, $a=1,\ldots,m$, parallel transported along
the reference geodesic. Then $\cB_{ab}=g(E_a,\nabla_{E_b}U)$,and
\begin{equation}
 V_A=E_aV^a{}_A,\qquad
 \dot V^a{}_A=\cB^a{}_bV^b{}_A.
 \label{eq:Jacobi-matrix-evolution}
\end{equation}
The components $V^a{}_A$ describe the transverse separations,
while $\cB^a{}_b$ describes their instantaneous deformation.
The dot denotes differentiation at fixed labels. This evolution
law applies to the deviation fields $V_A$. The basis vectors
instead satisfy $\nabla_UE_a=0$.

\paragraph{Expansion, shear, and rotation.}
Along the evolution an initially circular, marked cross-section can change its size,
become elongated, or rotate, as illustrated in
\cref{fig:congruence-deformation}. These changes correspond to the
trace, symmetric traceless, and antisymmetric parts of the
deformation tensor:
\begin{equation}
 \cB_{ab}=\frac{\theta}{m}h_{ab}+\sigma_{ab}+\varpi_{ab},
 \label{eq:decomposition}
\end{equation}
\begin{equation}
 \begin{aligned}
 \theta&=h^{ab}\cB_{ab},\\
 \sigma_{ab}&=\cB_{(ab)}-\frac{\theta}{m}h_{ab},\\
 \varpi_{ab}&=\cB_{[ab]}.
 \end{aligned}
\end{equation}
Here $h_{ab}=\delta_{ab}$ in the chosen basis. Orthogonal changes of
screen axes do not mix these three tensor sectors. This gives the
decomposition a basis-independent physical meaning.

As illustrated in \cref{fig:congruence-deformation},  the \emph{expansion} $\theta$ is the fractional rate of change of
transverse volume. Its isotropic part changes the size while
preserving the shape. The \emph{shear} $\sigma$ describes unequal
stretching, changing shape without contributing to the instantaneous
fractional volume change. The antisymmetric \emph{twist} $\varpi$
rotates separation vectors relative to a parallel frame without
directly changing their lengths or inner products. Their evolutions
can nevertheless be coupled. 

\paragraph{The label metric and transverse volume.}
The label metric \eqref{eq:label-metric} compares outputs at equal arclength and includes both transverse and longitudinal displacements. We now separate these contributions and relate the transverse metric to the deformation of the bundle. Collect the components of the transverse separation fields $V_A=E_aV^a{}_A$ into the $m\times q$ matrix $ V=(V^a{}_A)$, whose columns correspond to the different labels. Then the transverse counterpart is given by
\begin{equation}
 \begin{aligned}
 \cG_{\perp,AB}
   &=g(\eta_A^\perp,\eta_B^\perp)
     =\cG_{AB}-a_Aa_B,\\
 \cG_\perp&=V^TV.
 \end{aligned}
 \label{eq:screen-label-metric}
\end{equation}
The two coincide for fully transverse families of geodesics, including angular
variations of a point-emitted bundle. Equation
\eqref{eq:Jacobi-matrix-evolution} gives
\begin{equation}
 \dot\cG_\perp
 =2V^T\left(\frac{\theta}{m}h+\sigma\right)V.
 \label{eq:Gdot-orthonormal}
\end{equation}
Twist drops out of this instantaneous derivative, though it can
influence later evolution. Since $\dot a_A=0$, also
$\dot\cG=\dot\cG_\perp$.

For a full screen-spanning family, $q=m$ and $V$ is invertible. The volume of a fixed infinitesimal label cross-section is proportional to $\sqrt{\det\cG_\perp}=|\det V|$. So
\begin{equation}
 \theta=\partial_s\ln|\det V|
       =\partial_s\ln\sqrt{\det\cG_\perp}.
 \label{eq:theta-qfi-volume}
\end{equation}
The matrix and determinant identities are
derived in \cref{app:re-screen,app:re-label-metric}. For pure states, four times the transverse label-space metric $4\cG_\perp$ defines the transverse quantum Fisher information (\cref{sec:fidelity-qfi-signatures}).

\paragraph{Caustics and conjugate points.}
As a geodesic bundle evolves, its transverse cross-section can contract and collapse in one or more directions. Such a collapse is called a \emph{caustic}: neighboring paths lose an independent direction of transverse separation. The collapse can be partial in that the entire bundle need not converge to a single state.

For a family spanning the full screen, this occurs when the
separation matrix $ V$ loses rank, so
$\det\cG_\perp=0$ and the transverse volume vanishes.
If $k$ independent separation amplitudes vanish linearly
at $s=s_*$, then on approach to the caustic,
\begin{equation}
 \sqrt{\det\cG_\perp}\propto(s_*-s)^k,
 \qquad
 \theta=-\frac{k}{s_*-s}+O(1).
\end{equation}
The expansion therefore tends to $-\infty$, expressing the
diverging fractional contraction rate of the shrinking
transverse volume. Negative expansion indicates contraction, but does not by itself
guarantee a caustic at finite distance. 

For a point-emitted family, all separations are transverse,
so $\cG_\perp=\cG$. At a caustic away from the source, a
nonzero combination of initial-direction changes produces
a Jacobi field that vanishes at both endpoints:
$\eta_c(0)=\eta_c(s_*)=0$. The endpoint is then called
\emph{conjugate} to the source \cite{Ilmavirta2020}.

In a nontransverse family, a collapsing transverse separation
can retain a longitudinal component, $\eta_c=a_cU\ne0$.
The full ray map $(s,\lambda)\mapsto P(s,\lambda)$ is then
singular even though its equal-arclength label map can remain
regular. Such a collapse alone does not establish conjugacy
to the chosen initial point.

At both types of collapse, the paths and Jacobi fields can remain smooth. The inverse in $\cB=\dot V V^{-1}$ ceases to exist, so the linear Jacobi equation, rather than the deformation tensor, supplies the continuation through the caustic.

\subsection{Raychaudhuri-Ehlers equations}

We now determine how the bundle's expansion, shear, and rotation
evolve. Each transverse separation field $V_A$ obeys the
geodesic-deviation equation $\nabla_U^2V_A+R(V_A,U)U=0$.Writing $V_A=E_aV^a{}_A$ in the parallel-transported screen basis, the covariant derivatives act only on its components. With the tidal tensor $\cE_{ab}=g(E_a,R(E_b,U)U)$, these equations become
\begin{equation}
 \ddot V+\cE V=0,
\end{equation}
where $V=(V^a{}_A)$ is the separation matrix, with one column
for each label. Differentiating $\dot V=\cB V$ gives
$\ddot V=(\dot\cB+\cB^2)V$, and hence
\begin{equation}
 \dot\cB+\cB^2+\cE=0.
 \label{eq:riccati}
\end{equation}
This Riccati equation expresses geodesic deviation in terms
of the deformation tensor. The quadratic term arises because
$\cB$ measures rates of change relative to separations that
are themselves evolving.

The trace of the tidal tensor $\cE$ gives the total transverse curvature:
\begin{equation}
 \begin{gathered}
 \Ric(X,Y)=\sum_i g(R(e_i,X)Y,e_i),\\
 \Tr_S\cE=\Ric(U,U).
 \end{gathered}
 \label{eq:ricci-definition}
\end{equation}

The trace, symmetric trace-free, and antisymmetric parts give the
Raychaudhuri-Ehlers equations
\cite{Raychaudhuri1955,Ehlers1993}:
\begin{equation}
 \dot\theta=-\frac{\theta^2}{m}-\norm\sigma^2+\norm\varpi^2
             -\Ric(U,U),
 \label{eq:raychaudhuri-intrinsic}
\end{equation}
\begin{align}
 \dot\sigma_{ab}={}&-\frac{2\theta}{m}\sigma_{ab}
 -(\cE_{ab})_{\TF}\nonumber\\
 &-\bigl(\sigma_a{}^c\sigma_{cb}
         +\varpi_a{}^c\varpi_{cb}\bigr)_{\TF},
 \label{eq:shear-intrinsic}\\
 \dot\varpi_{ab}={}&-\frac{2\theta}{m}\varpi_{ab}
 -\bigl(\sigma_a{}^c\varpi_{cb}
         +\varpi_a{}^c\sigma_{cb}\bigr).
 \label{eq:vorticity-intrinsic}
\end{align}
Here $T_{\TF}=T-(\Tr_ST/m)h$ for a symmetric screen tensor.
See \Cref{app:raychaudhuri-ehlers} for detailed derivations.

Positive Ricci curvature and shear drive the expansion evolution towards focussing whereas
twist enters with an opposite sign with a tendency towards defocussing. Positive Ricci curvature alone
therefore gives no common focusing bound for arbitrary twisting
families. The trace-free tidal tensor directly generates shear, while its
symmetry prevents a direct twist source. An initially twist-free
congruence stays twist-free. Existing twist
can affect subsequent focusing through its coupling to the other
components. These equations require only the metric and initial
family, and apply equally to an induced state manifold when its
own connection and curvature are used.

\section{Inevitable shear and caustics from the complex geometry of quantum states} 
\label{sec:cp-exact}

The congruence equations relate the deformation of a geodesic bundle to the curvature sampled by its velocity and transverse separations. We now evaluate that curvature for the full pure-state space $\CP^{N-1}$, whose screen has dimension
$m=2N-3$. The complex structure selects a direction of enhanced tidal curvature. We determine how this anisotropy generates shear and selective focusing, and how the resulting deformation appears in fidelity susceptibility and QFI.

\subsection{Sectional curvature and complex tidal anisotropy}
The compatibility relations $\omega(X,Y)=g(JX,Y)$ and
$g(JX,JY)=g(X,Y)$ make $JU$ a unit orthogonal partner of $U$.
The complex structure $JX=-i[P,X]$ maps the mixing block
$Z=QXP$ to $iZ$. Thus $X$ and $JX$ describe the two
quadratures of the same mixing amplitude, differing by
a phase of $\pi/2$. Their span is a \emph{holomorphic plane},
preserved by $J$. The K\"ahler property $\nabla J=0$ preserves this
pairing under parallel transport \cite{AshtekarSchilling1999}.

For projective state space, the metric and complex structure enter the
curvature tensor as \cite{Helgason2001}
\begin{equation}
\begin{aligned}
 R(X,Y)Z={}&g(Y,Z)X-g(X,Z)Y\\
 &+g(JY,Z)JX-g(JX,Z)JY\\
 &+2g(X,JY)JZ.
\end{aligned}
\label{eq:CP-curvature}
\end{equation}
For orthonormal tangent vectors $X,Y$, the \emph{sectional curvature}
$K(X,Y)=g(R(X,Y)Y,X)$ measures the curvature in plane spanned by the two vectors.
Equation~\eqref{eq:CP-curvature} then gives
\begin{equation}
 K(X,Y)=1+3\omega(X,Y)^2,\qquad 1\le K\le4.
\label{eq:CP-sectional-curvature}
\end{equation}
The upper bound is attained for $Y=\pm JX$. The lower bound is attained
when $Y\perp JX$, so the plane has zero symplectic area and is called
a \emph{totally real} plane. The upper bound is attained by planes spanned by $X$ and $JX$.
For $N\ge3$, the lower bound is attained by planes spanned by
orthonormal $X,Y$ with $g(JX,Y)=0$.

In the Jacobi equation, curvature determines the relative
acceleration through $\nabla_U^2V=-\cE(V)$. For a unit geodesic
velocity $U$, the transverse tidal operator is
\begin{equation}
 \cE(V):=R(V,U)U=V+3g(JU,V)JU.
 \label{eq:CP-Jacobi-operator}
\end{equation}
The symplectic overlap $g(JU,V)=\omega(U,V)$ therefore selects
one enhanced direction:
\begin{equation}
 \cE(JU)=4JU,\qquad \cE(E)=E,
 \quad E\perp\{U,JU\}.
 \label{eq:CP-tidal-spectrum}
\end{equation}
These channels have multiplicities $1$ and
$m-1=2N-4$, respectively. Initially equal separations along them
experience inward accelerations in the ratio $4{:}1$ but their instantaneous
contraction rates also depend on the initial data.

\begin{figure}[tb]
 \centering
 % Exact polar plot r=1+3 cos^2(alpha), alpha measured from +JU.
% All radius markers lie at their true curvature values.
\begin{tikzpicture}[x=.52cm,y=.52cm,>=Stealth,font=\footnotesize]
  \draw[black!35,dashed] (0,0) circle[radius=4];
  \draw[black!35,dashed] (0,0) circle[radius=1];
  \draw[->,thin] (-4.65,0)--(4.65,0);
  \draw[->,thin] (0,0)--(0,4.6);
  \draw[->,thin] (0,0)--(0,-4.6);
  \draw[black,very thick,domain=0:360,samples=241,smooth,variable=\t]
    plot ({(1+3*cos(\t)^2)*sin(\t)},
          {(1+3*cos(\t)^2)*cos(\t)});
  \foreach \x/\y in {0/4,0/-4,1/0,-1/0}
    \fill (\x,\y) circle[radius=1.7pt];
  \fill (0,0) circle[radius=1.3pt];
  \draw[->,dashed] (0,0)--({4.35*sin(52)},{4.35*cos(52)})
    node[above right,inner sep=2pt] {$V_\alpha$};
  \draw[->] (0,1.35) arc[start angle=90,end angle=38,radius=1.35];
  \node at ({1.65*sin(26)},{1.65*cos(26)}) {$\alpha$};
  \node[above,align=center,inner sep=3pt] at (0,4.65)
    {$JU$\\$K=4$};
  \node[below,align=center,inner sep=3pt] at (0,-4.65)
    {$-JU$\\$K=4$};
  \node[right,align=left,inner sep=3pt] at (4.65,0)
    {$E$\\$K=1$};
  \node[left,align=right,inner sep=3pt] at (-4.65,0)
    {$-E$\\$K=1$};
  \node[align=center,anchor=south west,font=\scriptsize]
    at (2.2,3.7) {$K(U,V_\alpha)$\\$=1+3\cos^2\alpha$};
\end{tikzpicture}

 \caption{Directional tidal curvature in a two-dimensional
 screen slice, for $N\ge3$. The polar radius is
 $K(U,V_\alpha)=1+3\cos^2\alpha$, with $\alpha$ measured
 from $JU$. The dashed circles mark the eigenvalues $1$ and $4$.
 The velocity $U$ is perpendicular to the slice, so each
 radial direction specifies a plane containing $U$.}
 \label{fig:CP-tidal-anisotropy}
\end{figure}
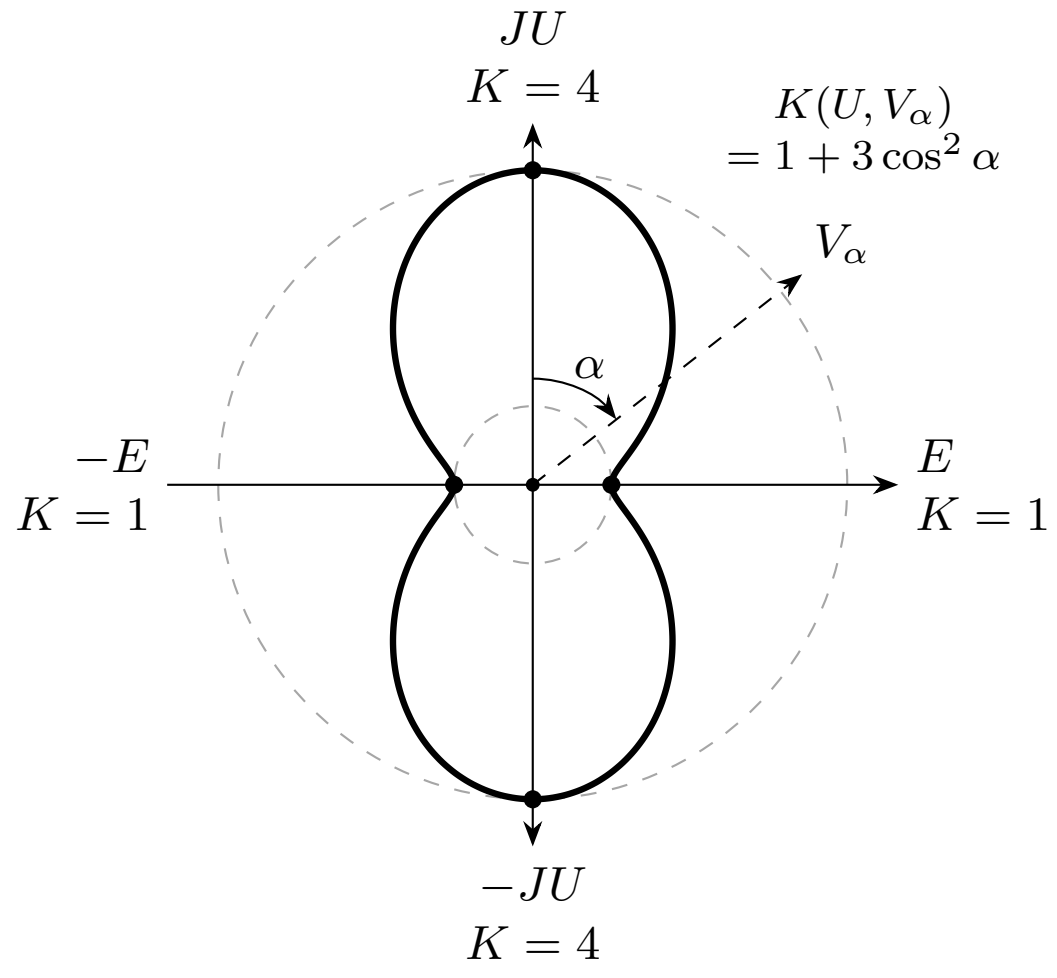

For $N\ge3$, a unit transverse direction can be written as
\begin{equation}
 \begin{aligned}
 V_\alpha&=\cos\alpha\,JU+\sin\alpha\,E,\\
 K(U,V_\alpha)
 &=g(\cE(V_\alpha),V_\alpha)=1+3\cos^2\alpha,
 \end{aligned}
 \label{eq:CP-angular-curvature}
\end{equation}
where $E$ is unit and perpendicular to both $U$ and $JU$.
This sectional curvature measures the inward acceleration
component along the separation. Its angular dependence is
shown in \cref{fig:CP-tidal-anisotropy}.

In the screen frame $E_a$, let us define $v_a=g(E_a,JU)$.
The trace-free part of the tidal tensor is then
\begin{equation}
 (\cE_{ab})_{\TF}=3\left(v_av_b-\frac{h_{ab}}m\right).
\label{eq:CP-tidal-TF}
\end{equation}

\begin{figure*}[t]
  \centering
  \resizebox{\textwidth}{!}{% Requires tikz and \usetikzlibrary{arrows.meta}.
% Shape of an initially circular angular perturbation, divided by sin(s).
% The vertical semiaxis is cos(s); the horizontal semiaxis is one.
\begin{tikzpicture}[x=1cm,y=1cm,
  font=\small,>=Stealth,
  axis/.style={thin,->},
  outline/.style={line width=0.85pt},
  heading/.style={align=center,font=\small}]
  \path[use as bounding box] (-1.50,-1.38) rectangle (14.18,4.03);

  % Arclength increases from left to right.
  \node[anchor=west] at (-1.35,3.65) {$s=0$};
  \draw[->,line width=0.7pt] (-0.15,3.65) -- (10.00,3.65)
    node[midway,above,fill=white,inner sep=3pt] {$s$};
  \node[anchor=west] at (10.15,3.65)
    {$s=\pi/2$ (first caustic)};

  % Common scale: the ordinary transverse spreading sin(s) is divided out.
  \foreach \x/\letter/\title/\subtitle/\ratio in {
    0/{(i)}/{Initial shape}/{$s\to0^+$}/1,
    4.1/{(ii)}/{Shear develops}/{$s=\pi/3$}/0.5,
    8.2/{(iii)}/{Near caustic}/{$s\lesssim\pi/2$}/0.16,
    12.3/{(iv)}/{At caustic}/{$s=\pi/2$}/0}{
    \begin{scope}[shift={(\x,0)}]
      \node[heading] at (0,2.62) {\letter\ \title\\[3pt]\subtitle};
      \draw[axis] (-1.25,0) -- (1.43,0);
      \draw[axis] (0,-1.14) -- (0,1.42)
        node[above] {$JU$};
      \node[anchor=north east,inner sep=1pt] at (1.44,-0.10) {$E_a$};
      \ifdim\ratio pt>0pt
        \draw[outline] (0,0) ellipse [x radius=1cm,y radius=\ratio cm];
      \else
        \draw[line width=1.8pt] (-1,0) -- (1,0);
      \fi
      \fill (0,0) circle[radius=1.6pt];
    \end{scope}
  }
\end{tikzpicture}
}
  \caption{Complex-structure-driven shear and the first caustic for
  unit-speed geodesics from a common rank-one projector in
  $\mathbb{CP}^{N-1}$, with $N\geq3$.
  A two-dimensional section of the transverse bundle is shown in the
  parallel-transported directions $JU$ and $E_a$, where
  $E_a\perp U,JU$.
  An initially circular set of angular variations develops semiaxes
  $\sin(2s)/2$ along $JU$ and $\sin s$ along $E_a$.
  To display the change of shape, each section is divided by the common
  factor $\sin s$: the plotted semiaxes are therefore $\cos s$ and $1$.
  Panel (i) shows the isotropic limit as $s\to0^+$. 
  Panels (ii)--(iii) show the increasing relative compression along $JU$.
  At $s=\pi/2$, this Jacobi separation along $JU$  collapses while the ordinary
  transverse directions remain finite, producing the first caustic.}
  \label{fig:complex-caustic-shape}
\end{figure*}
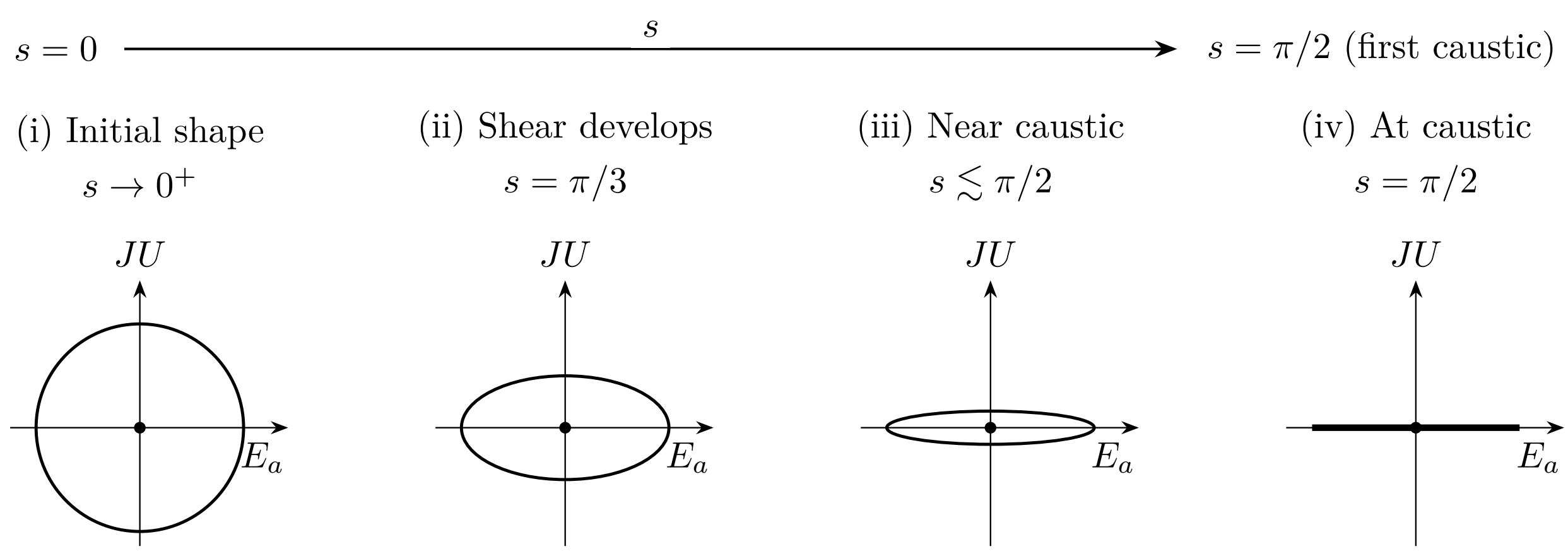

Consequently, \cref{eq:shear-intrinsic} gives, at a given point where
the deformation is initially isotropic,
\begin{equation}
\left.\dot\sigma_{ab}\right|_{\sigma=\varpi=0}
 =-3\left(v_av_b-\frac{h_{ab}}m\right).
\label{eq:shear-initial-source}
\end{equation}

This leads us to one of the striking consequences of the complex structure in the congruence kinematics.
\emph{For $N\ge3$, any irrotational, isotropic deformation immediately develops shear:
the rate along $JU$ decreases more rapidly than the other rates.
The complex structure fixes a preferred axis even when the initial
deformation has none.} Figure~\ref{fig:complex-caustic-shape} shows the
resulting relative compression for a point-emitted bundle.
For $N=2$, $m=1$ and the source vanishes, consistently with the absence
of shear in a one-dimensional screen.

\subsection{Positive Ricci curvature and finite-distance caustic}

The Raychaudhuri equation now gives a focusing result directly
from the geometry of quantum state space. In general relativity,
focusing theorems combine vanishing twist with a curvature
condition obtained from an energy condition on the matter obtained through
Einstein's equations. In contrast, here no field equation or assumption
about a microscopic Hamiltonian is required: the quantum mechanical state-space structure itself supplies the positive curvature,
$\Ric(U,U)=2N$ for unit $U$.

With the initially vanishing twist remaining zero,
\begin{equation}
 \dot\theta
 =-\frac{\theta^2}{m}-\norm{\sigma}^2-2N
 \le -\frac{\theta^2}{m}-2N,
 \qquad m=2N-3.
 \label{eq:CP-focusing-inequality}
\end{equation}
Thus even an initially expanding bundle must eventually
contract and focus. Integrating this inequality gives the
following bound.

\begin{theorem}[Curvature-driven focusing in quantum state space]
Consider a unit-speed geodesic congruence in $\CP^{N-1}$
whose deviation fields span the full transverse screen.
Suppose the bundle is regular at $s_0$, with expansion
$\theta_0$ and vanishing twist. Its transverse volume
collapses at a finite arclength $s_*>s_0$, with
\begin{equation}
 s_*-s_0\le\frac1{\sqrt{k}}
 \left[\frac\pi2+
 \arctan\!\left(\frac{\theta_0}{m\sqrt{k}}\right)\right],
 \qquad k=\frac{2N}{m}.
 \label{eq:CP-focusing-bound}
\end{equation}
For $\theta_0<0$, the weaker bound
$s_*-s_0\le m/|\theta_0|$ also holds.
\end{theorem}

\begin{proof}
Writing $y=\theta/m$, inequality
\eqref{eq:CP-focusing-inequality} becomes
$\dot y\le-y^2-k$. Hence, while the bundle remains regular,
\begin{equation}
 \arctan\!\frac{y(s)}{\sqrt{k}}
 \le
 \arctan\!\frac{y(s_0)}{\sqrt{k}}
 -\sqrt{k}(s-s_0).
\end{equation}
Let $S$ denote the endpoint specified by the bound
\eqref{eq:CP-focusing-bound}. Suppose, for contradiction, 
that the transverse volume remains nonzero throughout
$[s_0,S]$. Smooth Jacobi evolution then ensures that
the deformation tensor and its trace $\theta$ remain
finite on this interval. Evaluating the integrated
inequality at $S$ gives
\begin{equation}
 \arctan\!\frac{\theta(S)}{m\sqrt{k}}\le-\frac{\pi}{2},
\end{equation}
which is impossible for finite $\theta(S)$.
The transverse volume must therefore vanish at some
$s_*\in(s_0,S]$, establishing the stated focusing bound.

 The loss of
regularity thus occurs through collapse of the transverse
volume, with $\theta\to-\infty$ on approach from the
incoming branch. If $\theta_0<0$, integrating the weaker
inequality $\dot\theta\le-\theta^2/m$ gives
$s_*-s_0\le m/|\theta_0|$.
\end{proof}

This bound uses only the Ricci curvature, which sums over transverse directions. The collapse concerns the transverse separation map, while individual geodesics continue smoothly in the complete projective state space. For a complete point-emitted family, a caustic identifies a conjugate point to the source. A segment extended beyond that conjugate point is no longer locally length minimizing with fixed endpoints. A caustic of a more general congruence does not, by itself, imply this loss of minimality or geodesic incompleteness.

\subsection{Exact deformation of a point-emitted geodesic bundle}

The focusing theorem applies to any regular,
full screen-spanning, twist-free geodesic congruence with specified initial expansion. It bounds the distance to collapse using the Ricci curvature, without resolving the individual transverse directions.

We now specialize to a complete family emitted from a common state. For this family, the full tidal operator determines the separation in each direction and hence the exact expansion, shear, and first focusing distance. The source itself is singular as a congruence: the separation vectors vanish there. We therefore impose the radial initial data directly on the Jacobi equation.

A variation of the initial direction produces a Jacobi field
with
\begin{equation}
 \eta(0)=0,\qquad
 (\nabla_U\eta)(0)=E(0),
\end{equation}
where $E(0)$ is a unit vector orthogonal to the initial
velocity. The second condition normalizes the angular separation of
neighboring paths. Allowing $E(0)$ to span the screen
describes the complete bundle. These variations remain
transverse, so $\cG=\cG_\perp$, and the bundle has zero twist.

Choose a parallel-transported orthonormal screen frame
with $E_{\mathbb C}=JU$ and
$E_a\perp\{U,JU\}$, $a=1,\ldots,m-1$.
Because the complex structure is parallel,
$\nabla J=0$, these directions retain their tidal
eigenvalues $4$ and $1$ along the geodesic. The Jacobi
equation therefore reduces to independent scalar equations of the form
$\ddot f_{\mathbb C}+4f_{\mathbb C}=0$ and
$\ddot f_\perp+f_\perp=0$, with $f(0)=0$ and $\dot f(0)=1$.
Their solutions give the separation fields
\begin{align}
 \eta_{\mathbb C}(s)
 &=\frac{\sin2s}{2}E_{\mathbb C}(s),
 \label{eq:CP-J-complex}\\
 \eta_a(s)&=\sin s\,E_a(s).
 \label{eq:CP-J-ordinary}
\end{align}
The derivation is detailed in \cref{app:CP-jacobi}.
Both amplitudes initially grow as $s$: the bundle opens
isotropically near its source, and curvature subsequently
produces unequal separations. The Jacobi field along $JU$, governed by sectional curvature $4$, also appears in geometric accounts of quantum uncertainty \cite{Benvegnu2007,Spera2011}.

\paragraph{Volume, expansion, and shear.}
Multiplying the separation amplitudes gives the transverse
volume per unit initial angular volume. Its logarithmic
derivative gives the expansion:
\begin{align}
 \cV(s)&=\frac{\sin2s}{2}(\sin s)^{m-1},
 \label{eq:CP-volume}\\
 \theta(s)&=2\cot2s+(m-1)\cot s.
 \label{eq:CP-expansion}
\end{align}
These expressions indicate the interval
$0<s<\pi/2$ over which the paths are regular.

The principal deformation rates are the logarithmic
derivatives of the individual amplitudes:
$b_{\mathbb C}=2\cot2s$ and $b_\perp=\cot s$.
Their difference, $b_{\mathbb C}-b_\perp=-\tan s$,
shows that the complex direction expands more slowly
and begins contracting earlier. Subtracting the mean
rate $\theta/m$ gives
\begin{equation}
 \begin{aligned}
 \sigma_{\mathbb C}&=-\frac{m-1}{m}\tan s,
 &\sigma_\perp&=\frac1m\tan s,\\
 \norm{\sigma}^2&=\frac{m-1}{m}\tan^2s.
 \end{aligned}
 \label{eq:CP-shear-exact}
\end{equation}
The ordinary transverse sector is absent for $m=1$.
Near the source,
$\sigma=-s\cE_{\TF}/3+O(s^3)$: shear tends to zero
at emission but develops immediately when $N\ge3$.
Although this shear limit is finite, the deformation
tensor itself diverges as $\cB\sim I_m/s$.
A point-emitted bundle therefore differs from a bundle
with finite isotropic deformation at a regular point,
as considered in \eqref{eq:shear-initial-source}.

\begin{proposition}[Universal shear of point-emitted bundles]
For a complete unit-speed geodesic family emitted from
one state in $\CP^{N-1}$, with Hilbert-space dimension
$N\ge3$, the shear satisfies
\begin{equation*}
 \norm{\sigma}^2=\frac{m-1}{m}\tan^2s>0, \quad 0<s<\pi/2,
\end{equation*}
where $m=2N-3.$
Thus a point-emitted family of geodesics necessarily develops shear
because the complex and ordinary transverse directions
experience unequal tidal curvature.
\label{prop:CP-universal-shear}
\end{proposition}

\begin{proposition}[First conjugate point in pure-state space]
Every unit-speed geodesic in $\CP^{N-1}$, $N\ge2$,
has its first conjugate point at $s_*=\pi/2$.
The complete point-emitted family loses exactly one
independent transverse direction there in the sense that the $JU$
Jacobi field vanishes, while the remaining $2N-4$
fields stay nonzero.
\label{prop:CP-first-conjugate}
\end{proposition}

This classical conjugate-locus result
\cite{Helgason2001,Berceanu1997} follows directly from the first zeros in \cref{eq:CP-J-complex,eq:CP-J-ordinary}. The endpoint is orthogonal to the initial ray. The segment from the source remains length minimizing up to
this endpoint. Extending that segment beyond the first
conjugate point destroys its local minimizing property
with fixed endpoints.

The collapsing direction has a simple
interpretation. Consider the neighboring geodesics
\[
 |\psi(s,\alpha)\rangle
 =\cos s\,|u_0\rangle
  +e^{i\alpha}\sin s\,|v\rangle,
 \qquad \langle u_0|v\rangle=0,
\]
where both $|u_0\rangle$ and $|v\rangle$ are normalized. The label $\alpha$ changes the relative phase between the two components. At $s=\pi/2$, this phase becomes an overall phase, so varying $\alpha$ no longer changes the endpoint projector. For $N\ge3$, variations of the target ray $|v\rangle$ within the orthogonal complement of $|u_0\rangle$ remain distinguishable. For $N=2$, that complement contains only one ray.
The Fisher-information calculation below quantifies this selective loss of sensitivity.

\subsection{Information-theoretic signatures: fidelity and quantum Fisher information}
\label{sec:fidelity-qfi-signatures}

We now apply the fidelity and QFI relations
\eqref{eq:label-fidelity} and \eqref{eq:label-qfi} to the point-emitted projective family. Its connecting fields are transverse, so $\cG=\cG_\perp$: the geometric deformation directly describes changes in sensitivity to the initial geodesic direction. For more general families, the full and transverse sensitivities must be distinguished.

For a regular pure-state family spanning the full screen,
$\cI_\perp=4\cG_\perp$  and 
$\theta=\frac12\partial_s\ln\det\cI_\perp.$
Thus expansion measures the fractional change of transverse
Fisher volume. More locally, in fixed label coordinates, the
eigenvalues of $\tfrac12\cG_\perp^{-1}\dot\cG_\perp$ give the
principal instantaneous stretching rates and subtracting their
mean $\theta/m$ gives the shear eigenvalues. Extracting shear
therefore requires the evolution of the metric, rather than only the
Fisher eigenvalues at two endpoints. 

For the point-emitted projective family, the Jacobi amplitudes
in \cref{eq:CP-J-complex,eq:CP-J-ordinary} give
\begin{equation}
 F^{\mathrm Q}_{\mathbb C}=\sin^2(2s),\qquad
 F^{\mathrm Q}_\perp=4\sin^2s.
 \label{eq:CP-QFI-eigenvalues}
\end{equation}
These eigenvalues use the orthonormal initial angular variations
specified above. Their zeros and rank loss are invariant under
nonsingular relabeling. The corresponding fidelity susceptibilities
are one quarter of these values.

At a caustic of a point-emitted pure-state family, the QFI loses rank:
at least one label combination no longer changes the output
state to first order. But a higher-order sensitivity can remain,
and states at finite label separation need not coincide.

A nontransverse family can instead have
\begin{equation}
 \begin{gathered}
 \eta_c^\perp(s_*)=0,\qquad \eta_c(s_*)=a_cU\ne0,\\
 c^A\cI_{AB}c^B=4a_c^2>0.
 \end{gathered}
\end{equation}
This is a transverse caustic: its transverse metric is
singular, while its fixed-$s$ QFI can remain nonsingular.
An infinitesimal arclength adjustment cancels the remaining
longitudinal displacement. This distinction is realized by
the geodesic bundles on the phase diagram of the Transverse field XY spin chain in \cref{sec:xy}.

\section{Schr\"odinger evolution under a common Hamiltonian}
\label{sec:schrodinger-rigidity}

The focusing established above raises a natural question:
how is it consistent with unitary evolution, which preserves
overlaps between states? The distinction lies in how the
neighboring paths are generated. Each geodesic can be
implemented by the mixing Hamiltonian
$H_{\rm mix}$ in \cref{eq:mixing-CD}, but this Hamiltonian
generally depends on the initial direction. The bundle
therefore compares different quantum transformations.
We now consider a family of initial states evolving under
the same prescribed Hamiltonian and show how its rigid
transport is expressed in the congruence equations.

For a time- and label-independent $H=H^\dagger$, set
$\tau=t/\hbar$ and write
\begin{equation}
 K_H(P)=-\ii[H,P],\qquad
 \Phi_\tau(P)=\ee^{-\ii H\tau}P\ee^{\ii H\tau}.
 \label{eq:schrodinger-killing-flow}
\end{equation}
Unitary conjugation preserves the metric, symplectic form,
and complex structure. Its generator $K_H$ is therefore
a Killing field: its flow preserves geometric distances.
For an initial family $P_0(\lambda)$,
$P(\tau,\lambda)=\Phi_\tau(P_0(\lambda))$, and every
label derivative transforms by the same conjugation.
Cyclicity of the trace then gives
\begin{equation}
 \cG_{AB}(\tau)=\cG_{AB}(0).
 \label{eq:unitary-qfi-rigidity}
\end{equation}
The pure-state QFI is consequently unchanged.
This conclusion also holds for a common time-dependent
unitary propagator: a nonsingular initial family cannot
develop an equal-time QFI caustic.

\paragraph{Rigid transport in the congruence equations.}
To understand this rigidity locally, return to the
time-independent Hamiltonian and normalize its velocity
away from stationary projectors:
\begin{equation}
 \begin{aligned}
 f&=\norm{K_H},\qquad \hat U=K_H/f,\\
 f^2&=\Tr(PHQHP)=(\Delta_P H)^2.
 \end{aligned}
 \label{eq:normalized-killing-field}
\end{equation}
Here $Q=I-P$. The speed is constant along each trajectory,
$K_H(f)=0$, although different trajectories can have
different speeds.

Schr\"odinger trajectories in general need not be geodesics.
Accordingly, define the deformation tensor by projecting
the change of $\hat U$ onto its transverse screen:
\begin{equation*}
 \cB(V)=\mathsf \Pi\nabla_V\hat U,\quad
 \mathsf \Pi X=X-g(X,\hat U)\hat U,\quad V\perp\hat U.
\end{equation*}
Differentiating the normalization $1/f$ contributes only
along $\hat U$, which the screen projection removes.
The Killing equation then makes the symmetric part of
$\cB$ vanish:
\begin{equation}
 \theta=0,\qquad \sigma_{ab}=0.
 \label{eq:killing-kinematics}
\end{equation}
Thus the transverse bundle neither expands nor changes
shape, although it can rotate.

The trajectories can nevertheless accelerate within
state space. Using
$\nabla_{K_H}K_H=-\tfrac12\nabla f^2$ gives
\begin{equation}
 a:=\nabla_{\hat U}\hat U=-\nabla\ln f,
 \qquad
 \nabla_\mu a^\mu=-\Delta\ln f,
 \label{eq:killing-acceleration}
\end{equation}
where $\Delta=\nabla_\mu\nabla^\mu$.
The gradient is transverse because $\hat U(f)=0$.
This acceleration must be included when applying the
Raychaudhuri equation:
\begin{equation}
 \begin{aligned}
 \hat U(\theta)={}&-\frac{\theta^2}{m}
 -\norm{\sigma}^2+\norm{\varpi}^2\\
 &-\Ric(\hat U,\hat U)+\nabla_\mu a^\mu,
 \end{aligned}
 \label{eq:accelerated-raychaudhuri}
\end{equation}
with $m=2N-3$ and
$\theta=\nabla_\mu\hat U^\mu$.
This expression is independent of the screen frame.

Substituting the vanishing expansion and shear yields
the exact balance
\begin{equation}
 \norm{\varpi}^2
 =\Ric(\hat U,\hat U)+\Delta\ln f.
 \label{eq:twist-rigidity}
\end{equation}
The positive Ricci curvature is therefore balanced by
acceleration and twist, consistently with rigid
unitary transport.

For example, rotation about the Bloch sphere's $z$ axis
has $f\propto\sin\vartheta$ in the metric
$ds^2=(d\vartheta^2+\sin^2\vartheta\,d\phi^2)/4$.
Away from the poles, $\Delta\ln f=-4$ and
$\Ric(\hat U,\hat U)=4$. The screen is one-dimensional
and has no twist, so acceleration alone supplies the
balance.

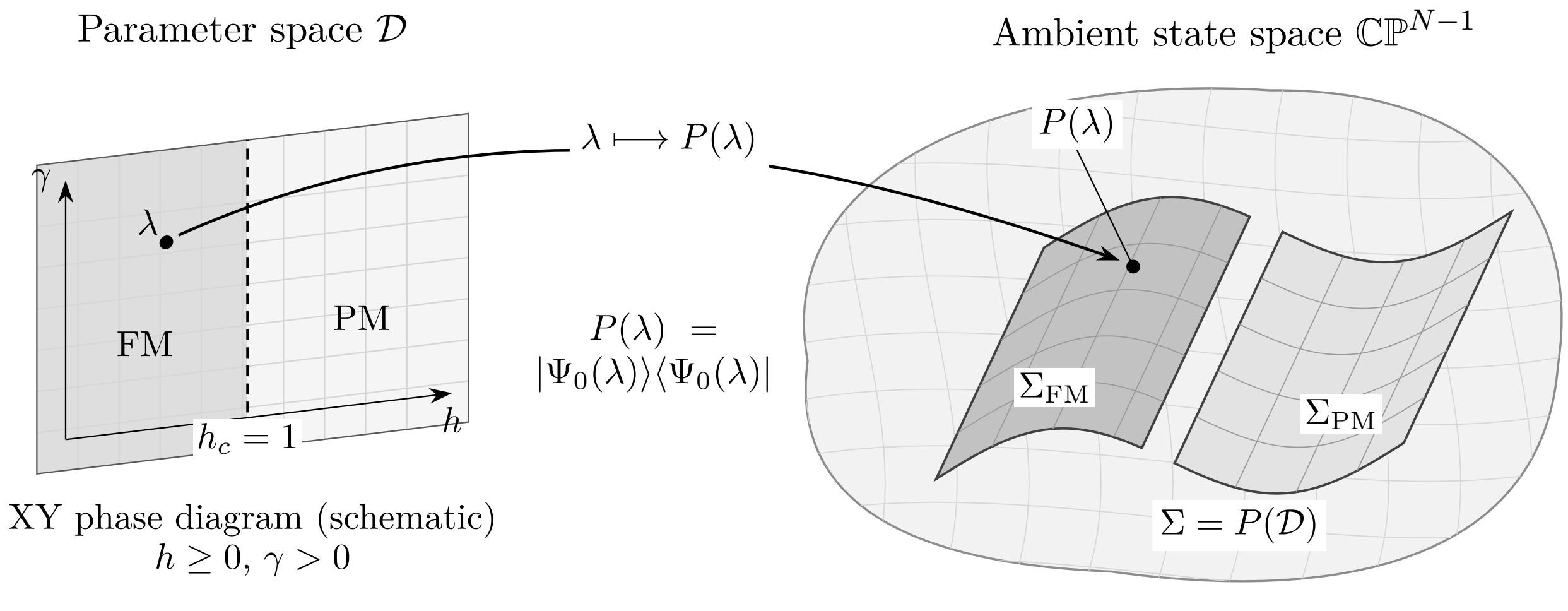
\begin{figure*}[t]
 \centering
 \resizebox{\textwidth}{!}{
% Requires tikz and \usetikzlibrary{arrows.meta,calc}.
\begin{tikzpicture}[x=1cm,y=1cm,font=\small,
 >={Stealth[length=2.2mm,width=1.5mm]},
 declare function={
 fx(\u,\v)=11.55+4.55*(\u-0.5)+1.05*(\v-0.5);
 fy(\u,\v)=2.55+2.25*(\v-0.5)+0.35*(\u-0.5)+0.40*sin(360*\u);
 }]
\path[use as bounding box] (-0.5,-0.30) rectangle (15.20,6.00);
\node[align=center,font=\normalsize] at (2.0,5.60)
 {Parameter space $\mathcal D$};
\node[align=center,font=\normalsize] at (11.65,5.60)
 {Ambient state space $\mathbb{CP}^{N-1}$};

% Ambient manifold: subdued curved grid.
\begin{scope}
 \path[fill=black!5,draw=black!45,line width=0.6pt]
 (7.5,2.4) .. controls (7.15,4.25) and (9.4,5.2) .. (12.0,5.03)
 .. controls (14.0,5.05) and (15.05,4.0) .. (14.80,2.35)
 .. controls (14.65,0.4) and (12.6,0.05) .. (10.45,0.35)
 .. controls (8.7,0.42) and (7.3,0.9) .. (7.5,2.4) -- cycle;
 \begin{scope}
 \clip (7.5,2.4) .. controls (7.15,4.25) and (9.4,5.2) .. (12.0,5.03)
 .. controls (14.0,5.05) and (15.05,4.0) .. (14.80,2.35)
 .. controls (14.65,0.4) and (12.6,0.05) .. (10.45,0.35)
 .. controls (8.7,0.42) and (7.3,0.9) .. (7.5,2.4) -- cycle;
 \foreach \y in {0.3,0.85,...,5.3}
  \draw[black!16,line width=0.3pt]
  (7.1,\y) .. controls (9.6,{\y+0.6}) and (12.8,{\y-0.6}) .. (15.1,\y);
 \foreach \x in {7.3,8.0,...,15.0}
  \draw[black!16,line width=0.3pt]
  (\x,0) .. controls ({\x+0.8},1.7) and ({\x-0.6},3.5) .. (\x,5.3);
 \end{scope}
\end{scope}

% Images of two regular parameter regions, separated at the boundary.
\foreach \lo/\hi/\shade in {0/0.44/24,0.51/1/11}{
 \path[fill=black!\shade,draw=black!75,line width=0.65pt]
 plot[domain=\lo:\hi,samples=30,variable=\t] ({fx(\t,0)},{fy(\t,0)}) --
 plot[domain=0:1,samples=25,variable=\t] ({fx(\hi,\t)},{fy(\hi,\t)}) --
 plot[domain=\hi:\lo,samples=30,variable=\t] ({fx(\t,1)},{fy(\t,1)}) --
 plot[domain=1:0,samples=25,variable=\t] ({fx(\lo,\t)},{fy(\lo,\t)}) -- cycle;
 \foreach \v in {0.2,0.4,0.6,0.8}{
  \draw[black!40,line width=0.3pt]
  plot[domain=\lo:\hi,samples=30,variable=\t] ({fx(\t,\v)},{fy(\t,\v)});
 }
}
\foreach \u in {0.12,0.25,0.38,0.64,0.77,0.90}
 \draw[black!40,line width=0.3pt]
 plot[domain=0:1,samples=25,variable=\t] ({fx(\u,\t)},{fy(\u,\t)});
\node[fill=white,inner sep=1.2pt] at ({fx(0.21,0.19)},{fy(0.21,0.19)}) {$\Sigma_{\rm FM}$};
\node[fill=white,inner sep=1.2pt] at ({fx(0.79,0.33)},{fy(0.79,0.33)}) {$\Sigma_{\rm PM}$};
\coordinate (Pimg) at ({fx(0.26,0.70)},{fy(0.26,0.70)});
\fill (Pimg) circle (1.9pt);
\draw[thin] (Pimg) -- (10.12,4.45);
\node[anchor=south,fill=white,inner sep=1.8pt] at (10.12,4.45) {$P(\lambda)$};
\node[align=center,fill=white,inner sep=2pt] at (11.7,0.80)
 {$\Sigma=P(\mathcal D)$};

% Parameter plane, with the same grid and phase partition as the source.
\begin{scope}[shift={(0,1.30)},cm={1,0.12,0,1,(0,0)}]
 \fill[black!13] (0,0) rectangle (2.05,3.0);
 \fill[black!4] (2.05,0) rectangle (4.20,3.0);
 \foreach \x in {0.4,0.8,...,4.0} \draw[black!16,thin] (\x,0)--(\x,3.0);
 \foreach \y in {0.4,0.8,...,2.8} \draw[black!16,thin] (0,\y)--(4.20,\y);
 \draw[black!65] (0,0) rectangle (4.20,3.0);
 \draw[densely dashed,line width=0.7pt] (2.05,0)--(2.05,3.0);
 \draw[->] (0.28,0.30)--(4.04,0.30) node[below] {$h$};
 \draw[->] (0.28,0.30)--(0.28,2.82) node[left] {$\gamma$};
 \node[below,fill=white,inner sep=1pt] at (2.05,0.29) {$h_c=1$};
 \node at (1.05,1.14) {FM};
 \node at (3.16,1.14) {PM};
 \coordinate (Lam) at (1.26,2.10);
 \fill (Lam) circle (1.9pt);
 \node[above left,inner sep=2pt] at (Lam) {$\lambda$};
\end{scope}
\node[align=center] at (2.10,0.67)
 {XY phase diagram (schematic)\\[-1pt] $h\geq0$, $\gamma>0$};

% Map from a regular point of the parameter plane to its projector.
\draw[->,line width=0.8pt]
 ($(Lam)+(0.12,0.07)$) .. controls (4.55,5.04) and (7.6,4.40) .. ($(Pimg)+(-0.13,0.06)$);
\node[fill=white,inner sep=3pt,align=center] at (6.15,4.55) {$\lambda\longmapsto P(\lambda)$};
\node[align=center,text width=3.0cm] at (6.0,2.48)
 {$P(\lambda)=\vert\Psi_0(\lambda)\rangle\langle\Psi_0(\lambda)\vert$};
\end{tikzpicture}
}
 \caption{A physical parameter space and its image in quantum state space.
 The left panel illustrates the ferromagnetic (FM) and paramagnetic (PM)
 regions of the transverse-field XY phase diagram for $h\geq0$ and
 $\gamma>0$. A chosen ground-state branch assigns a rank-one projector
 $P(\lambda)$ to each regular parameter point. Its image forms a
 parametrised state manifold $\Sigma\subset\mathbb{CP}^{N-1}$. The induced
 quantum metric and Berry curvature describe variations along this
 image, while its second fundamental form describes bending within
 the ambient state space. }
 \label{fig:ground-state-embedding}
\end{figure*}

\section{Extrinsic geometry in congruence kinematics}
\label{sec:extrinsic}

The preceding focusing results concern geodesics of the full pure-state space. Physical parameters usually select a restricted family of states, and paths constrained to that family follow its induced metric. We now ask how the curvature governing their separation differs from the ambient projective curvature. The Gauss equation answers this question by adding the contribution from the bending of the parametrised family.

For example, a Hamiltonian $H(\lambda)$ selects a ground-state family through its fields and couplings, while crystal momentum parameterizes the Bloch states of an isolated band. On a smooth region with a chosen nondegenerate state, these parameters define a map into quantum state space,
\begin{equation}
 P:\mathcal D\longrightarrow\CP^{N-1},
 \qquad \lambda\longmapsto P(\lambda).
 \label{eq:physical-state-map}
\end{equation}
For a nondegenerate ground state,
$P(\lambda)=\ket{\Psi_0(\lambda)}\bra{\Psi_0(\lambda)}$ has rank one.

The geometry below applies on regular regions where $P(\lambda)$ is
smooth and its tangent map has full rank, so the induced quantum
metric is nondegenerate. After restricting to a locally embedded
patch, we identify that parameter region with its image $\Sigma$ in
state space (Figure~\ref{fig:ground-state-embedding}). This is a locally restricted construction and we have not addressed the problems when there is no smooth continuation through a phase boundary or a
rank-changing degeneracy.

Suppose we have a family of projectors \(P(\lambda)\), where \(\lambda=(\lambda^1,\lambda^2,\ldots)\) denotes a set of physical parameters. A small change in these parameters produces the tangent displacement \(dP=(\partial_\mu P)\,d\lambda^\mu\). Evaluating the metric of Eq. \ref{eq:RiemannMetric} and symplectic form \ref{eq:SymplecticForm} on such displacements defines the counterparts on the parameter space, a construction known as the \emph{pullback}:
\begin{align}
g_{\mu\nu}&=\frac12\Tr(\partial_\mu P\,\partial_\nu P),
  \label{eq:qmetric-projector}\\
  \cF_{\mu\nu}&=\ii\Tr\!\left(P[\partial_\mu P,\partial_\nu P]\right),
  \label{eq:berry-projector}\\
  \omega_{\mu\nu}&=-\frac12\cF_{\mu\nu}.
  \label{eq:omega-half-F}
\end{align}
The one-band quantum geometric tensor is consequently
\begin{equation}
  \cQ_{\mu\nu}=g_{\mu\nu}-\frac{\ii}{2}\cF_{\mu\nu}.
  \label{eq:qgt-convention}
\end{equation}
Here $\cF$ is the well-known Berry curvature, while $\omega=-\cF/2$ is the normalized Fubini-Study K\"ahler form. The induced metric determines the \emph{intrinsic geometry} of the embedding of the image of the parameter space in the full state-space. The Berry curvature is additional geometric data inherited from its embedding in projective space

A geodesic of the induced metric locally minimizes length among
paths confined to $\Sigma$, but need not be a geodesic of the full
quantum state space. The reason is that the parametrised state family
can bend within the ambient space: a direction tangent to $\Sigma$
can change, as we move along the family, partly towards directions
outside its tangent space. The \emph{extrinsic geometry} describes how $\Sigma$ bends within
the ambient quantum state-space. This bending is measured locally by the \emph{second fundamental form}, which captures the normal component of the change of tangent vectors \cite{DajczerTojeiro2019}.

\subsection{The second fundamental form of the parametrised state manifold}

On the sub-manifold of $P(\lambda)$, the derivatives
$e_\mu=\partial_\mu P$ span the tangent space of $\Sigma$.
This is a subspace of the tangent space of the full pure-state
manifold $\mathbb{CP}^{N-1}\simeq\Gr(1,N)$. At each
$P=P(\lambda)$, the latter splits orthogonally as
\begin{equation}
 T_P\mathbb{CP}^{N-1}
 =
 \underbrace{\operatorname{span}\{e_\mu\}}_{T_P\Sigma}
 \oplus
 \underbrace{N_P\Sigma}_{\text{normal to }\Sigma}.
 \label{eq:tangent-normal-splitting}
\end{equation}
The normal directions are valid infinitesimal changes of the
quantum state, but cannot be produced by an infinitesimal change
of the available parameters. The Gauss formula separates the change of a tangent vector into
its tangential and normal parts:
\begin{equation}
\begin{aligned}
 \nabla^{\rm FS}_{\mu}e_\nu
 &=P(\partial_\mu\partial_\nu P)Q
   +Q(\partial_\mu\partial_\nu P)P\\
 &=\Gamma^\rho_{\mu\nu}e_\rho+\Pi_{\mu\nu},
 \qquad Q=I-P.
\end{aligned}
\label{eq:gauss-formula}
\end{equation}
Here $\nabla^{\rm FS}_{\mu}$ denotes the ambient Fubini-Study
covariant derivative along $e_\mu$, and
$\Gamma^\rho_{\mu\nu}$ is the Levi-Civita connection of the
induced metric $g_{\mu\nu}=g(e_\mu,e_\nu)$.
The first term describes the change within the tangent space
of $\Sigma$. The remainder
$\Pi_{\mu\nu}\in N_P\Sigma$ is its second fundamental form,
symmetric in $\mu,\nu$, which is along the normal and describes the bending of $\Sigma$ within quantum state space. In particular, for an intrinsic geodesic with velocity $U$,
\begin{equation}
 \nabla^{\rm FS}_U U=\Pi(U,U).
 \label{eq:geodesic-normal-acceleration}
\end{equation}
Thus a path that is geodesic within $\Sigma$ can still have a normal acceleration when viewed in the full
state space. The projector-based calculation is detailed in
\cref{app:projector}.

\subsection{Extrinsic and Berry curvature effects in Raychaudhuri-Ehlers kinematics}

Now we consider the contribution of extrinsic geometric effects to the curvature structure and further to the congruence kinematics. Particularly interesting is the question whether extrinsic effects preserve or spoil the inevitable focusing result that we have obtained in the previous sections.

The curvature experienced by paths confined to $\Sigma$ depends
both on the curvature of the surrounding quantum state space
and on how $\Sigma$ bends within it. The Gauss equation relates
these contributions. Let $e_\mu=\partial_\mu P$, and define curvature components by
$R^\Sigma_{\mu\nu\rho\sigma}
=g(R^\Sigma(e_\mu,e_\nu)e_\sigma,e_\rho)$.
Then
\begin{equation}
\begin{aligned}
 R^\Sigma_{\mu\nu\rho\sigma}
 ={}&\widehat R_{\mu\nu\rho\sigma}
 +\langle\Pi_{\mu\rho},\Pi_{\nu\sigma}\rangle\\
 &-\langle\Pi_{\mu\sigma},\Pi_{\nu\rho}\rangle.
\end{aligned}
\label{eq:Gauss}
\end{equation}
Here
$\widehat R_{\mu\nu\rho\sigma}
=g_{\rm FS}(R^{\rm FS}(e_\mu,e_\nu)e_\sigma,e_\rho)$
is the ambient Fubini-Study curvature evaluated on tangent
directions of the parametrised family, and
$\langle\cdot,\cdot\rangle$ is the Fubini-Study inner product
on its normal vectors. The two contractions of $\Pi$ describe
the curvature contribution from bending.

In our metric normalization, the ambient curvature operator is
\begin{equation}
\begin{aligned}
 R^{\rm FS}(X,Y)Z={}&g_{\rm FS}(Y,Z)X-g_{\rm FS}(X,Z)Y\\
 &+g_{\rm FS}(JY,Z)JX-g_{\rm FS}(JX,Z)JY\\
 &+2g_{\rm FS}(X,JY)JZ.
\end{aligned}
\label{eq:FS-curvature-operator}
\end{equation}
Its restriction to the tangent space of $\Sigma$ is
therefore determined by the induced metric
$g_{\mu\nu}=g_{\rm FS}(e_\mu,e_\nu)$ and the pulled-back
symplectic form
$\omega_{\mu\nu}=g_{\rm FS}(Je_\mu,e_\nu)
=-\cF_{\mu\nu}/2$:
\begin{equation}
\begin{aligned}
 \widehat R_{\mu\nu\rho\sigma}
 ={}&g_{\mu\rho}g_{\nu\sigma}
     -g_{\mu\sigma}g_{\nu\rho}\\
 &+\omega_{\mu\rho}\omega_{\nu\sigma}
     -\omega_{\mu\sigma}\omega_{\nu\rho}\\
 &+2\omega_{\mu\nu}\omega_{\rho\sigma}.
\end{aligned}
\label{eq:FS-curvature-pullback}
\end{equation}
The first two terms have the form of a space with constant
sectional curvature $1$. The remaining terms encode the
direction-dependent enhancement associated with the complex
structure, expressed here through the Berry curvature.
The Gauss equation adds the bending contribution to obtain
the intrinsic curvature governing geodesic deviation within
$\Sigma$.

Let $U$ be the unit velocity of a geodesic in $\Sigma$, and
let $\{E_a\}_{a=1}^m$ be an orthonormal screen, with
$d=\dim\Sigma$, $m=d-1$, and $h_{ab}=g(E_a,E_b)=\delta_{ab}$.
The ambient complex direction $JU$ need not be tangent to
$\Sigma$. Its accessible components are
\begin{equation}
 v_a=g_{\rm FS}(JU,E_a)=\omega(U,E_a)
     =-\frac12\cF(U,E_a).
 \label{eq:v-Berry}
\end{equation}
Here $\omega$ and $\cF$ are evaluated on tangent vectors of
$\Sigma$. The sign follows from our conventions
$JX=-i[P,X]$ and $\cF(X,Y)=i\Tr(P[X,Y])$.

Evaluating the Gauss equation on the velocity and screen
directions gives the intrinsic tidal tensor
$\cE_{ab}=g(R^\Sigma(E_b,U)U,E_a)$:
\begin{equation}
 \begin{aligned}
 \cE_{ab}={}&h_{ab}+3v_av_b
 +\langle\Pi_{ab},\Pi_{UU}\rangle\\
 &-\langle\Pi_{aU},\Pi_{bU}\rangle.
 \end{aligned}
 \label{eq:tidal-extrinsic}
\end{equation}
We write $\Pi_{ab}=\Pi(E_a,E_b)$,
$\Pi_{aU}=\Pi(E_a,U)$, and $\Pi_{UU}=\Pi(U,U)$.
This decomposition identifies how Berry curvature and
extrinsic bending enter the relative acceleration of
neighboring geodesics.

The ambient contribution consists of an isotropic term
$h_{ab}$ and a \emph{Berry-curvature enhancement} $3v_av_b$.
For $v\ne0$, its eigenvalues are
\begin{equation}
\begin{aligned}
 1+3|v|^2
 &\quad \text{along }v^aE_a,\\
 1
 &\quad \text{orthogonal to }v^aE_a.
\end{aligned}
\label{eq:Berry-tidal-enhancement}
\end{equation}
Here $|v|^2=h^{ab}v_av_b$. Since $J$ preserves the ambient
norm and $g_{\rm FS}(JU,U)=0$,
\begin{equation}
 |v|^2
 =\norm{\mathsf P_\Sigma(JU)}_{\rm FS}^2
 \le1.
 \label{eq:Berry-enhancement-bound}
\end{equation}
The enhancement is therefore at most three, reaching its
maximum when $JU$ is tangent to $\Sigma$. It vanishes when
$JU$ is normal to $\Sigma$. 

Extrinsic bending supplies the remaining two terms in
\cref{eq:tidal-extrinsic}. The contraction
$\langle\Pi_{ab},\Pi_{UU}\rangle$ has no definite sign,
whereas $-\langle\Pi_{aU},\Pi_{bU}\rangle$ is negative
semidefinite. Together they can strengthen, weaken, or cancel
the ambient tide and can change its principal directions.

Tracing over the full accessible screen gives
\begin{equation}
\begin{aligned}
 \Ric^\Sigma(U,U)
 ={}&m+3|v|^2+\langle\mathbf H,\Pi_{UU}\rangle\\
 &-\sum_{i=0}^{d-1}\norm{\Pi(U,e_i)}^2,
\end{aligned}
\label{eq:Ricci-extrinsic}
\end{equation}
where $\{e_i\}=\{U,E_1,\ldots,E_m\}$ and
$\mathbf H=\sum_{i=0}^{d-1}\Pi(e_i,e_i)$ is the unaveraged
trace of the second fundamental form.
The sum includes $e_0=U$: its contribution accounts for
$\sum_a\Pi(E_a,E_a)=\mathbf H-\Pi_{UU}$.
The ambient contribution $m+3|v|^2$ is the trace of
the ambient tidal tensor over the accessible screen
$T_P\Sigma\cap U^\perp$.

Let the screen basis be parallel transported along the
geodesic using the intrinsic connection of $\Sigma$.
Substituting the tidal tensor into the Raychaudhuri-Ehlers equations then gives
\begin{equation}
\begin{aligned}
 \dot\theta={}&-\frac{\theta^2}{m}
 -\norm{\sigma}^2+\norm{\varpi}^2\\
 &-m-3|v|^2-\langle\mathbf H,\Pi_{UU}\rangle\\
 &+\sum_{i=0}^{d-1}\norm{\Pi(U,e_i)}^2
\end{aligned}
\label{eq:raychaudhuri-extrinsic}
\end{equation}
The term $-3|v|^2$ strengthens the ambient curvature contribution toward decreasing expansion, giving a Berry-curvature enhancement of focusing. Extrinsic bending can reinforce or counteract this contribution, and can even make the intrinsic Ricci curvature negative. The focusing guarantee for complete point-emitted
families in the full projective state space therefore does not automatically extend to the parameter-accessible manifold $\Sigma$.
Its embedding can enhance focusing or favor defocusing, also depending on the initial congruence and its shear and twist.

\begin{equation}
\begin{aligned}
 \dot\sigma_{ab}={}&-\frac{2\theta}{m}\sigma_{ab}
 -3(v_av_b)_{\TF}\\
 &-\bigl(\sigma_a{}^c\sigma_{cb}
         +\varpi_a{}^c\varpi_{cb}\bigr)_{\TF}\\
 &-\bigl(\langle\Pi_{ab},\Pi_{UU}\rangle\bigr)_{\TF}\\
 &+\bigl(\langle\Pi_{aU},\Pi_{bU}\rangle\bigr)_{\TF}.
\end{aligned}
\label{eq:shear-extrinsic}
\end{equation}
The trace-free operation acts on the screen indices $a,b$. For $m>1$ and $v\ne0$, the Berry contribution has a nonzero
trace-free part but the extrinsic bending can nevertheless cancel
this anisotropy. For initially isotropic, twist-free data,
\cref{eq:shear-extrinsic} reduces initially to
$\dot\sigma_{ab}=-(\cE_{ab})_{\TF}$: it is the total
trace-free tide that determines the onset of shear. As a particularly interesting case, in Sec. \ref{sec:xy}, we
identify the cancellation of the collective ambient Berry
term by part of the extrinsic part. Thus, the onset of shear could be completely sourced by the extrinsic terms. The twist equation remains \cref{eq:vorticity-intrinsic},
because the tidal tensor is symmetric and supplies no direct
twist source.

\subsection{Extrinsic geometry and the variation of the mixing generator}
\label{sec:generator-bending}

The same second fundamental form that enters the tidal
curvature also determines how the mixing generator must change
to follow a parametrised geodesic. Let $P(s)$ be a unit-speed
geodesic of $\Sigma$, with $U=\partial_sP$. Its anti-Hermitian
mixing generator is
\begin{equation}
\begin{aligned}
 X&=[U,P],\qquad U=[X,P],\\
 \norm{X}_{\rm ah}^2
 &:=-\tfrac12\Tr(X^2)=1.
\end{aligned}
\label{eq:horizontal-mixing-generator}
\end{equation}
This generator contains only mixing between the selected state
and its orthogonal complement and we
relate its variation to the bending of the parametrised family.

Along an intrinsic geodesic, the Gauss formula gives
$\nabla^{\rm FS}_U U=\Pi(U,U)$. The ordinary matrix
acceleration $\partial_s^2P$ additionally contains diagonal
blocks fixed by the projector constraint, which commute
with $P$. Differentiating $X=[\partial_sP,P]$
therefore gives
\begin{equation}
 \begin{aligned}
  \partial_sX&=[\Pi(U,U),P],\\
  \norm{\partial_sX}_{\rm ah}^2
  &=\norm{\Pi(U,U)}_{\rm FS}^2.
 \end{aligned}
 \label{eq:generator-bending}
\end{equation}
The norm equality follows from
$-\tfrac12\Tr[Z,P]^2=\tfrac12\Tr Z^2$
for any off-diagonal Hermitian tangent vector $Z$.

Equation~\eqref{eq:generator-bending} gives bending a direct
transport interpretation. An ambient projective geodesic has
a constant mixing generator. An intrinsic geodesic with
$\Pi(U,U)\ne0$ instead requires a changing generator to remain
within $\Sigma$, even though it locally minimizes length among
accessible paths. The bending norm is precisely the rate of
change of this generator per unit arclength.

The relation extends to variations across the whole family.
Define $X_\nu=[\partial_\nu P,P]$. Projecting its
derivative onto the off-diagonal anti-Hermitian blocks and
subtracting the induced connection gives
\begin{equation}
\begin{aligned}
 \mathcal D_\mu X_\nu
 :={}&P(\partial_\mu X_\nu)Q
      +Q(\partial_\mu X_\nu)P\\
 &-\Gamma^\rho_{\mu\nu}X_\rho\\
 ={}&[\Pi_{\mu\nu},P].
\end{aligned}
\label{eq:generator-bending-tensor}
\end{equation}
The term $[\partial_\nu P,\partial_\mu P]$ arising on
differentiation is block diagonal and is removed by this
projection. Consequently,
\begin{equation}
\begin{aligned}
 \langle\Pi_{\mu\nu},\Pi_{\rho\sigma}\rangle
 =-\tfrac12\Tr\bigl[
  (\mathcal D_\mu X_\nu)
  (\mathcal D_\rho X_\sigma)\bigr].
\end{aligned}
\label{eq:generator-normal-contractions}
\end{equation}
Thus the normal contractions entering the Gauss equation also quantify the covariant variation of the mixing generators. 

For a physical-time protocol $s=s(t)$, the minimal
counterdiabatic representative in \cref{eq:mixing-CD} is
$H_{\rm cd}=i\hbar\dot s\,X(s)$ \cite{DemirplakRice2003,Berry2009}.
Since $X$ has constant norm,
$X$ and $\partial_s X$ are orthogonal in the
anti-Hermitian inner product. Hence
\begin{equation}
 \frac{\Tr[(\partial_tH_{\rm cd})^2]}{2\hbar^2}
 =\ddot s^{\,2}
  +\dot s^{\,4}\norm{\Pi(U,U)}_{\rm FS}^2.
 \label{eq:CD-rate-bending}
\end{equation}
The first term comes from changing the speed along the path, while
the second persists at constant speed and measures the
change of the transport Hamiltonian required by bending.
For the XY-spin chain we explicitly obtain both the bending and its generator interpretation in
\cref{eq:XY-bending-exact,eq:XY-CD-rate}.

\section{Transverse-field XY spin chain: Extrinsic - Ambient cancellations and Fold catastrophes}
\label{sec:xy}

We now apply the ambient-extrinsic decomposition to an exactly solvable many-body state family. The XY ground state's product representation allows the normal derivatives and their tidal contributions to be evaluated explicitly. We first establish a general product-state cancellation and resolve the surviving tide at the zero-field Ising point at finite size. We then use the metric in thermodynamic limit to follow geodesic families over finite distances and identify their folds and fidelity signatures.

The spin-$1/2$ XY chain in a transverse field is a paradigmatic
exactly solvable quantum many-body model, providing a common
setting for studying quantum phase transitions, entanglement,
and nonequilibrium dynamics.
\cite{LiebSchultzMattis1961,BarouchMcCoyDresden1970,BarouchMcCoy1971,Dutta2015}.
 Early studies of ground-state overlaps and free-fermion fidelity showed how enhanced sensitivity to parameter changes diagnoses quantum criticality
\cite{ZanardiPaunkovic2006,CozziniGiordaZanardi2007}. 
Geometric-phase studies have also probed the same critical structure
\cite{CarolloPachos2005,Zhu2006,ChengGongGuoZhou2017}.
Its quantum metric, curvature, geodesics, and critical scaling have
subsequently been studied from several complementary perspectives
\cite{KumarMahapatraPhukonSarkar2012,VenutiZanardi2007,
Kolodrubetz2013,MaityMahapatraSarkar2015,LiskaGritsev2021,
MeraPaunkovicAminVieira2022}. Geodesic lengths in the ground-state quantum metric have also
been studied as diagnostics of quantum criticality, under the
name Fubini-Study complexity
\cite{JaiswalGautamSarkar2021}.

Consider a periodic chain with $L$ sites ($L$ is even), restricted to its conserved
even parity sector:
\begin{equation}
\begin{aligned}
H(\gamma,h,0)
 &=-\sum_j\Bigg[
 \frac{1+\gamma}{2}\sigma_j^x\sigma_{j+1}^x\\
 &\qquad+\frac{1-\gamma}{2}\sigma_j^y\sigma_{j+1}^y
 +h\sigma_j^z\Bigg],\\
H(\gamma,h,\phi)&=e^{-i\phi\hat G}H(\gamma,h,0)e^{i\phi\hat G}.
\end{aligned}
\label{eq:XY-H}
\end{equation}
Here $\hat G=\hat N_f/2=L I/4-\sum_j\sigma_j^z/4$.
The scalar term $L I/4$ in $\hat G$ contributes only an overall
phase to the rotated state vector. It leaves both the state
projector and the conjugated Hamiltonian unchanged. The parameter $\phi$ specifies the orientation of the exchange
anisotropy in the $xy$ plane: changing $\phi$ rotates its principal
axes through an angle $\phi/2$
\cite{CarolloPachos2005,Zhu2006,ChengGongGuoZhou2017,
Kolodrubetz2013}. This leaves the spectrum unchanged but generally changes the ground-state projector relative to fixed spin axes. It provides an
isospectral direction enlarging the ground state manifold to three dimensions, allowing a two-dimensional screen with nontrivial shear. 

The geometric object of interest is the rank-one projector onto the even-parity sector ferromagnetic many-body ground state. We take the ambient state space to be
$\CP^{2^L-1}$, the space of pure states of the full $L$-spin
Hilbert space. 

Varying $(h,\gamma,\phi)$ traces out a three-dimensional
ground-state manifold within the ambient projective space.
Its bending is measured by the second fundamental form $\Pi$.

\subsection{Ground-state quantum metric for finite-size and thermodynamic limit.}

The selected ground-state family admits an exact product
representation over momentum pairs after the Jordan-Wigner
and Bogoliubov transformations:
\begin{equation}
\begin{aligned}
\ket{\Omega(\gamma,h,\phi)}=\bigotimes_{k>0}\bigg[ &\cos\frac{\vartheta_k}{2}\ket{00}_{k,-k} \\
 &+e^{-i\phi}\sin\frac{\vartheta_k}{2}\ket{11}_{k,-k}\bigg].
 \end{aligned}
 \label{eq:XY-product-state}
\end{equation}
Here $\ket{00}_{k,-k}$ and $\ket{11}_{k,-k}$ denote
the empty and doubly occupied fermionic momentum pair,
respectively in the Jordan-Wigner representation. The Bogoliubov angle is fixed by
\begin{equation}
 \begin{aligned}
 \cos\vartheta_k&=\frac{h-\cos k}{\epsilon_k}, \quad \sin\vartheta_k=\frac{\gamma\sin k}{\epsilon_k},\\
 \epsilon_k&=\sqrt{(h-\cos k)^2+\gamma^2\sin^2k}.
 \end{aligned}
 \label{eq:XY-Bogoliubov-angle}
\end{equation}
Each factor in the product \ref{eq:XY-product-state} is therefore a two-level state on a Bloch
sphere, whose position depends on the same parameters
$(\gamma,h,\phi)$. This product representation simplifies
the calculation of the many-body quantum geometry. Parameter variations mix the
occupied and unoccupied eigenstates within each Bogoliubov-de
Gennes block, and the resulting metric contributions add to give
the total many-body quantum metric $g^{\mathrm{tot}}$.

To distinguish the sensitivity of the full state from its value
per spin, we define
\begin{equation}
  g^{(L)}=\frac{g^{\mathrm{tot}}}{L},
  \qquad
  g=\lim_{L\to\infty}g^{(L)}.
\end{equation}
Here $g^{(L)}$ is the finite-size metric per site, while $g$ is
its thermodynamic limit. Exact fidelity and fidelity-susceptibility
analyses provide complementary results at finite size and in the
thermodynamic limit
\cite{RamsDamski2011,Damski2013,LuoZhaoWang2018}.
The finite-size mode sums and the derivatives required for our calculation are given in \cref{app:XY-finite}.

In the ferromagnetic region $|h|<1$, $\gamma>0$, the coordinates
$\gamma=\tan^2v$ and $h=\sin u$ put the thermodynamic-limit metric per
site in the form
\cite{ZanardiGiordaCozzini2007,Kolodrubetz2013}
\begin{equation}
\begin{aligned}
ds^2&=\frac14dv^2+\frac{\cot^2v}{16}du^2
     +\frac{\sin^2v}{8}d\phi^2,\\
&\qquad 0<v<\pi/2,\quad |u|<\pi/2.
\end{aligned}
\label{eq:XY-metric-vuphi}
\end{equation}
Its sectional curvatures in the coordinate planes and its scalar
curvature are
\begin{equation}
\begin{aligned}
K_{v\phi}&=4,\qquad K_{vu}=-8\csc^2v,\\
K_{u\phi}&=4\csc^2v,\qquad R=-8/\gamma.
\end{aligned}
\label{eq:XY-sectional-curvatures}
\end{equation}
The positive and negative sectional curvatures describe different
tidal responses depending on the geodesic direction and the
transverse displacement.

When comparing these quantities with the finite-size many-body
geometry, the normalization must be kept explicit. At any fixed
$L$, the total and per-site metrics obey the exact scaling relations
\begin{equation}
  s_{\mathrm{tot}}=\sqrt{L}\,s_{(L)},
  \qquad
  K^{\mathrm{tot}}=\frac{K^{(L)}}{L}.
\end{equation}
Taking $g^{(L)}$ to its thermodynamic limit $g$ is a separate step and
the curvatures in \eqref{eq:XY-sectional-curvatures} refer to
that limiting metric.

\subsection{ Ambient-Extrinsic cancellations in the tidal tensor of the product state.}

Having specified the ground-state metric, we now resolve its
tidal curvature into the ambient and extrinsic contributions
identified by the Gauss equation \cref{eq:tidal-extrinsic}. This reveals how the bending
of the ferromagnetic ground-state manifold modifies the focusing
and shear expected from the full quantum state space.

The product structure
$\ket{\Omega}=\bigotimes_{k>0}\ket{w_k}$ leads to an exact
cancellation between the ambient tide and part of the
extrinsic contribution. We first establish this
identity, then evaluate the surviving terms explicitly at
the zero-field Ising point, where exact finite-size results
are available.

The main task is to calculate the second fundamental form that has double derivatives of the product state. The product rule gives terms containing either $\partial_\mu\partial_\nu\ket{w_k}$, with both derivatives acting on one momentum-pair factor, or $\partial_\mu\ket{w_k}\otimes\partial_\nu\ket{w_\ell}$ with $k\ne\ell$, where two distinct factors change. The remaining factors are unchanged. The full expansion and its horizontal projection are given in \eqref{eq:XY-NC-product} of \cref{app:XY-SFF}. After removing the overall-phase component and subtracting the tangent component within the ground-state family, these give two orthogonal contributions,
\begin{equation}
 \Pi=\Pi^{\rm single}+\Pi^{\rm two}.
\end{equation}
Their orthogonality means that the normal contractions
in the Gauss equation \eqref{eq:tidal-extrinsic} split
into single-pair and two-pair contributions, with no
cross terms. The explicit product-state derivatives, their normal
projections, and the orthogonality of the two contributions
are derived in \cref{app:XY-SFF}.

This separation also has a geometric interpretation.
Allowing every momentum-pair state to vary independently
gives a product of Bloch spheres, whereas the physical
ground-state family $\Sigma$ ties their motion to the parameters $(\gamma,h,\phi)$:
\begin{equation}
 \Sigma\subset\prod_{k>0}\CP^1_k\subset\CP^{2^L-1}.
 \label{eq:XY-nested-embedding}
\end{equation}
Thus $\Pi^{\rm single}$ measures how the ground-state family
bends within the product of spheres, while $\Pi^{\rm two}$
measures how that product itself bends within the full
many-body state space.

First use the total Fubini-Study metric, with unit velocity $U$
and orthonormal screen $E_a$. Let $g_k$ and $\omega_k$ be the
metric and symplectic contributions of pair $k$, so that
$g^{\rm tot}=\sum_k g_k$ and $\omega^{\rm tot}=\sum_k\omega_k$.
With $v_a=\omega^{\rm tot}(U,E_a)$ as in \cref{eq:v-Berry}, the
product rule gives
\begin{equation}
\begin{aligned}
 &\langle\Pi^{\rm two}_{ab},\Pi^{\rm two}_{UU}\rangle
 -\langle\Pi^{\rm two}_{aU},\Pi^{\rm two}_{bU}\rangle\\
 &\qquad=-h_{ab}-3v_av_b
       +4\sum_{k>0}\omega_k(U,E_a)\omega_k(U,E_b).
\end{aligned}
\label{eq:XY-sector-tide-definition}
\end{equation}
Here $\Pi^\mathrm{two}_{ab}=\Pi^\mathrm{two}(E_a,E_b)$,
and likewise for the other components. Substituting into
\cref{eq:tidal-extrinsic} cancels the explicit ambient terms:
\begin{equation}
\begin{aligned}
 \cE^\Sigma_{ab}
 ={}&4\sum_{k>0}\omega_k(U,E_a)\omega_k(U,E_b)\\
 &+\langle\Pi^{\rm single}_{ab},\Pi^{\rm single}_{UU}\rangle
 -\langle\Pi^{\rm single}_{aU},\Pi^{\rm single}_{bU}\rangle.
\end{aligned}
\label{eq:XY-product-tide}
\end{equation}
The cancellation concerns the difference of the two normal
contractions. The remaining sum is the curvature contribution of the individual Bloch spheres, derived in \cref{app:XY-normal-contractions}.

The distinction is between the collective product
$v_av_b=(\sum_k\omega_k(U,E_a))(\sum_\ell\omega_\ell(U,E_b))$
and the sum of products within individual pairs in
\eqref{eq:XY-product-tide}. Thus $v_a=0$ does not remove the
individual-pair contribution: their signed overlaps can cancel
while their squares remain nonzero. The trace-free part of
\eqref{eq:XY-product-tide} is the curvature source of shear.

\subsection{Exact finite-size result on purely extrinsic shear at the Ising-point}

We now evaluate the obtained \ref{eq:XY-product-tide} at the zero-field Ising point, where one can calculate exact results for system size $L\geq8$. We now use the metric per site, dividing both ambient and induced
metrics by $L$. Detailed derivations are given in \cref{app:XY-exact-proof}. At $(\gamma,h)=(1,0)$,
$g^{(L)}=\operatorname{diag}(1/64,1/16,1/16)$ in $(\gamma,h,\phi)$,
with orthonormal frame
$(e_\gamma,e_h,e_\phi)=(8\partial_\gamma,4\partial_h,4\partial_\phi)$
and $e_\phi=Je_h$. A general unit velocity is
\begin{equation}
\begin{aligned}
 U&=c_0\,e_\gamma+4Q_u\,e_h+4Q_\phi\,e_\phi,\\
 c_0^2&=1-16Q_u^2-16Q_\phi^2,
\end{aligned}
\label{eq:XY-general-Ising-velocity}
\end{equation}
with $Q_A=g^{(L)}(\partial_A,U)$ for $A=u,\phi$ and
$\partial_u=\partial_h$.
For $c_0\ne0$, use the projected coordinate vectors
$V_A=\partial_A-Q_AU$ and their screen metric
\begin{equation}
 h_{AB}=g^{(L)}(V_A,V_B)=\tfrac1{16}\delta_{AB}-Q_AQ_B.
 \label{eq:XY-Ising-charge-screen}
\end{equation}
Unlike $E_a$, these vectors are not orthonormal. At $c_0=0$ this basis degenerates, and an orthonormal
screen gives the regular limiting tide.

In this normalization and basis, the ambient contribution is
\begin{equation}
\begin{aligned}
 \cE^{\rm amb}_{AB}&=L(h_{AB}+3v_Av_B),\\
 v_A&=\omega^{(L)}(U,V_A)=(-Q_\phi,\,Q_u),
\end{aligned}
\label{eq:XY-general-ambient-tide}
\end{equation}
where $\omega^{(L)}=\omega^{\rm tot}/L$. Its Berry weight is
\begin{equation}
 |v|^2=h^{AB}v_Av_B=16(Q_u^2+Q_\phi^2)=1-c_0^2.
 \label{eq:XY-directional-Berry-weight}
\end{equation}
It vanishes only for $U=\pm e_\gamma$, where $JU$ is normal to
$\Sigma$. Field- and phase-directed velocities have $|v|=1$ and 
for $U=\cos\alpha\,e_\gamma+\sin\alpha\,e_j$ ($j=h,\phi$),
the Berry weight is $\sin^2\alpha$.

The appendix \cref{app:XY-exact-proof} evaluates the individual-pair sum and the single-pair
normal contractions separately in \cref{eq:XY-charge-tides}.
Their sum gives
\begin{equation}
\begin{gathered}
 \cE^\Sigma=\begin{pmatrix}
 -1+16Q_u^2+24Q_\phi^2&-8Q_uQ_\phi\\
 -8Q_uQ_\phi&1/4+4Q_u^2-4Q_\phi^2
 \end{pmatrix},\\
 \Ric^\Sigma(U,U)=h^{AB}\cE^\Sigma_{AB}=-12+64Q_u^2+384Q_\phi^2.
\end{gathered}
\label{eq:XY-general-intrinsic-tide}
\end{equation}
These values are exact for every even $L\ge8$: at the Ising point
$\epsilon_k=1$, and the surviving sums contain Fourier harmonics
through order six, whose discrete averages are exact for these
sizes. They agree with the metric
curvatures \eqref{eq:XY-sectional-curvatures} at $v=\pi/4$.
Both surviving contributions are $O(1)$ in per-site units, or
$O(1/L)$ in total-metric orthonormal units.

For $Q_u=Q_\phi=0$, $U=e_\gamma$ and the orthonormal screen is
$(e_h,e_\phi)=4(V_u,V_\phi)$. Then
\begin{equation}
 \cE^{(L)}=\begin{pmatrix}-16&0\\0&4\end{pmatrix},\qquad
 (\cE^{(L)})_{\TF}=\begin{pmatrix}-10&0\\0&10\end{pmatrix}.
 \label{eq:XY-density-tide}
\end{equation}
The ambient tide is isotropic here, so the whole shear source is
extrinsic relative to the full many-body projective space. The
single-pair normal contractions give $\operatorname{diag}(-16,-4)$,
while the individual-pair sum in \eqref{eq:XY-product-tide} gives
$\operatorname{diag}(0,8)$ after normalization conversion.
Thus the collective Berry enhancement vanishes, but the separate
pairs still contribute to shear. For a complete
point-emitted family the initial shear is
\begin{equation}
 \sigma^{(L)}(s)=\frac{10s}{3}\operatorname{diag}(1,-1)+O(s^2),
 \qquad\varpi^{(L)}=0,
 \label{eq:XY-density-shear-onset}
\end{equation}
with $s$ the arclength per site. Field- and phase-directed velocities
instead have trace-free eigenvalues $\pm12$ and $\pm2$, after their
$O(L)$ Berry anisotropy has cancelled. Thus, at the zero-field Ising point and for an
anisotropy-directed geodesic, the entire trace-free tide
is extrinsic relative to the full many-body projective
space.

\paragraph{Normal acceleration and transport-generator variation.}
The same normal derivatives also determine how the transport
Hamiltonian changes along an intrinsic geodesic. As shown in
\cref{sec:generator-bending}, $\Pi(U,U)$ measures the variation
of the minimal mixing generator required to keep the path
within the ground-state family.

At the Ising point, consider the anisotropy-directed unit
velocity $U_{\rm tot}=e_\gamma/\sqrt L$ in the total metric.
The single-pair and two-pair contributions to its squared
normal acceleration are $16/L$ and $2-6/L$, respectively, giving
\begin{equation}
 \norm{\Pi(U_{\rm tot},U_{\rm tot})}^2
 =2+\frac{10}{L}.
 \label{eq:XY-bending-exact}
\end{equation}
This expression is exact for every even $L\ge10$: the
calculation involves Fourier harmonics through order eight,
which raises the size threshold compared with the tidal
contractions above. The derivation and the exceptional
$L=8$ value are given in \cref{app:XY-normal-contractions}.

For constant physical speed $\upsilon=ds_{\rm tot}/dt$,
the corresponding minimal transport Hamiltonian obeys
\begin{equation}
 \frac{\Tr[(\partial_tH_{\rm cd})^2]}{2\hbar^2}
 =\upsilon^4\left(2+\frac{10}{L}\right).
 \label{eq:XY-CD-rate}
\end{equation}
Here $H_{\rm cd}$ transports the many-body rank-one projector.
Its variation remains finite as $L\to\infty$, with the
nonzero limit supplied entirely by derivatives acting on
two distinct momentum pairs. These expressions are exact at the zero-field Ising point. Away from it, the normal acceleration and the corresponding generator variation must be evaluated using the parameter-dependent geometry.

We now follow geodesic families over finite distances using the metric in the thermodynamic limit. For families through the Ising point, the
calculation above gives the local origin of their tide.

\subsection{Fold catastrophes of ground-state geodesic families}
\label{sec:XY-two-charge}

The family of geodesics on the ground-state metric in the thermodynamic limit provides an explicit example
of a caustic inside a regular quantum phase. We construct a family
of geodesics which develops
an $A_2$ fold, the simplest stable catastrophe familiar from ray
optics \cite{Arnold1990,BerryUpstill1980}. We then associate a quantum information theoretic meaning to transverse collapse through fidelity response and quantum Fisher information. All lengths and responses below use the metric per site.

\begin{figure*}[t]
 \centering
 \includegraphics[width=\linewidth]{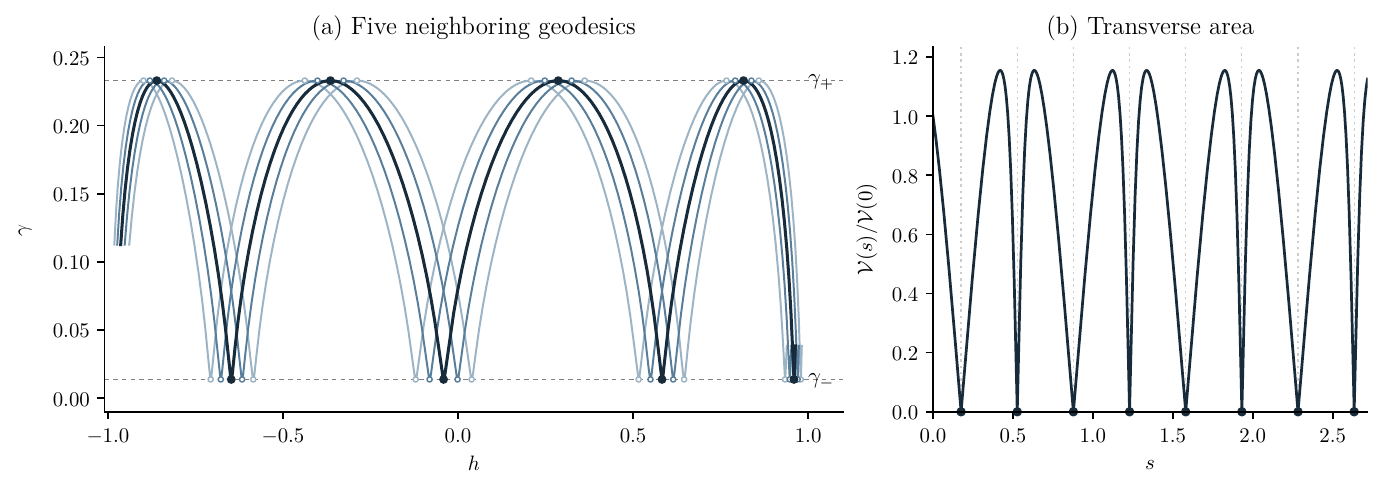}
 \caption{Neighboring geodesics and transverse folds in the quantum metric of the Ferromagnetic ground state, with $Q_u=1/2$, $Q_\phi=1/25$ and
 $z_0=-\pi/2$.
 (a) Five members of the fixed-momentum family, projected onto the $(h,\gamma)$ plane. Their initial labels are $u_0=-13/10+\delta u_0$, with $\delta u_0=-0.08,-0.04,0,0.04,0.08$. The dark curve is the
 reference path. All are shown over the same arclength interval, ending when the reference path reaches $u=13/10$.
 Each encounters eight folds, marked by circles at the
 dashed levels $\gamma_\pm$. These levels are the projections of the caustic surfaces. Other crossings elsewhere in this projection do not indicate state coincidence. (b) The normalized transverse area of the full two-label family vanishes at the eight common turning arclengths. The area is independent of the translation labels and continues on successive regular branches. Every displayed trajectory lies strictly inside $|h|<1$ with finite positive $\gamma$.}
 \label{fig:XY-K-recurrence}
\end{figure*}

\paragraph{A family with two conserved momenta.}
We obtain the geodesics by making the metric action
$\frac12\int g_{\mu\nu}\dot\lambda^\mu\dot\lambda^\nu\,ds$
stationary. For the XY metric, write
$a^2(v)=\cot^2v/16$ and $b^2(v)=\sin^2v/8$.
The corresponding geodesic Lagrangian is
\begin{equation}
 \mathcal L_{\rm geo}
 =\frac{\dot v^2}{8}
  +\frac{a^2(v)}2\dot u^2
  +\frac{b^2(v)}2\dot\phi^2.
 \label{eq:XY-geodesic-Lagrangian}
\end{equation}
Dots denote differentiation with respect to arclength.
Since $u$ and $\phi$ do not appear explicitly, their
conjugate momenta are conserved. Together with unit speed,
$2\mathcal L_{\rm geo}=1$, this gives
\begin{equation}
\begin{aligned}
 Q_u&=a^2\dot u,\qquad Q_\phi=b^2\dot\phi,\\
 \dot v^2=F(v)
 &=4[1-16Q_u^2\tan^2v-8Q_\phi^2\csc^2v].
\end{aligned}
 \label{eq:XY-K-first-integrals}
\end{equation}
These first integrals reduce the trajectory calculation
to motion in the single coordinate $v$
\cite{Kolodrubetz2013,LiskaGritsev2021}.

Let us fix nonzero $Q_u,Q_\phi$, an initial $v_0$ with $F(v_0)>0$,
and a sign of $\dot v(0)$. Varying the starting positions $(u_0,\phi_0)$
then gives neighboring geodesics with separations
$\eta_u=\partial_u$ and $\eta_\phi=\partial_\phi$.
Its motion in the anisotropy coordinate $v$ is restricted by
$\dot v^2=F(v)\ge0$. A simple zero of $F$ is a turning point:
$\dot v$ vanishes there and changes sign, while the geodesic
continues smoothly in $u$ and $\phi$. For nonzero $Q_u,Q_\phi$,
there are two such turning values $v_-<v_+$ inside
$0<v<\pi/2$ provided
\begin{equation}
 8Q_\phi^2+16\sqrt2\,|Q_uQ_\phi|<1.
 \label{eq:XY-K-allowed}
\end{equation}
The coordinate $v$ oscillates between these values for as long
as the trajectory remains within the physical phase.
We show below how these reversals produce `folds' of the
geodesic family. The exact trajectories and their physical
domain are derived in \cref{app:XY-K-integration}.

\paragraph{Turning point as a fold catastrophe}
The geodesic family acts as a map of arclength `$s$' and initial parameters $(u_0,\phi_0)$ to the accessible set of states,
represented locally by
\begin{equation}
 \mathscr R(s,u_0,\phi_0)
 =\bigl(v(s),u_0+\bar u(s),\phi_0+\bar\phi(s)\bigr),
 \label{eq:XY-ray-map}
\end{equation}
where $\bar u(s)=u(s)-u_0$ and
$\bar\phi(s)=\phi(s)-\phi_0$ are given explicitly in
\cref{eq:XY-K-primitives}.

The Jacobian determinant of this map is
\begin{equation}
 \det\frac{\partial(v,u,\phi)}
              {\partial(s,u_0,\phi_0)}
 =\det\begin{pmatrix}
 \dot v&0&0\\
 \dot u&1&0\\
 \dot\phi&0&1
 \end{pmatrix}
 =\dot v.
\end{equation}
The map therefore loses rank when a geodesic reaches either
turning value $v_\pm$. Denote the corresponding arclength by
$s_*$ and write $v_*=v(s_*)$. At a simple turn,
$\dot v(s_*)=0$ but $c_*:=\ddot v(s_*)\ne0$, so
\begin{equation}
 v(s)=v_*+\tfrac12c_*(s-s_*)^2+O((s-s_*)^4).
 \label{eq:XY-turn-expansion}
\end{equation}

For $v$ close to $v_*$ on the allowed side, the quadratic
relation gives two nearby arclengths, one before and one
after the turn. At either arclength, the starting labels
$u_0=u-\bar u(s)$ and $\phi_0=\phi-\bar\phi(s)$ can be chosen
to reach the same target state $(v,u,\phi)$. Thus two local
members of the family reach that state at different arclengths.
As $v\to v_*$, both arclengths approach $s_*$ and their
starting labels coincide. This merging of two solutions
defines a caustic which is a ``fold catastrophe".\Cref{app:XY-K-ray-fold} establishes its $A_2$ classification
through the corresponding cubic generating function
\cite{Arnold1990,BerryUpstill1980}.

Successive folds are separated by
$\Delta s=\pi/[4\sqrt{1+16Q_u^2}]$. The monotonic drift in $u$
eventually reaches the Ising critical line $|u|=\pi/2$, beyond which the ferromagnetic family is not defined.
Figure~\ref{fig:XY-K-recurrence} shows five neighboring paths and
eight interior folds. The exact count is given in
\cref{app:XY-K-count-proof}.

\begin{figure*}[t]
 \centering\includegraphics[width=\linewidth]{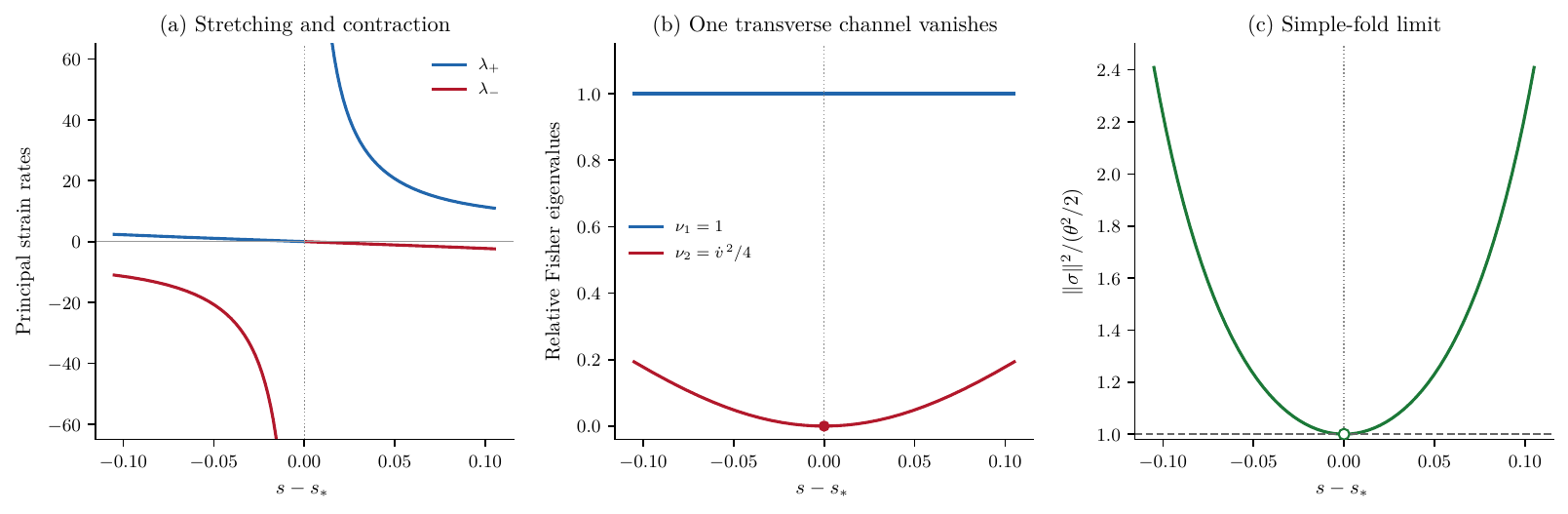}
 \caption{Analytic congruence data near the first upper fold of
 Fig.~\ref{fig:XY-K-recurrence}.
 (a) Principal stretching rates: one is positive and the other negative on
 each regular sheet.
 (b) Generalized eigenvalues of the transverse Fisher matrix relative
 to the equal-arclength Fisher matrix, Eq.~\eqref{eq:XY-K-relativeFisher}.
 One vanishes quadratically, while the equal-arclength label metric
 stays nonsingular. (c) The shear-to-expansion ratio approaches one
 from above showing that the collapse involves one transverse direction.}
 \label{fig:XY-K-strain}
\end{figure*}

\paragraph{Transverse collapse and shear.}
The deviation vectors have longitudinal components
$g(U,\eta_A)=Q_A$. Their transverse metric is therefore given by
\begin{equation}
\begin{aligned}
 \cG_\perp&=\cG-\boldsymbol Q\boldsymbol Q^T,\qquad
 \cG=\operatorname{diag}(a^2,b^2),\\
 \det\cG_\perp&=\frac{\cos^2v}{512} \dot v^2,
 \qquad \boldsymbol Q=(Q_u,Q_\phi)^T.
\end{aligned}
\label{eq:XY-K-H}
\end{equation}
The transverse area is proportional to $|\dot v|$ and therefore
vanishes at each turning point:
\begin{equation}
\begin{aligned}
 \frac{\cV(s)}{\cV(0)}
 &=\frac{\cos v(s)\,|\dot v(s)|}
         {\cos v_0\,|\dot v(0)|},\\
 \theta&=\frac{\ddot v}{\dot v}-\tan v\,\dot v.
\end{aligned}
 \label{eq:XY-K-area}
\end{equation}
On each regular branch the twist vanishes, and the exact identities
\begin{equation}
 \det\cB=-K_{u\phi},\qquad
 \norm\sigma^2=\frac{\theta^2}{2}+2K_{u\phi}
 \label{eq:XY-K-geometric}
\end{equation}
require simultaneous stretching and contraction because
$K_{u\phi}=4\csc^2v>0$. Expansion decreases throughout each branch,
even where the Ricci term favors defocusing
(\cref{app:XY-K-geometry}). Approaching a fold,
$\theta\sim1/(s-s_*)\to-\infty$ and
$\norm\sigma^2/(\theta^2/2)\to1$: one transverse direction collapses
while the other remains open. Figure~\ref{fig:XY-K-strain}(a) shows the divergent collapsing rate and the vanishing rate in the surviving transverse direction near a fold, while panel~(b) shows the
quadratic suppression of one transverse sensitivity channel.
These are complementary signatures of the same collapse:
the relative deformation rate diverges as the transverse
response vanishes, although the paths remain smooth and the
equal-arclength label metric stays regular.

\paragraph{Fidelity and quantum Fisher signatures.}
The equal-arclength label QFI per site, $4\cG$, remains nonsingular at the fold.  To identify the collapsing direction, consider a small shift
$(\delta u_0,\delta\phi_0)=\epsilon n$ of the starting labels, where
\begin{equation}
 n=(\dot u_*,\dot\phi_*)
   =\left(\frac{Q_u}{a_*^2},\frac{Q_\phi}{b_*^2}\right).
 \label{eq:XY-K-null}
\end{equation}
At the fold, $\dot v_*=0$, so $n^A\eta_A=U_*$:
this label shift moves the state along the geodesic,
leaving no transverse separation. Thus $\cG_{\perp,*}n=0$,
although the equal-arclength response remains finite.
Normalizing the transverse response by the equal-arclength
metric gives
\begin{equation}
 \operatorname{eig}(\cG^{-1/2}\cG_\perp\cG^{-1/2})
 =\{1,\dot v^{\,2}/4\}.
 \label{eq:XY-K-relativeFisher}
\end{equation}
One channel is unchanged by the transverse projection and 
the other vanishes quadratically because
$\dot v=c_*(s-s_*)+O((s-s_*)^3)$ near the fold.

The transverse collapse has a fidelity signature, but it requires
distinguishing two comparisons. At equal arclength, the label shift
$\epsilon n$ produces the nonzero displacement $\epsilon U_*$,
so the states remain distinguishable. To isolate their transverse
difference, compare the shifted path at $s_*-\epsilon$ with the
reference path at $s_*$. This removes the first-order displacement
along the geodesic, leaving a mismatch of order $\epsilon^2$.

For the normalized even-parity ground state
$\ket{\Omega_L(\lambda)}$, define the logarithmic fidelity per site,
\begin{equation}
 \mathfrak f(\lambda,\lambda')
 =-\lim_{L\to\infty}\frac1L
 \ln|\langle\Omega_L(\lambda)|\Omega_L(\lambda')\rangle|^2.
 \label{eq:XY-fidelity-density}
\end{equation}
Here $\lambda=(v,u,\phi)$ denotes the state parameters.
For nearby states,
$\mathfrak f=g_{\mu\nu}\delta\lambda^\mu\delta\lambda^\nu
+O(|\delta\lambda|^3)$.
The two comparisons therefore give
\begin{equation}
\begin{aligned}
 \mathfrak f_{\rm eq}&=\epsilon^2+O(\epsilon^3),\\
 \mathfrak f_{\rm comp}&=\frac{c_*^2}{16}\epsilon^4+o(\epsilon^4).
\end{aligned}
 \label{eq:XY-K-overlap}
\end{equation}
Thus the fold suppresses the quadratic fidelity response only
after the longitudinal displacement is removed. The remaining
quartic coefficient follows from
$\Delta v=c_*\epsilon^2/2$ and $g_{vv}=1/4$ (see
its derivation in \cref{app:XY-K-overlap-proof}).

Compensation has a natural Hamilton-Jacobi interpretation.
On each regular branch, the stationary action satisfies
$dS=U^\flat$, whose pullback to the family is
$dS=ds+Q_A\,d\lambda^A$. Comparing states on the same
geometrical wavefront therefore requires
$\delta s=-Q_A\delta\lambda^A$ and leaves precisely the
transverse separation $\eta_A^\perp\delta\lambda^A$.
At the fold, $Q_An^A=1$, so the label shift $\epsilon n$
is compensated by $\delta s=-\epsilon$.

The XY family thus realizes a
fold catastrophe of the quantum-state ray map within a regular phase,
with transverse focusing and a calculable compensated-fidelity signature.

\section{A qutrit illustration of shear and selective focusing}
\label{sec:operational}
\label{sec:qutrit-realization}

The shear and selective focusing discussed in
\cref{sec:cp-exact} can be illustrated in a qutrit.
Two preparation labels probe the complex direction and
one ordinary transverse direction of the point-emitted
bundle. Their different separation rates reveal the
curvature anisotropy, and only the complex direction
collapses at the first conjugate point. The paths are
geodesics of the full pure-state space $\CP^2$, so no
extrinsic-curvature correction enters their deviation
equation.

Take orthonormal states $\ket0,\ket1,\ket2$ and define
\begin{equation}
\begin{aligned}
 \ket{v_{\alpha\beta}}
 &=e^{i\alpha}\cos\beta\ket1+\sin\beta\ket2,\\
 H(\alpha,\beta)
 &=i\hbar\nu\left(\ket{v_{\alpha\beta}}\bra0
                 -\ket0\bra{v_{\alpha\beta}}\right),\\
 \ket{\psi(s;\alpha,\beta)}
 &=\cos s\ket0+\sin s\ket{v_{\alpha\beta}},
 \qquad s=\nu t .
\end{aligned}
\label{eq:qutrit-protocol}
\end{equation}
Here $\nu>0$ is the coupling rate. For each choice of labels,
$H(\alpha,\beta)$ generates a unit-speed geodesic from $\ket0$,
with $s$ its Fubini-Study arclength.

At $\alpha=\beta=0$, the label derivatives are
$\partial_\alpha\ket\psi=i\sin s\ket1$ and
$\partial_\beta\ket\psi=\sin s\ket2$.
Using the pure-state quantum Fisher information
\cite{BraunsteinCaves1994,Facchi_2010},
\begin{equation}
 \cI_{AB}
 =4\operatorname{Re}\left[
 \braket{\partial_A\psi|\partial_B\psi}
 -\braket{\partial_A\psi|\psi}
  \braket{\psi|\partial_B\psi}
 \right],
\end{equation}
gives
\begin{equation}
 \cI_{\alpha\alpha}=\sin^2(2s),\qquad
 \cI_{\beta\beta}=4\sin^2s,\qquad
 \cI_{\alpha\beta}=0 .
\label{eq:qutrit-protocol-QFI}
\end{equation}
The phase label $\alpha$ probes the complex direction $JU$,
while $\beta$ probes an ordinary transverse direction.
For $0<s<\pi/2$, their separation amplitudes are
$\sqrt{\cI_{\alpha\alpha}}/2=\sin(2s)/2$ and
$\sqrt{\cI_{\beta\beta}}/2=\sin s$.
The corresponding principal deformation rates satisfy
\begin{equation}
 b_{\mathbb C}-b_\perp
 =2\cot(2s)-\cot s=-\tan s,
\end{equation}
exhibiting the shear described in
\cref{eq:CP-shear-exact}.

At $s=\pi/2$ and $\beta=0$, the state is
$e^{i\alpha}\ket1$: $\alpha$ has become a global phase,
and $\cI_{\alpha\alpha}=0$. Varying $\beta$ still changes
the target ray by mixing $\ket1$ with $\ket2$, so
$\cI_{\beta\beta}=4$. This is the selective collapse at
the first conjugate point.

The robustness of this sensitivity contrast against
dephasing, population relaxation, and preparation errors
could be studied separately. Possible platforms include
ultracold atomic qutrits \cite{Lindon2023} and superconducting
transmons \cite{Blok2021,Morvan2021}, where coherent qutrit
control has been demonstrated. These controls could be used
to prepare the state family above and examine its label
dependence, even when the pulse sequence used for preparation
does not itself follow the geodesic.

Optical settings offer another possibility: geodesic evolution in
$\CP^2$ has been proposed using three-channel interferometry
\cite{Sanders2001}, and geometric-phase effects have been measured
in a two-photon polarisation qutrit \cite{Ogawa2015}.
Such settings could also be explored for observing the unequal
separation rates and selective loss of label sensitivity described here.

\section{Discussion and conclusions}
\label{sec:discussion}
\label{sec:conclusions}

The key motivation of this work is to understand how the curvature of quantum state space manifests itself in the kinematics of neighbouring geodesics. The differential-geometric identities used here and the curvature structure of complex projective space are indeed established mathematical results. We delineate their consequences for quantum-state congruences and connect them to state preparation and distinguishability. In general relativity, geodesic deviation and the Raychaudhuri-Ehlers equations relate curvature to the relative motion of neighboring particle trajectories. Here, neighboring paths describe families of quantum-state preparations, and their separation determines sensitivity to preparation labels. Fidelity susceptibility and quantum Fisher information thus give physical meaning to the expansion, shear, and focusing of these families.

The intrinsic quantum metric on the parameter space, together with the initial family, already determines intrinsic geodesic deviation. The extrinsic analysis explains how this behavior arises from the embedding of the accessible states in the full
pure-state space. Through the Gauss equation, we separate the tidal curvature into an isotropic ambient contribution, a Berry-curvature-dependent enhancement, and contractions of the second fundamental form. This identifies how bending reinforces or counteracts the ambient contribution to focusing and shear. The relation between the second fundamental form and the variations of the unitary transport generator further connects this decomposition to the changes in the generator required to follow an intrinsic geodesic \cref{eq:generator-bending,eq:generator-bending-tensor}.

The XY chain makes these distinctions explicit. Its ground-state product structure yields an exact cancellation of collective ambient tidal terms, including their Berry-curvature enhancement, against extrinsic contributions. At the zero-field Ising point, the finite-size calculation shows that the shear source for anisotropy-directed motion is entirely extrinsic relative to the full many-body pure-state space. Separately, the geodesic families studied in the thermodynamic limit develop repeated $A_2$ folds within the ferromagnetic phase. These are caustics of the specified family map, rather than a determination of the conjugate locus of the parameter manifold. Their fidelity signatures address Berry's question about the physical meaning of caustics of quantum-metric geodesics \cite{Berry1989FiveYears}.

Several extensions follow naturally. Higher-rank projectors would allow the same questions to be asked for occupied band subspaces and degenerate state manifolds. Their Grassmannian geometry admits several independent mixing directions, making multiband systems a setting in which to investigate competing tidal channels and their physical significance. Topological phases and ideal Chern bands offer a particularly structured application: relations between the quantum metric, Berry curvature, and complex structure \cite{MeraOzawa2021} invite a study of how these constraints restrict extrinsic bending and geodesic deformation. The question is which features of the congruence are fixed by these geometric conditions and which retain information about the microscopic realization.

Within the XY chain, the disorder circle $h^2+\gamma^2=1$ offers a tractable extension of the local Ising-point calculation. Its factorized ground states may provide further analytical control over bending and shear \cite{ChitovGadgeTimonin2022,RossignoliCanosaMatera2008}. Future work could investigate the relation of our results to the quantum caustics in Fock-space dynamics and quenched spin chains\cite{Mumford2019,Kirkby2019}. Beyond solvable chains, interacting ground-state families and Fermi liquids pose the question of how collective low-energy degrees of freedom shape the parametrised geometry. For gapless systems, this requires specifying a smooth family and controlling the finite-size and thermodynamic limits before applying the congruence equations.

Dissipative systems require further extension from pure states to density matrices, with a metric such as the Bures metric that retains a direct relation to fidelity and QFI. Geodesics of this metric and trajectories generated by an open-system evolution are generally different. A useful development would formulate their collective deformation while separating curvature effects from the contraction of distinguishability induced by noise. This would place the rigid common-unitary evolution discussed here within a broader account of quantum-state transport.

The relation to entanglement is another substantive open direction. Although a wavefunction may factorize across momentum sectors, it can still exhibit entanglement with respect to spatial or other degrees of freedom. Geometric deformation of quantum-state congruences does not, by itself, characterize such correlations. Connecting bending or caustics to entanglement could therefore be explored through the geometry of reduced density matrices, variations of the entanglement spectrum, or other geometric descriptions of many-body correlations. This could clarify how the geometry of quantum-state families encodes information about the organization and evolution of quantum correlations.

These directions share a common aim: to understand quantum
geometry as a source of constraints on families of paths followed by quantum states. The present work establishes this viewpoint
through universal projective-space deformation and exact
many-body examples, while the qutrit construction supplies
a simple measurement illustration. Its broader promise is
a framework connecting the local geometry of quantum states
to the collective behavior of paths through quantum matter.

\section*{Acknowledgments}
This project was supported by the ANRF Prime Minister Early Career Grant
(PM-ECRG), project ANRF/ECRG/2024/002083/PMS.

\paragraph{Author Contributions}
SH formulated the research problem, set up the calculations, derived the results on state-space congruences and extrinsic geometry, and prepared the manuscript. NK worked through the results and performed most of the XY spin-chain calculations and helped in preparation of the manuscript. AB worked through the results, contributed to the section on qutrits and helped in preparation of the manuscript.

\paragraph{AI assistance.}
The research formulation, preliminary calculations, and initial results were developed before we started to use AI. AI assistance subsequently accelerated some of the differential-geometric and spin-chain calculations.

ChatGPT (OpenAI)  and Claude (Anthropic)  were used to assist in the writing and revising of the text,
organizing the manuscript, preparing \LaTeX{} and figures, identifying references, and developing and checking analytical derivations and numerical code.  The authors assume full responsibility for the scientific claims, interpretations and final content of this work.

\appendix

\section{Derivation of the Riccati and Raychaudhuri-Ehlers equations}
\label[appendix]{app:raychaudhuri-ehlers}

This appendix derives the evolution equations for the deformation of a
unit-speed geodesic bundle. We first obtain the operator Riccati equation
from geodesic deviation, and then take its trace, symmetric trace-free,
and antisymmetric parts. This is the geometric content of the
Raychaudhuri-Ehlers construction \cite{Raychaudhuri1955,Ehlers1993}.
 Lowercase indices
$a,b=1,\ldots,m$ label screen directions, while uppercase indices
$A,B=1,\ldots,q$ label the geodesics.

\subsection{Screen geometry and the deformation operator}
\label[appendix]{app:re-screen}

Let $P(s,\lambda^A)$ be a smooth family of unit-speed geodesics, with
$U=\partial_sP$ and $\eta_A=\partial_AP$. On the Grassmannian,
\begin{equation}
 \begin{aligned}
 g(X,Y)&=\frac12\operatorname{Tr}(XY),\\
 \nabla_UX&=P\frac{dX}{ds}Q+Q\frac{dX}{ds}P.
 \end{aligned}
 \label{eq:app-re-projector-connection}
\end{equation}
The latter formula applies to tangent fields along an ambient
Grassmannian path. For geodesics confined to a ground-state submanifold,
$\nabla$ instead denotes its induced Levi-Civita connection, including
the projection onto that submanifold's tangent space. The derivation
below applies to either geometry, using its own metric and curvature.

The screen is $S=U^\perp$, with metric $h=g|_S$. Its dimension is
$m=2r(N-r)-1$ on the full Grassmannian, or $m=\dim\mathcal M-1$ on an
induced state manifold $\mathcal M$. Since $\nabla_UU=0$, parallel
transport along the reference geodesic preserves the screen:
\begin{equation}
\begin{aligned}
 g(\nabla_UV,U)&=\frac{d}{ds}g(V,U)-g(V,\nabla_UU)\\
 &=0\qquad(V\in S).
\end{aligned}
 \label{eq:app-re-screen-preservation}
\end{equation}
We may therefore choose a parallel orthonormal basis $\{e_a\}$ satisfying
$\nabla_Ue_a=0$ and $h_{ab}=g(e_a,e_b)=\delta_{ab}$.

On a regular congruence patch, $U$ is a smooth unit vector field.
Its deformation operator is the pointwise linear map
\begin{equation}
 \cB(V)=\nabla_VU,\qquad
 \cB_{ab}=g(e_a,\nabla_{e_b}U).
 \label{eq:app-re-B-definition}
\end{equation}
It maps $S$ to $S$, because $g(U,\nabla_VU)=\tfrac12V[g(U,U)]=0$.
For a connecting field, the torsion-free identity and the commuting
family coordinates give
\begin{equation}
 [U,\eta_A]_{\mathrm{Lie}}=0,
 \qquad \nabla_U\eta_A=\nabla_{\eta_A}U.
 \label{eq:app-re-commuting}
\end{equation}
The equality $\cB(V)=\nabla_UV$ thus applies to commuting connecting
fields, not to arbitrary screen fields. In particular, the derivative
of a freely chosen parallel basis vector is zero even when $\cB(e_a)$
is nonzero.

The longitudinal components $a_A=g(U,\eta_A)$ are constant along a
unit-speed geodesic family:
\begin{equation}
 \begin{aligned}
 \dot a_A
 &=g(U,\nabla_U\eta_A)
   =g(U,\nabla_{\eta_A}U)\\
 &=\tfrac12\partial_Ag(U,U)=0.
 \end{aligned}
 \label{eq:app-re-longitudinal}
\end{equation}
Consequently, $\eta_A^\perp=\eta_A-a_AU$ obeys
\begin{equation}
 \nabla_U\eta_A^\perp
 =\nabla_{\eta_A}U
 =\nabla_{\eta_A^\perp}U
 =\cB(\eta_A^\perp).
 \label{eq:app-re-screen-transport}
\end{equation}
Writing $\eta_A^\perp=e_aV^a{}_A$ in the parallel screen basis gives
\begin{equation}
 \dot V^a{}_A=\cB^a{}_bV^b{}_A,
 \qquad \dot V=\cB V.
 \label{eq:app-re-Ldot}
\end{equation}
Thus $V$ contains the transverse separations, while $\cB$ gives their
relative rates of change. The matrix inverse used below requires a
full screen-spanning family, $q=m$, with $V$ invertible on the regular
segment under consideration.

\subsection{From the Jacobi equation to the Riccati equation}
\label[appendix]{app:re-riccati}

Using the curvature convention in Eq.\eqref{eq:curvature-convention},
applying this identity to $U$ and $\eta_A$, and using
Eq.~\eqref{eq:app-re-commuting} and $\nabla_UU=0$, yields
\begin{align}
 \nabla_U^2\eta_A
 &=\nabla_U\nabla_{\eta_A}U\nonumber\\
 &=R(U,\eta_A)U
 =-R(\eta_A,U)U.
 \label{eq:app-re-Jacobi-derivation}
\end{align}
Because $a_A$ is constant and $U$ is parallel, the projected fields obey
the same Jacobi equation. Define the screen tidal operator by
\begin{equation}
 \cE(V)=R(V,U)U,\qquad
 \cE_{ab}=g(e_a,R(e_b,U)U).
 \label{eq:app-re-E-definition}
\end{equation}
Its symmetry, established below, makes this component definition equal
to $g(R(e_a,U)U,e_b)$ used in the main text. The projected Jacobi equation
is therefore
\begin{equation}
 \nabla_U^2\eta_A^\perp+\cE(\eta_A^\perp)=0,
 \qquad \ddot V+\cE V=0.
 \label{eq:app-re-Jacobi-matrix}
\end{equation}
Differentiating Eq.~\eqref{eq:app-re-Ldot},
\begin{equation}
 \ddot V=\dot\cB V+\cB\dot V
         =(\dot\cB+\cB^2)V,
\end{equation}
and multiplying the resulting Jacobi equation by $V^{-1}$ gives
\begin{equation}
\dot\cB+\cB^2+\cE=0.
 \label{eq:app-re-riccati}
\end{equation}
The dot here is the derivative of matrix components in the parallel
screen frame. The nonlinearity $\cB^2$ appears when the linear equation
for separations is rewritten as an equation for their relative rates.

For completeness, the basis-independent derivative of a screen
endomorphism $T$ is
\begin{equation}
 (D_sT)(V)=\nabla_U(TV)-T(\nabla_UV).
 \label{eq:app-re-operator-derivative}
\end{equation}
Hence $\nabla_U^2\eta_A^\perp=
(D_s\cB+\cB^2)\eta_A^\perp$, and the operator form of
Eq.~\eqref{eq:app-re-riccati} is $D_s\cB+\cB^2+\cE=0$.
If one uses instead $\mathcal R_U(V)=R(U,V)U$, then
$\mathcal R_U=-\cE$ and the same equation reads
$D_s\cB=-\cB^2+\mathcal R_U$. In particular,
$\operatorname{tr}_S\mathcal R_U=-\operatorname{Ric}(U,U)$
with the present Ricci convention.

\subsection{Adjoints, curvature, and the quadratic deformation term}
\label[appendix]{app:re-decomposition}

The adjoint $T^{\dagger_h}$ is defined by
$h(TV,W)=h(V,T^{\dagger_h}W)$ on the real screen space. It is distinct
from the Hilbert-space adjoint of projector matrices. Write
\begin{equation}
 \cB=\frac{\theta}{m}I_S+\sigma+\varpi,
 \qquad \theta=\operatorname{tr}_S\cB,
 \label{eq:app-re-decomposition}
\end{equation}
where
\begin{equation}
 \sigma^{\dagger_h}=\sigma,\quad
 \operatorname{tr}_S\sigma=0,\quad
 \varpi^{\dagger_h}=-\varpi.
\end{equation}
The matrix adjoint is simply the transpose in the chosen orthonormal
frame. Screen traces below are distinguished from the Hilbert-space
trace $\operatorname{Tr}$ in Eq.~\eqref{eq:app-re-projector-connection}.

\paragraph{Differentiation preserves the three tensor sectors.}
Metric compatibility implies
\begin{equation}
 D_s(T^{\dagger_h})=(D_sT)^{\dagger_h},\qquad
 \partial_s\operatorname{tr}_S T=\operatorname{tr}_S(D_sT).
 \label{eq:app-re-adjoint-derivative}
\end{equation}
To verify the first identity, choose arbitrary screen fields $V,W$
and differentiate the adjoint relation. Subtracting the terms in which
the derivative acts on $V$ or $W$ gives
\begin{align}
 h((D_sT)V,W)
 &=\partial_s h(TV,W)-h(T\nabla_UV,W)\nonumber\\
 &\quad-h(TV,\nabla_UW)\nonumber\\
 &=h(V,(D_sT^{\dagger_h})W).
\end{align}
The trace identity follows by taking components in the parallel frame.
Since $D_sI_S=0$, the derivative of the deformation splits as
\begin{equation}
 D_s\cB=\frac{\dot\theta}{m}I_S+D_s\sigma+D_s\varpi,
 \label{eq:app-re-Bdot-split}
\end{equation}
with $D_s\sigma$ still symmetric and trace-free and $D_s\varpi$ still
antisymmetric.

\paragraph{The tidal operator is symmetric.}
Define $R(X,Y,Z,W)=g(R(X,Y)Z,W)$. Its usual pair and antisymmetry
identities give
\begin{align}
 h(\cE V,W)
 &=R(V,U,U,W)\nonumber\\
 &=-R(V,U,W,U)=-R(W,U,V,U)\nonumber\\
 &=R(W,U,U,V)=h(V,\cE W).
\end{align}
Also $g(\cE V,U)=R(V,U,U,U)=0$, so $\cE$ maps the screen to itself.
Tracing over $\{e_a\}$ and noting that the missing $U$ term vanishes,
\begin{equation}
 \operatorname{tr}_S\cE
 =\sum_{a=1}^m g(R(e_a,U)U,e_a)
 =\operatorname{Ric}(U,U).
 \label{eq:app-re-E-trace}
\end{equation}
Thus
\begin{equation}
 \cE=\frac{\operatorname{Ric}(U,U)}{m}I_S+\cE_{\mathrm{TF}},
 \qquad \cE_{\mathrm{anti}}=0.
 \label{eq:app-re-E-split}
\end{equation}
For positive sectional curvature in a plane containing $U$,
$g(\cE V,V)>0$ for a nonzero vector $V$ in that plane orthogonal to $U$.
The minus sign in the Jacobi equation then gives a focusing acceleration.

\paragraph{Decomposition of $\cB^2$.}
Expanding the operator product, without assuming that $\sigma$ and
$\varpi$ commute, gives
\begin{align}
 \cB^2={}&\frac{\theta^2}{m^2}I_S
 +\frac{2\theta}{m}(\sigma+\varpi)\nonumber\\
 &+\sigma^2+\varpi^2+\sigma\varpi+\varpi\sigma.
 \label{eq:app-re-B-square}
\end{align}
The identity $(T_1T_2)^{\dagger_h}=T_2^{\dagger_h}T_1^{\dagger_h}$ implies
\begin{align}
 (\sigma^2)^{\dagger_h}&=\sigma^2,
 & (\varpi^2)^{\dagger_h}&=\varpi^2,\nonumber\\
 (\sigma\varpi+\varpi\sigma)^{\dagger_h}
 & =-(\sigma\varpi+\varpi\sigma).
 \label{eq:app-re-product-symmetries}
\end{align}
These identities concern squares and the indicated anticommutator.
A product of two different self-adjoint operators is not generally
self-adjoint: that would additionally require them to commute.

Define the nonnegative squared norms, with no factor of $1/2$, by
\begin{equation}
 \|\sigma\|^2=\sigma_{ab}\sigma^{ab},\qquad
 \|\varpi\|^2=\varpi_{ab}\varpi^{ab}.
\end{equation}
Their relation to operator traces is
\begin{align}
 \operatorname{tr}_S\sigma^2
 &=\sigma_a{}^b\sigma_b{}^a=\|\sigma\|^2,\nonumber\\
 \operatorname{tr}_S\varpi^2
 &=\varpi_a{}^b\varpi_b{}^a=-\|\varpi\|^2.
 \label{eq:app-re-norm-signs}
\end{align}
The mixed trace vanishes because
$\operatorname{tr}_S(\sigma\varpi)
=\operatorname{tr}_S[(\sigma\varpi)^{\dagger_h}]
=-\operatorname{tr}_S(\varpi\sigma)
=-\operatorname{tr}_S(\sigma\varpi)$.
Consequently,
\begin{align}
 \operatorname{tr}_S\cB^2
 &=\frac{\theta^2}{m}+\|\sigma\|^2-\|\varpi\|^2,
 \label{eq:app-re-square-trace}\\
 (\cB^2)_{\mathrm{sym,TF}}
 &=\frac{2\theta}{m}\sigma+(\sigma^2+\varpi^2)_{\mathrm{TF}},
 \label{eq:app-re-square-shear}\\
 (\cB^2)_{\mathrm{anti}}
 &=\frac{2\theta}{m}\varpi+\sigma\varpi+\varpi\sigma.
 \label{eq:app-re-square-twist}
\end{align}
For any symmetric operator $T$, $T_{\mathrm{TF}}=
T-(\operatorname{tr}_ST/m)I_S$. 

\subsection{Expansion, shear, and twist evolution}
\label[appendix]{app:re-equations}

\paragraph{Expansion.}
Taking the trace of Eq.~\eqref{eq:app-re-riccati} and inserting
Eqs.~\eqref{eq:app-re-E-trace} and \eqref{eq:app-re-square-trace} yields
\begin{equation}
\dot\theta=-\frac{\theta^2}{m}
 -\sigma_{ab}\sigma^{ab}+\varpi_{ab}\varpi^{ab}
 -\operatorname{Ric}(U,U).
 \label{eq:app-re-raychaudhuri}
\end{equation}
The positive twist contribution follows from
$-\operatorname{tr}_S(\varpi^2)=\|\varpi\|^2$. Positive Ricci curvature and shear
contribute negatively to $\dot\theta$, while twist can oppose focusing.
The equation alone does not imply a common finite focusing bound for
arbitrary twisting congruences.

\paragraph{Shear.}
Taking the symmetric trace-free part gives
\begin{align}
 \dot\sigma_{ab}={}&-\frac{2\theta}{m}\sigma_{ab}
 -(\cE_{ab})_{\mathrm{TF}}\nonumber\\
 &-\left(\sigma_a{}^c\sigma_{cb}
       +\varpi_a{}^c\varpi_{cb}\right)_{\mathrm{TF}}.
 \label{eq:app-re-shear}
\end{align}
The trace subtraction can be made explicit:
\begin{align}
 \left(\sigma_a{}^c\sigma_{cb}+\varpi_a{}^c\varpi_{cb}\right)_{\mathrm{TF}}
 ={}&\sigma_a{}^c\sigma_{cb}+\varpi_a{}^c\varpi_{cb}\nonumber\\
 &-\frac{\|\sigma\|^2-\|\varpi\|^2}{m}h_{ab},
 \label{eq:app-re-shear-trace-subtraction}\\
 (\cE_{ab})_{\mathrm{TF}}
 ={}&\cE_{ab}-\frac{\operatorname{Ric}(U,U)}{m}h_{ab}.
\end{align}
Every term in Eq.~\eqref{eq:app-re-shear} is symmetric and trace-free.
The traceless tide directly sources directional deformation. In
particular, if $\sigma=\varpi=0$ at a regular point, then
$\dot\sigma=-\cE_{\mathrm{TF}}$ there.

\paragraph{Twist.}
The antisymmetric projection has no curvature contribution and gives
\begin{align}
 \dot\varpi_{ab}={}&-\frac{2\theta}{m}\varpi_{ab}\nonumber\\
 &-\left(\sigma_a{}^c\varpi_{cb}
          +\varpi_a{}^c\sigma_{cb}\right).
 \label{eq:app-re-twist}
\end{align}
Equivalently, $D_s\varpi=-(2\theta/m)\varpi-\{\sigma,\varpi\}$,
where the braces denote the operator anticommutator. Neither curvature nor shear alone provides an additive twist source. Expansion and shear just transport existing twist through the flow. Since this equation is homogeneous
in $\varpi$, uniqueness on a regular segment implies
$\varpi(s_0)=0\Rightarrow\varpi(s)=0$ throughout that segment. All index contractions in these expressions use $h^{ab}$.

\subsection{Relation to quantum Fisher information}
\label[appendix]{app:re-label-metric}

The same calculation connects the evolution equations to the measurable
separations of the labeled states. In the parallel orthonormal screen
basis, $\cG_\perp=V^TV$ and Eq.~\eqref{eq:app-re-Ldot} gives
\begin{equation}
 \dot\cG_\perp
 =V^T(\cB^T+\cB)V
 =2V^T\left(\frac{\theta}{m}I_S+\sigma\right)V.
 \label{eq:app-re-Gram-evolution}
\end{equation}
Here $\cG_{\perp,AB}=g(\eta_A^\perp,\eta_B^\perp)$. The full label
metric obeys $\cG_{AB}=\cG_{\perp,AB}+a_Aa_B$.
Twist drops out of the instantaneous Gram-matrix derivative, but may
affect its later evolution through the coupled equations above.

For square invertible $V$, Jacobi's determinant identity yields
\begin{align}
 \partial_s\ln\sqrt{\det\cG_\perp}
 &=\partial_s\ln|\det V|\nonumber\\
 &=\operatorname{tr}_S(\dot V V^{-1})=\theta.
 \label{eq:app-re-volume}
\end{align}
This is why the trace equation governs the transverse-volume change.
If the family samples fewer than $m$ directions, $V$ is rectangular:
its Gram matrix remains meaningful, but its volume rate need not equal
the full screen expansion, nor does its curvature trace equal the full
$\operatorname{Ric}(U,U)$ in general.

At a caustic, $V$ loses rank, so $\dot V V^{-1}$ and the logarithmic
volume formula no longer define regular quantities. Thus the evolution equations derived here are exact
on each regular branch. Whether transverse collapse is also a loss of
the full label metric is determined separately by the longitudinal
components $a_A$.

\section{Radial Jacobi fields and congruences}
\label[appendix]{app:CP-jacobi}

We derive the radial Jacobi fields used in
\cref{eq:CP-J-complex,eq:CP-J-ordinary}. Throughout, the
Fubini-Study metric has the normalization
$g(X,Y)=\tfrac12\Tr(XY)$, and the reference geodesic has unit
velocity $U$. The essential simplification is that a
parallel-transported eigenbasis of the tidal operator reduces
the vector Jacobi equation to independent scalar equations.

\paragraph{A parallel frame adapted to the tidal curvature.}
The Jacobi equation is
\begin{equation}
  \nabla_U^2\eta+\cE(\eta)=0,
  \qquad
  \cE(V)=R(V,U)U.
  \label{eq:app-CP-jacobi}
\end{equation}
For a transverse vector $V\perp U$, the projective curvature
formula gives
\begin{equation}
  \cE(V)=V+3g(JU,V)JU.
\end{equation}
Thus the complex partner $JU$ is a tidal eigenvector with
eigenvalue $4$, while every vector orthogonal to both $U$
and $JU$ has eigenvalue $1$.

Choose $E_{\mathbb C}=JU$. This vector is parallel transported
along the reference geodesic:
\begin{equation}
  \nabla_U E_{\mathbb C}
  =(\nabla_UJ)U+J\nabla_UU=0,
\end{equation}
because the complex structure is covariantly constant and the
reference path is geodesic. Complete it to an orthonormal
screen frame with parallel-transported vectors $E_a$,
$a=1,\ldots,m-1$, where $m=2N-3$. Their orthogonality to
$U$ and $JU$ is preserved along the path. Consequently,
\begin{equation}
  \cE(E_{\mathbb C})=4E_{\mathbb C},
  \qquad
  \cE(E_a)=E_a
\end{equation}
holds at every arclength. For $N=2$, the screen contains only
$E_{\mathbb C}$.

\paragraph{Reduction to scalar equations.}
Expand a transverse Jacobi field in this frame:
\begin{equation}
  \eta(s)=j_{\mathbb C}(s)E_{\mathbb C}(s)
          +\sum_{a=1}^{m-1}j_a(s)E_a(s).
\end{equation}
Since the basis vectors are parallel transported, covariant
differentiation acts only on the coefficients:
\begin{equation}
  \nabla_U^2\eta
  =\ddot j_{\mathbb C}E_{\mathbb C}
   +\sum_{a=1}^{m-1}\ddot j_aE_a.
\end{equation}
Substitution into \cref{eq:app-CP-jacobi} therefore gives
\begin{equation}
  \ddot j_{\mathbb C}+4j_{\mathbb C}=0,
  \qquad
  \ddot j_a+j_a=0.
  \label{eq:app-CP-scalar-jacobi}
\end{equation}
The two tidal eigenvalues thus become squared frequencies
of the separation coefficients, with arclength playing
the role of the evolution parameter.

\paragraph{Radial initial conditions and their normalization.}
In a radial family, all geodesics start at the same projector
$P_0$, while their labels vary the initial direction.
The initial separation therefore vanishes. For a variation
along a chosen unit screen vector $E(0)$, normalize the
angular label so that the change of initial velocity is
$E(0)$. The corresponding initial conditions are
\begin{equation}
  \eta(0)=0,
  \qquad
  \left.\nabla_U\eta\right|_{s=0}=E(0).
  \label{eq:app-CP-radial-data}
\end{equation}
For its scalar coefficient, these become
$j(0)=0$ and $\dot j(0)=1$.

The solution of
$\ddot j+\lambda j=0$ with these initial conditions is
\begin{equation}
  j(s)=\frac{\sin(\sqrt{\lambda}\,s)}{\sqrt{\lambda}}.
  \label{eq:app-CP-mode-solution}
\end{equation}
Using $\lambda=4$ and $\lambda=1$ yields
\begin{equation}
  \begin{aligned}
    \eta_{\mathbb C}(s)
      &=\frac{\sin2s}{2}E_{\mathbb C}(s),\\
    \eta_a(s)&=\sin s\,E_a(s).
  \end{aligned}
  \label{eq:app-CP-radial-fields}
\end{equation}
The factor $1/2$ in the complex channel ensures that its
initial relative velocity has the same unit normalization
as the other channels. Indeed,
\begin{equation}
  \begin{aligned}
    j_{\mathbb C}(s)&=s-\frac23s^3+O(s^5),\\
    j_a(s)&=s-\frac16s^3+O(s^5).
  \end{aligned}
\end{equation}
Both separations initially grow as $s$. Their subsequent
difference is generated by curvature: the stronger tidal
curvature along $JU$ suppresses that separation more rapidly
relative to its initial linear growth.

These sine solutions apply specifically to the normalized
radial initial data in \cref{eq:app-CP-radial-data}.
A general congruence can have nonzero initial separation
and different initial relative velocities, leading to
different combinations of sine and cosine solutions.
While the deformation description
$\cB=\dot V V^{-1}$ becomes singular there, the parallel frame and the Jacobi fields themselves remain regular through their zeros.

Near the caustic, let $\varepsilon=s-\pi/2$. Then
\begin{equation}
 \begin{aligned}
 F^{\mathrm Q}_{\mathbb C}&=4\varepsilon^2+O(\varepsilon^4),
 &F^{\mathrm Q}_\perp&=4-4\varepsilon^2+O(\varepsilon^4),\\
 b_{\mathbb C}&=\varepsilon^{-1}+O(\varepsilon),
 &b_\perp&=-\varepsilon+O(\varepsilon^3).
 \end{aligned}
\label{eq:CP-channel-caustic-scaling}
\end{equation}

\paragraph{Consistency with the Raychaudhuri equation.}
Away from the Jacobi zeros, the principal deformation rates are
\begin{equation}
  b_{\mathbb C}=\frac{\dot j_{\mathbb C}}{j_{\mathbb C}}
              =2\cot2s,
  \qquad
  b_\perp=\frac{\dot j_a}{j_a}=\cot s.
\end{equation}
Writing $p=m-1=2N-4$, the expansion and shear satisfy
\begin{equation}
  \theta=b_{\mathbb C}+p\,b_\perp,
  \qquad
  \frac{\theta^2}{m}+\norm{\sigma}^2
       =b_{\mathbb C}^2+p\,b_\perp^2.
\end{equation}
The scalar Jacobi equations imply
$\dot b_{\mathbb C}+b_{\mathbb C}^2=-4$ and
$\dot b_\perp+b_\perp^2=-1$. Hence
\begin{equation}
  \dot\theta+\frac{\theta^2}{m}
  +\norm{\sigma}^2+2N
  =-4-p+2N=0.
\end{equation}
This is precisely the Raychaudhuri equation for the
twist-free radial family, with $\Ric(U,U)=2N$.

\section{Projector derivatives and normal contractions}
\label[appendix]{app:projector}

We derive the projector formula for the second fundamental form
used in \cref{sec:extrinsic}.  Differentiating $P^2=P$ shows that $\partial_\mu P$ is
off-diagonal with respect to $P$ and $Q=I-P$. In contrast, further differentiation $\partial_\mu \partial_\nu P$ determines diagonal blocks in $P$ and $Q$, while its off-diagonal blocks contain the ambient covariant derivative :
\begin{equation}
\begin{aligned}
 P(\partial_\mu\partial_\nu P)P
 &=-P\{\partial_\mu P,\partial_\nu P\}P,\\
 Q(\partial_\mu\partial_\nu P)Q
 &=Q\{\partial_\mu P,\partial_\nu P\}Q.
\end{aligned}
\label{eq:projector-Hessian-blocks}
\end{equation}

Subtracting these blocks and also the components tangent to $\Sigma$ then gives
\begin{equation}
\begin{aligned}
 \Pi_{\mu\nu}
 ={}&\partial_\mu\partial_\nu P
 +P\{\partial_\mu P,\partial_\nu P\}P\\
 &-Q\{\partial_\mu P,\partial_\nu P\}Q
 -\Gamma^\rho_{\mu\nu}\partial_\rho P.
\end{aligned}
\label{eq:B-expanded}
\end{equation}
This is the expanded form of the main-text Gauss formula. $\Pi_{\mu\nu}$ therefore measures bending within projective
state space.

The normal contractions entering the Gauss equation can be
evaluated without choosing a normal basis. Writing
$T_{ab}=[P,[P,\partial_a\partial_bP]]$ and
$\Gamma_{e,ab}=g_{ef}\Gamma^f_{ab}
=\tfrac12\Tr[(\partial_eP)(\partial_a\partial_bP)]$ gives
\begin{equation}
 \langle\Pi_{ab},\Pi_{cd}\rangle
 =\tfrac12\Tr(T_{ab}T_{cd})
  -g_{ef}\Gamma^e_{ab}\Gamma^f_{cd}.
 \label{eq:normal-projector-contraction}
\end{equation}
The second term subtracts the tangential contribution, leaving
precisely the normal contraction required for the extrinsic
curvature. This provides a direct projector-based expression
\cite{AvdoshkinPopov2023,Mitscherling2025}.
Ordinary derivatives here are taken in a fixed ambient Hilbert-space basis.

\section{XY ground-state geometry at finite size}
\label[appendix]{app:XY-finite}

\subsection{Ground-state representation and conventions}

Under Jordan-Wigner transformation, with $\sigma_j^z=1-2c_j^\dagger c_j$, even  parity gives antiperiodic
fermion momenta
$k_m=(2m+1)\pi/L$, $m=0,\ldots,L/2-1$ in the positive half zone \cite{MbengRussomannoSantoro2024}.
A fixed momentum-dependent phase convention for the fermions brings each
paired block to
\begin{equation}
\begin{aligned}
H_k&=-2\begin{pmatrix}
 d_k&\gamma\sin k\,e^{i\phi}\\
 \gamma\sin k\,e^{-i\phi}&-d_k
 \end{pmatrix},\\
d_k&=h-\cos k,\qquad \epsilon_k=\sqrt{d_k^2+\gamma^2\sin^2k}.
\end{aligned}
\label{eq:XY-mode-H}
\end{equation}
The ground and pair-excited energies are $-2\epsilon_k$ and
$+2\epsilon_k$.
The state and angular coordinates are
\begin{equation}
\begin{aligned}
|\Omega\rangle&=\bigotimes_{k>0}|w_k\rangle,\\
|w_k\rangle&=\cos\frac{\vartheta_k}{2}|00\rangle+
 e^{-i\phi}\sin\frac{\vartheta_k}{2}|11\rangle,\\
\cos\vartheta_k&=\frac{d_k}{\epsilon_k},\qquad
\sin\vartheta_k=\frac{\gamma\sin k}{\epsilon_k}.
\end{aligned}
\label{eq:XY-BCS}
\end{equation}
The generator $N_f/2$ has eigenvalues $0,1$ on these two states.  In spin
variables $[\hat G,\sigma_j^+]=-\sigma_j^+/2$, so
$e^{-i\phi\hat G}\sigma_j^+\sigma_{j+1}^+e^{i\phi\hat G}
=e^{i\phi}\sigma_j^+\sigma_{j+1}^+$.  This fixes both the sign and the factor of two in
\cref{eq:XY-mode-H,eq:XY-BCS}.  All calculations
follow the smooth parity-resolved state. If the global ground state
switches parity along a finite-size path, the selected family remains the
lowest state in its conserved sector, not the global ground state through the
crossing.

Take the orthogonal pair state
$|w_k^\perp\rangle=-\sin(\vartheta_k/2)|00\rangle+
 e^{-i\phi}\cos(\vartheta_k/2)|11\rangle$.  Its tangent amplitude is
\begin{equation}
\begin{aligned}
\langle w_k^\perp|\dot w_k\rangle
 &=\tfrac12\dot\vartheta_k-\tfrac i2\sin\vartheta_k\dot\phi,\\
g^{\rm tot}(\dot\lambda,\dot\lambda)
 &=\tfrac14\sum_{k>0}(\dot\vartheta_k^2+
 \sin^2\vartheta_k\dot\phi^2).
\end{aligned}
\label{eq:XY-mode-mixing}
\end{equation}
A derivative of the product state excites one such pair at a time, and these
states are mutually orthogonal.  This proves additivity of the squared
interband-mixing amplitudes without an extensivity assumption.

Writing $a,b\in\{\gamma,h,\phi\}$ and
$\vartheta_{k,a}=\partial_a\vartheta_k$, the metric and
Berry curvature are
\begin{align}
 g^{\rm tot}_{ab}&=\frac14\sum_{k>0}
 (\vartheta_{k,a}\vartheta_{k,b}+
 \sin^2\vartheta_k\delta_{a\phi}\delta_{b\phi}),
 \label{eq:XY-finite-metric}\\
 \cF^{\rm tot}_{a\phi}&=\frac12\sum_{k>0}
 \sin\vartheta_k\vartheta_{k,a},\qquad
 \cF^{\rm tot}_{\gamma h}=0,
 \label{eq:XY-finite-Berry}\\
 \vartheta_{k,h}&=-\frac{\gamma\sin k}{\epsilon_k^2},\qquad
 \vartheta_{k,\gamma}=\frac{d_k\sin k}{\epsilon_k^2}.
 \label{eq:XY-angle-derivatives}
\end{align}
Here $\delta_{a\phi}$ is the Kronecker delta, equal to
one for $a=\phi$ and zero otherwise. Since
$\vartheta_k$ is independent of $\phi$, the phase
contribution appears only in $g^{\rm tot}_{\phi\phi}$,
and $g^{\rm tot}_{\gamma\phi}=g^{\rm tot}_{h\phi}=0$.

\subsection{Metric derivatives and the many-body normal tensor}

The derivatives of the quantum metric follow from analytic derivatives of the Bogoliubov angle.
For $r_k=d_k+i\gamma\sin k$, let
$r_{k,a}=(i\sin k,1,0)_a$.  Since $\vartheta_k=\arg r_k$,
\begin{equation}
\begin{aligned}
\vartheta_{k,a}&=\Im\frac{r_{k,a}}{r_k},\qquad
 \vartheta_{k,ab}=\Im\frac{-r_{k,a}r_{k,b}}{r_k^2},\\
\vartheta_{k,abc}&=\Im\frac{2r_{k,a}r_{k,b}r_{k,c}}{r_k^3}.
\end{aligned}
\label{eq:XY-third-jet}
\end{equation}
Substitution in \cref{eq:XY-finite-metric} gives $g$, $\partial g$, and
$\partial^2g$ analytically.  For instance,
\begin{equation}
\begin{aligned}
\partial_cg_{ab}^{\rm tot}&=\tfrac14\sum_k
 (\vartheta_{k,ac}\vartheta_{k,b}+\vartheta_{k,a}\vartheta_{k,bc}),\\
\partial_d\partial_cg_{ab}^{\rm tot}&=\tfrac14\sum_k
 \Big(\vartheta_{k,acd}\vartheta_{k,b}+\vartheta_{k,ac}\vartheta_{k,bd}\\
 &\qquad+\vartheta_{k,ad}\vartheta_{k,bc}+\vartheta_{k,a}\vartheta_{k,bcd}\Big).
\end{aligned}
\label{eq:XY-metric-jets}
\end{equation}
for $a,b$ in the real parameter plane. Differentiating
$\tfrac14\sum\sin^2\vartheta_k$ handles the $\phi\phi$ component.

We now calculate the second fundamental form directly from the
many-body wavefunction. This requires separating the change of
the state into its phase, its motion along the ground-state
manifold, and its bending away from that manifold.

Write the normalized ground state as $\ket{\psi(\lambda)}$,
with $\lambda^a=(\gamma,h,\phi)$, and denote its derivatives by
$\ket{\psi_a}=\partial_a\ket{\psi}$ and
$\ket{\psi_{ab}}=\partial_a\partial_b\ket{\psi}$.
Normalization implies that
$A_a=\braket{\psi|\psi_a}$ is purely imaginary.
The component $A_a\ket{\psi}$ therefore describes an
infinitesimal phase change. Removing it gives $\ket{\xi_a}
 =\ket{\psi_a}-A_a\ket{\psi},
 \braket{\psi|\xi_a}=0$.
These vectors represent physical changes of the state, and
their real overlaps give the many-body quantum metric,$
 g_{ab}=\Re\braket{\xi_a|\xi_b}$.
All metric quantities in this calculation use the total,
rather than per-site, normalization.

To determine bending, we must differentiate these physical
tangent vectors once more. The ordinary second derivative
$\ket{\psi_{ab}}$ also contains phase-dependent contributions.
The covariant derivative in the ambient projective space $P=\ket{\psi}\bra{\psi}$
removes these contributions:
\begin{equation}
\begin{aligned}
 \ket{t_{ab}}
 &=(1-P)(\partial_a-A_a)\ket{\xi_b}\\
 &=(1-P)\ket{\psi_{ab}}
   -A_a\ket{\xi_b}-A_b\ket{\xi_a}.
\end{aligned}
\end{equation}
Thus $\ket{t_{ab}}$ describes how the physical tangent vector
changes within the full space of quantum states. It still
contains a component tangent to the ground-state manifold.

The tangent component is $\Gamma^c_{ab}\ket{\xi_c}$, where
\begin{equation}
 \Gamma^c_{ab}
 =g^{cd}\Re\braket{\xi_d|t_{ab}}.
\end{equation}
Subtracting it leaves the normal component,
\begin{equation}
 \ket{b_{ab}}
 =\ket{t_{ab}}-\Gamma^c_{ab}\ket{\xi_c},
 \qquad
 \Re\braket{\xi_d|b_{ab}}=0.
 \label{eq:XY-horizontal-B}
\end{equation}
This is the wavefunction representation of the second
fundamental form $\Pi_{ab}$: it measures the part of the
change in a tangent direction that points outside the
parametrised ground-state manifold. 
To express this normal vector in projector notation, recall
that a horizontal wavefunction variation $\ket{\delta\psi}$
corresponds to
$\delta P=\ket{\delta\psi}\bra\psi+
\ket\psi\bra{\delta\psi}$.
Applying the same correspondence to $\ket{b_{ab}}$ gives
\begin{equation}
 \Pi_{ab}
 =\ket{b_{ab}}\bra\psi+\ket\psi\bra{b_{ab}}.
\end{equation}
Since $\braket{\psi|b_{ab}}=0$, the projector metric yields
\begin{equation}
 \langle\Pi_{ab},\Pi_{cd}\rangle
 =\frac12\Tr(\Pi_{ab}\Pi_{cd})
 =\Re\braket{b_{ab}|b_{cd}}.
 \label{eq:NormalContractions}
\end{equation}
Thus the wavefunction and projector descriptions give
the same normal contractions entering the Gauss equation. The real part is essential because orthogonality here is
defined by the quantum metric.

The metric derivatives obtained from
\eqref{eq:XY-third-jet} determine the intrinsic curvature.
The normal vectors in \eqref{eq:XY-horizontal-B} independently
determine the extrinsic terms in the Gauss equation, allowing
the two calculations to be checked against one another.

\subsection{Second fundamental form for the ground state in XY spin chain.}
\label[appendix]{app:XY-SFF}

For the XY ground state,
$\ket{\psi}=\bigotimes_{k>0}\ket{w_k}$, the second derivative
contains two types of terms:
\begin{equation}
\begin{aligned}
 \ket{\psi_{ab}}
 ={}&\sum_k
 \ket{\partial_a\partial_b w_k}
 \bigotimes_{\ell\ne k}\ket{w_\ell}\\
 &+\sum_{k\ne\ell}
 \ket{\partial_a w_k}\otimes\ket{\partial_b w_\ell}
 \bigotimes_{j\ne k,\ell}\ket{w_j}.
\end{aligned}
\end{equation}
 The first sum differentiates
one momentum pair twice and the second differentiates two
distinct pairs. Both must be retained when calculating
$\ket{b_{ab}}$, even though the quantum metric itself is a
sum of separate pair contributions.

With $\ket{w_k^\perp}$ the normalized state orthogonal to $\ket{w_k}$, in order to describe derivatives of this product define
\begin{equation}
\begin{aligned}
 \ket{k}
 &:=\ket{w_k^\perp}\!
   \bigotimes_{\substack{q>0\\q\ne k}}\ket{w_q},\\
 \ket{k\ell}
 &:=\ket{w_k^\perp}\otimes\ket{w_\ell^\perp}\!
   \bigotimes_{\substack{q>0\\q\ne k,\ell}}\ket{w_q},
 \qquad k<\ell.
\end{aligned}
\end{equation}
The factors retain their original momentum ordering.
Thus $\ket{k}$ differs from $\ket\Psi$ only in the
$(k,-k)$ pair, while $\ket{k\ell}$ differs in both
the $(k,-k)$ and $(\ell,-\ell)$ pairs.
Since $\braket{w_q|w_q^\perp}=0$, these replacement
states are mutually orthonormal and orthogonal to
$\ket\Psi$. They label changes of entire two-level
pair states, not individual fermion occupations.

 For a parameter-space
vector $X$,
\begin{equation}
\begin{aligned}
 \alpha_k^X&=\braket{w_k|\partial_Xw_k}
       =-i\sin^2\frac{\vartheta_k}{2}\,X^\phi,\\
 \beta_k^X&=\braket{w_k^\perp|\partial_Xw_k}
       =\frac12\partial_X\vartheta_k
        -\frac i2\sin\vartheta_k\,X^\phi.
\end{aligned}
\label{eq:XY-NC-factor-derivatives}
\end{equation}
The first coefficient is a phase and the second moves the pair on its
Bloch sphere. Removing the many-body phase gives
$\ket{\xi_X}=\sum_k\beta_k^X\ket{k}$, and hence
\begin{equation}
 g^{(L)}(X,Y)+i\,\omega^{(L)}(X,Y)
 =\frac1L\sum_k\overline{\beta_k^X}\beta_k^Y.
\end{equation}
For later convenience, define $q_k(X,Y)=\overline{\beta_k^X}\beta_k^Y=g_k(X,Y)+i\omega_k(X,Y)$,
with $q=\sum_kq_k=g^{\rm tot}+i\omega^{\rm tot}$.

\paragraph{Computations of normals.}
The horizontal ambient second derivative of the product is
\begin{equation}
\begin{aligned}
 \ket{t_{XY}}&=\sum_k\tau_k^{XY}\ket{k}
             +\sum_{k<\ell}\kappa_{k\ell}^{XY}\ket{k\ell},\\
 \tau_k^{XY}&=\beta_k^{XY}
              -\alpha_k^X\beta_k^Y-\alpha_k^Y\beta_k^X,\\
 \kappa_{k\ell}^{XY}&=\beta_k^X\beta_\ell^Y
                      +\beta_\ell^X\beta_k^Y,
\end{aligned}
\label{eq:XY-NC-product}
\end{equation}
where $\beta_k^{XY}=\braket{w_k^\perp|\partial_X\partial_Yw_k}$.
The phases of the other factors cancel against the many-body phase
subtraction in \cref{eq:XY-horizontal-B}. Every tangent vector
$\ket{\xi_\mu}$ lies in the span of the $\ket{k}$, so the tangential
subtraction acts only on the single-pair sector:
\begin{equation}
\begin{aligned}
 \pi_k(X,Y)&=\tau_k^{XY}-\Gamma^\rho_{\mu\nu}X^\mu Y^\nu
                                  \beta_k^{\partial_\rho},\\
 \ket{b^{\rm single}_{XY}}&=\sum_k\pi_k(X,Y)\ket{k},\\
 \ket{b^{\rm two}_{XY}}&=\sum_{k<\ell}\kappa_{k\ell}^{XY}\ket{k\ell}.
\end{aligned}
\label{eq:XY-NC-normal}
\end{equation}
These are orthogonal horizontal representatives of
$\Pi^{\rm single}$ and $\Pi^{\rm two}$. The first is normal to
$\Sigma$ within the product of Bloch spheres and the second is normal to
the whole product.

\subsection{Exact cancellations in parts of ambient and extrinsic components of the Tidal tensor}
\label[appendix]{app:XY-bending-proof}
\label[appendix]{app:XY-normal-contractions}

For a geodesic congruence, let tangent $U$ and the screen $E_a$ to be unit vectors in the total metric. 
Using \eqref{eq:NormalContractions} and the orthonormality
of the two-pair replacement states gives
\begin{equation}
 \langle\Pi^{\rm two}(X,Y),\Pi^{\rm two}(Z,W)\rangle
 =\Re\sum_{k<\ell}
 \overline{\kappa_{k\ell}^{XY}}\kappa_{k\ell}^{ZW}.
\end{equation}
Substituting the coefficients from
\eqref{eq:XY-NC-product}, and writing
$q_k(X,Y)=\overline{\beta_k^X}\beta_k^Y$ and
$q(X,Y)=\sum_kq_k(X,Y)$, yields the two contractions
entering the Gauss equation:
\begin{equation}
\begin{aligned}
 &\langle\Pi^{\rm two}_{ab},\Pi^{\rm two}_{UU}\rangle\\
 &=2\Re[q(E_a,U)q(E_b,U)]\\
 &\quad-2\Re\sum_kq_k(E_a,U)q_k(E_b,U),\\
 &\langle\Pi^{\rm two}_{aU},\Pi^{\rm two}_{bU}\rangle\\
 &=\Re[q(E_a,E_b)q(U,U)]
       +\Re[q(E_a,U)q(U,E_b)]\\
 &\quad-2\Re\sum_kq_k(E_a,E_b)q_k(U,U).
\end{aligned}
\end{equation}

The subtracted single sums remove the equal-pair terms
introduced by the unrestricted double sums. We now simplify
the difference of the two contractions, which is the
combination entering the Gauss equation.

Recall that $q=g+i\omega$ and $v_a=\omega(U,E_a)$.
Since $U$ is unit and the screen vectors are transverse,
$q(E_a,U)=-iv_a, q(U,U)=1, \Re q(E_a,E_b)=h_{ab}.$
The products of unrestricted sums therefore contribute
$-2v_av_b$ to the first contraction and
$h_{ab}+v_av_b$ to the second. Their difference is
$-h_{ab}-3v_av_b$, exactly minus the ambient tide.
To simplify the remaining single sums, use the two-level
nature of each momentum pair. Its horizontal tangent space
has one complex dimension, so
$q_k(X,Y)=\overline{\beta_k^X}\beta_k^Y$ obeys $q_k(U,U)q_k(E_a,E_b) =q_k(E_a,U)q_k(U,E_b).$
Taking the real part gives
\begin{equation}
\begin{aligned}
 &g_k(U,U)g_k(E_a,E_b)
 -g_k(U,E_a)g_k(U,E_b)\\
 &\qquad=\omega_k(U,E_a)\omega_k(U,E_b).
\end{aligned}
\label{eq:XY-factor-tidal-form}
\end{equation}
Using this identity in the remaining terms yields
\begin{equation}
\begin{aligned}
 &\langle\Pi^{\rm two}_{ab},\Pi^{\rm two}_{UU}\rangle
 -\langle\Pi^{\rm two}_{aU},\Pi^{\rm two}_{bU}\rangle\\
 &\qquad=-h_{ab}-3v_av_b
 +4\sum_k\omega_k(U,E_a)\omega_k(U,E_b).
\end{aligned}
\label{eq:XY-two-pair-identity}
\end{equation}
Thus the two-pair contribution cancels the explicit
ambient tide, including its collective Berry enhancement.
What remains is a sum of curvature contributions from
individual momentum pairs, together with the single-pair
normal contractions. This is the Gauss equation for the product of Bloch
spheres inside projective space, the Segre embedding.\footnote{See
B.-Y. Chen, \emph{Segre embedding and related maps and immersions in
differential geometry}, Arab J. Math. Sci. \textbf{8}, 1-39 (2002),
\href{https://arxiv.org/abs/1307.0459}{arXiv:1307.0459}.}
The identity holds throughout a regular product-state family.

The remaining Ising-point calculation returns to the metric per
site and the velocity \eqref{eq:XY-general-Ising-velocity}.
For fixed parameter vectors the normal contractions are divided
by $L$, so the individual-pair term becomes
$4L^{-1}\sum_k\omega_k(U,V_A)\omega_k(U,V_B)$.

\subsection{Exact finite-size sums at the Ising point}
\label[appendix]{app:XY-exact-proof}

At $(\gamma,h)=(1,0)$, $\epsilon_k=1$ for every momentum and
$\vartheta_k=\pi-k$. With $S_k=\sin k$ and $C_k=\cos k$,
\begin{equation}
\begin{aligned}
 \vartheta_{k,h}&=-S_k,&\vartheta_{k,\gamma}&=-S_kC_k,\\
 \vartheta_{k,hh}&=-2S_kC_k,&
 \vartheta_{k,\gamma\gamma}&=2S_k^3C_k,\\
 \vartheta_{k,\gamma h}&=-S_k\cos(2k).
\end{aligned}
\label{eq:XY-Ising-angle-jets}
\end{equation}
Every mode sum below is therefore a finite trigonometric sum. The
antiperiodic momenta obey
\begin{equation}
 \frac1L\sum_{m=0}^{L-1}e^{i\ell(2m+1)\pi/L}=0,
 \qquad 0<|\ell|<L,
 \label{eq:XY-root-unity}
\end{equation}
so $\sum_{k>0}\cos(2nk)=0$ for $n=1,2,3$ and every even $L\ge8$.
Expanding into these harmonics gives
\begin{equation}
\begin{aligned}
 \sum_k S_k^2&=\frac L4,\qquad \sum_kS_k^4=\frac{3L}{16},\\
 \sum_k S_k^2C_k^2&=\frac L{16},\qquad
 \sum_k S_k^4C_k^2=\frac L{32},\\
 \sum_kS_k^2\bigl[\cos(2k)+\tfrac12\bigr]^2&=\frac L{16},\\
 \sum_kS_k^2\bigl(C_k^2-\tfrac14\bigr)^2&=\frac L{64}.
\end{aligned}
\label{eq:XY-single-sums}
\end{equation}
Sums with an odd power of $C_k$ and even powers of $S_k$ vanish under
$k\mapsto\pi-k$. In particular
$g^{\rm tot}=L\operatorname{diag}(1/64,1/16,1/16)$ and
$\cF^{\rm tot}_{\gamma a}=0$ for $a\in\{\gamma,h,\phi\}$. 
Although individual second derivatives of the metric in \cref{eq:XY-metric-jets}
can contain eighth harmonics, these contributions cancel
from the curvature combinations considered here.
The surviving contractions reduce to harmonics through
order six, giving the stated sectional curvatures \eqref{eq:XY-sectional-curvatures} 
exactly for every even $L\ge8$:
$(K^{\rm tot}_{\gamma h},K^{\rm tot}_{\gamma\phi},K^{\rm tot}_{h\phi})
=(-16,4,8)/L$. The normal contractions below reproduce these values
through the Gauss equation, which checks the extrinsic calculation
against the intrinsic one.

\paragraph{Single-pair bending.}
At the Ising point $\beta_k^X=-\tfrac12S_k(X^\gamma C_k+X^h+iX^\phi)$,
and \cref{eq:XY-single-sums} gives
$g^{(L)}=\operatorname{diag}(1/64,1/16,1/16)$ and the nonzero
Christoffel symbols
\begin{equation}
\begin{aligned}
 \Gamma^\gamma_{\gamma\gamma}&=-1,&
 \Gamma^\gamma_{hh}&=2,&\Gamma^\gamma_{\phi\phi}&=-1,\\
 \Gamma^h_{\gamma h}&=-\tfrac12,&
 \Gamma^\phi_{\gamma\phi}&=\tfrac14.
\end{aligned}
\label{eq:XY-NC-coordinate-connection}
\end{equation}

Differentiating a factor twice and removing its phase gives
\begin{equation}
\begin{aligned}
 \tau_{k,\mu\nu}
 &=\frac12\vartheta_{k,\mu\nu}
 -\frac i2\cos\vartheta_k
       (\delta_{\mu\phi}\vartheta_{k,\nu}
       +\delta_{\nu\phi}\vartheta_{k,\mu})\\
 &\quad-\frac12\delta_{\mu\phi}\delta_{\nu\phi}
                      \sin\vartheta_k\cos\vartheta_k.
\end{aligned}
\end{equation}
Subtracting \cref{eq:XY-NC-coordinate-connection} leaves
$\pi_{k,\mu\nu}:=\pi_k(\partial_\mu,\partial_\nu)$:
\begin{equation}
\begin{aligned}
 \pi_{k,\gamma\gamma}&=-\tfrac12S_kC_k\cos(2k),\\
 \pi_{k,\gamma h}&=-\tfrac12S_k[\cos(2k)+\tfrac12],\\
 \pi_{k,\gamma\phi}&=-\tfrac i2S_k(C_k^2-\tfrac14),\\
 \pi_{k,h\phi}&=-\tfrac i2S_kC_k,
 \qquad \pi_{k,hh}=\pi_{k,\phi\phi}=0.
\end{aligned}
\label{eq:XY-NC-coordinate-pi}
\end{equation}
For instance, $\tau_{k,hh}=-S_kC_k$ is cancelled by
$\Gamma^\gamma_{hh}\beta_k^{\partial_\gamma}$.

For the tide we use $T_u=\partial_h$ and $T_\phi=\partial_\phi$ in
place of the screen vectors $V_A=T_A-Q_AU$. By symmetry and
bilinearity of $\Pi^\chi$, this substitution leaves the tidal
difference in \eqref{eq:tidal-extrinsic} unchanged, although the
individual contractions change. By bilinearity,
\begin{equation}
\begin{aligned}
 \pi_k(T_u,U)&=-4c_0S_k[\cos(2k)+\tfrac12]-8iQ_\phi S_kC_k,\\
 \pi_k(T_\phi,U)&=-4ic_0S_k(C_k^2-\tfrac14)-8iQ_uS_kC_k,\\
 \pi_k(U,U)&=-32c_0^2S_kC_k\cos(2k)\\
 &\quad-128c_0Q_uS_k[\cos(2k)+\tfrac12]\\
 &\quad-128ic_0Q_\phi S_k(C_k^2-\tfrac14)\\
 &\quad-256iQ_uQ_\phi S_kC_k.
\end{aligned}
\end{equation}
With $L^{-1}\Re\sum_k\overline{\pi_k(X,Y)}\pi_k(Z,W)$ and
\cref{eq:XY-single-sums},
\begin{equation}
\begin{gathered}
 \langle\Pi^{\rm single}(T_A,T_B),\Pi^{\rm single}(U,U)\rangle_{(L)}\\
 =\begin{pmatrix}0&8Q_uQ_\phi\\8Q_uQ_\phi&0\end{pmatrix},\\
 \langle\Pi^{\rm single}(T_A,U),\Pi^{\rm single}(T_B,U)\rangle_{(L)}\\
 =\begin{pmatrix}
 1-16Q_u^2-12Q_\phi^2&4Q_uQ_\phi\\
 4Q_uQ_\phi&1/4-4Q_\phi^2
 \end{pmatrix}.
\end{gathered}
\label{eq:XY-single-normal-matrices}
\end{equation}
For example, the $uu$ entry of the second matrix is
$L^{-1}\sum_k\{16c_0^2S_k^2[\cos(2k)+\tfrac12]^2+64Q_\phi^2S_k^2C_k^2\}
=c_0^2+4Q_\phi^2$. The real-imaginary cross term vanishes. Each
contraction is a single sum divided by $L$, so this sector is $O(1)$.

\paragraph{Two-pair bending.}
Here $\beta_k^{T_u}=-S_k/2$, $\beta_k^{T_\phi}=-iS_k/2$, and
$\beta_k^U=-S_kF_k$ with $F_k=4c_0C_k+8Q_u+8iQ_\phi$. Hence
\begin{equation}
\begin{aligned}
 \kappa_{k\ell}^{T_uT_u}&=\tfrac12S_kS_\ell,&
 \kappa_{k\ell}^{T_uT_\phi}&=i\kappa_{k\ell}^{T_uT_u},\\
 \kappa_{k\ell}^{T_\phi T_\phi}&=-\kappa_{k\ell}^{T_uT_u},&
 \kappa_{k\ell}^{UU}&=2S_kS_\ell F_kF_\ell,\\
 \kappa_{k\ell}^{T_uU}&=\tfrac12S_kS_\ell(F_k+F_\ell),&
 \kappa_{k\ell}^{T_\phi U}&=i\kappa_{k\ell}^{T_uU}.
\end{aligned}
\end{equation}
Writing a sum over distinct pairs as the product of single sums minus
the equal-momentum terms, $2\sum_{k<\ell}f_kf_\ell=(\sum_kf_k)^2-\sum_kf_k^2$,
gives
\begin{equation}
\begin{aligned}
 \sum_{k<\ell}S_k^2S_\ell^2&=\frac{L(L-3)}{32},\\
 \sum_{k<\ell}S_k^2S_\ell^2C_kC_\ell&=-\frac L{64},\\
 \sum_{k<\ell}S_k^2S_\ell^2(C_k+C_\ell)^2&=\frac{L(L-4)}{64}.
\end{aligned}
\label{eq:XY-NC-double-sums}
\end{equation}
The product term scales as $L^2$, whereas the subtraction scales
as $L$. With the overall factor $1/L$, these yield the $O(L)$
and $O(1)$ contributions to the two-pair tide, respectively. The contractions are
\begin{equation}
\begin{aligned}
 &\langle\Pi^{\rm two}(T_A,T_B),\Pi^{\rm two}(U,U)\rangle_{(L)}\\
 &=2(L-3)\begin{pmatrix}
 Q_u^2-Q_\phi^2&2Q_uQ_\phi\\
 2Q_uQ_\phi&-Q_u^2+Q_\phi^2
 \end{pmatrix}\\
 &\quad-\frac{c_0^2}{4}\operatorname{diag}(1,-1),\\
 &\langle\Pi^{\rm two}(T_A,U),\Pi^{\rm two}(T_B,U)\rangle_{(L)}\\
 &=\left[\frac{c_0^2(L-4)}{16}
          +2(L-3)(Q_u^2+Q_\phi^2)\right]I_2.
\end{aligned}
\label{eq:XY-two-normal-matrices}
\end{equation}
The $uu$ entries follow respectively from
$L^{-1}\Re\sum_{k<\ell}S_k^2S_\ell^2F_kF_\ell$ and
$(4L)^{-1}\sum_{k<\ell}S_k^2S_\ell^2|F_k+F_\ell|^2$.
Terms linear in $C_k+C_\ell$ vanish by symmetry, and the mixed entry
of the second matrix vanishes because
$\kappa^{T_\phi U}=i\kappa^{T_uU}$.

\paragraph{Assembling the tide.}
In the projected screen $V_A=T_A-Q_AU$, the surviving terms are
\begin{equation}
\begin{aligned}
 &\langle\Pi^{\rm single}(V_A,V_B),\Pi^{\rm single}(U,U)\rangle_{(L)}\\
 &\quad-\langle\Pi^{\rm single}(V_A,U),\Pi^{\rm single}(V_B,U)\rangle_{(L)}\\
 &=\begin{pmatrix}
 -1+16Q_u^2+12Q_\phi^2&4Q_uQ_\phi\\
 4Q_uQ_\phi&-1/4+4Q_\phi^2
 \end{pmatrix},\\
 &\frac4L\sum_k\omega_k(U,V_A)\omega_k(U,V_B)\\
 &=\begin{pmatrix}
 12Q_\phi^2&-12Q_uQ_\phi\\
 -12Q_uQ_\phi&1/2+4Q_u^2-8Q_\phi^2
 \end{pmatrix}.
\end{aligned}
\label{eq:XY-charge-tides}
\end{equation}
For example, evaluating the $uu$ component of the two-pair
difference with $T_A$ gives
$L(-1/16+Q_u^2-3Q_\phi^2)+12Q_\phi^2$.
The term proportional to $L$ equals $-L(h_{uu}+3v_u^2)$
and cancels the ambient contribution, leaving $12Q_\phi^2$,
the $uu$ component of the individual-pair sum above. Adding the two surviving matrices
gives \cref{eq:XY-general-intrinsic-tide}. For $c_0\ne0$, their trace is obtained by contracting with
$h^{AB}=16\delta^{AB}+256Q_AQ_B/c_0^2$.
At $c_0=0$, the chosen basis becomes degenerate, and the trace
is instead evaluated in an orthonormal screen basis.

As a check, at $Q_u=Q_\phi=0$ the screen $(e_h,e_\phi)=4(T_u,T_\phi)$
gives single-pair and pair contributions $\operatorname{diag}(-16,-4)$
and $\operatorname{diag}(0,8)$ to \cref{eq:XY-density-tide}. More
generally, the single-pair and pair sectional curvatures in the
coordinate planes are
\begin{equation}
\begin{array}{c|rrr}
 &\gamma h&\gamma\phi&h\phi\\ \hline
 \text{Single-pair bending}&-16&-4&-4\\
 \text{Individual-pair curvature}&0&8&12\\ \hline
 K^\Sigma_{ij}&-16&4&8
\end{array}
\label{eq:XY-sector-curvatures}
\end{equation}
in per-site units, and the last row reproduces the metric curvatures
of \cref{app:XY-exact-proof}.

\paragraph{Normal acceleration.}
For the total-metric unit direction
$X=e_\gamma^{\rm tot}=8\partial_\gamma/\sqrt L$,
the single-pair amplitudes are
$\beta_k^X=-4\sin k\cos k/\sqrt L$ and
$\pi_k(X,X)=-8\sin(4k)/L$.
The two-pair coefficients are
$\kappa_{k\ell}^{XX}=2\beta_k^X\beta_\ell^X$.
Orthonormality of the replacement states therefore gives
\begin{equation}
\begin{aligned}
 \norm{\Pi^{\rm single}(X,X)}^2
 &=\frac{64}{L^2}\sum_{k>0}\sin^2(4k)
 =\frac{16}{L},\\
 \norm{\Pi^{\rm two}(X,X)}^2
 &=4\sum_{0<k<\ell}|\beta_k^X|^2|\beta_\ell^X|^2\\
 &=2\left(1-\sum_{k>0}|\beta_k^X|^4\right)
 =2-\frac6L.
\end{aligned}
\end{equation}
Here we used the unit-speed normalization
$\sum_{k>0}|\beta_k^X|^2=1$ and the exact sums
$\sum_{k>0}\sin^2(4k)=L/4$ and
$\sum_{k>0}|\beta_k^X|^4=3/L$.
The summands contain Fourier harmonics through order
eight, so these values hold for every even $L\ge10$.
Adding the two orthogonal contributions yields
$\norm{\Pi(X,X)}^2=2+10/L$, proving
\cref{eq:XY-bending-exact}.
At $L=8$, the eighth harmonic changes the single-pair
and two-pair contributions to $4$ and $3/2$,
respectively, giving $\norm{\Pi(X,X)}^2=11/2$.

\paragraph{Shear equation.}
For a twist-free bundle with a two-dimensional screen, $\sigma^2$ is
proportional to the identity, and in a parallel orthonormal screen
\begin{equation}
 \dot\sigma_{ab}=-\theta\sigma_{ab}-(\cE^\Sigma)_{\TF,ab}.
\end{equation}
Here $\cE^\Sigma$ is the sum of the explicit contributions in
\eqref{eq:XY-charge-tides}, transformed to this orthonormal screen.
In the projected basis $V_A$ the left side is the screen-covariant
derivative $(\nabla_U^\perp\sigma)_{AB}$. For a point-emitted family,
$V=sI-s^3\cE(0)/6+O(s^4)$ gives $\sigma=-s\cE_{\TF}(0)/3+O(s^2)$, and
\cref{eq:XY-density-tide} yields \cref{eq:XY-density-shear-onset}. At
$Q_u=Q_\phi=0$ the source $\operatorname{diag}(-10,10)$ consists of
$\operatorname{diag}(-6,6)$ from single-pair bending and
$\operatorname{diag}(-4,4)$ from pair curvature.

\subsection{Normalization}
\label[appendix]{app:XY-normalization}

Dividing the metric by $L$ leaves the connection and the curvature
map $R(X,Y)Z$ unchanged, but rescales unit vectors and arclength:
$s_{\rm tot}=\sqrt L\,s$ and $U_{(L)}=\sqrt L\,U_{\rm tot}$. In the
corresponding orthonormal frames,
\begin{equation}
\begin{aligned}
 \cE^{(L)}(s)&=L\,\cE^{\rm tot}(\sqrt L\,s),\\
 \sigma^{(L)}(s)&=\sqrt L\,\sigma^{\rm tot}(\sqrt L\,s).
\end{aligned}
 \label{eq:XY-normalization-laws}
\end{equation}
The conversion applies to each Gauss term separately when both the
ambient and the induced metric are divided by $L$. At the Ising point, the ambient and
two-pair terms grow with $L$, while their sum stays finite.

\section{Geodesic families of the quantum metric in ferromagnetic phase}
\label[appendix]{app:XY-congruence}

\subsection{Exact solutions and fold locations}
\label[appendix]{app:XY-K-proofs}
\label[appendix]{app:symmetry-Jacobi}

We use the metric per site throughout. Translations in
$u$ and $\phi$ preserve this metric and therefore map geodesics to
geodesics \cite{Kolodrubetz2013,LiskaGritsev2021}. Differentiating these
translations gives the connecting fields $\eta_A=\partial_A$, while
the Killing equation gives $\partial_s g(U,\partial_A)=0$.
Their conserved longitudinal components are $Q_A$, so the screen
fields are $\eta_A^\perp=\partial_A-Q_AU$ and
$\cG_{\perp,AB}=g_{AB}-Q_AQ_B$.

\paragraph{Exact trajectories and allowed momenta.}
\label[appendix]{app:XY-K-integration}

Starting from the geodesic Lagrangian
\eqref{eq:XY-geodesic-Lagrangian}, the Euler-Lagrange
equations for the two cyclic coordinates give
\begin{equation}
 \frac{d}{ds}(a^2\dot u)=0,\qquad
 \frac{d}{ds}(b^2\dot\phi)=0.
\end{equation}
Hence $a^2\dot u=Q_u$ and $b^2\dot\phi=Q_\phi$.
Substituting these into the unit-speed condition,
\begin{equation}
 \frac{\dot v^2}{4}
 +\frac{Q_u^2}{a^2}
 +\frac{Q_\phi^2}{b^2}=1,
\end{equation}
gives $\dot v^2=F(v)$ in
\eqref{eq:XY-K-first-integrals}.
The remaining Euler-Lagrange equation becomes
\begin{equation}
 \ddot v=\frac12F'(v)
 =-64Q_u^2\tan v\,\sec^2v
   +32Q_\phi^2\csc^2v\,\cot v.
\end{equation}
We first determine which momenta allow motion, requiring
$F(v)>0$ somewhere in $0<v<\pi/2$, and then integrate
the trajectories. For compactness, set
$\alpha=1+16Q_u^2$, $\beta=1+8Q_\phi^2$, and
$D=\beta^2-32\alpha Q_\phi^2$.

Motion is possible wherever $F(v)>0$. To determine the allowed
momenta, set $y=\tan^2v>0$, so that this condition becomes
\begin{equation}
 \min_{y>0}
 \left(16Q_u^2y+8Q_\phi^2+\frac{8Q_\phi^2}{y}\right)<1.
\end{equation}
For $Q_uQ_\phi\ne0$, the minimum occurs at
$y=|Q_\phi|/(\sqrt2|Q_u|)$ and gives
\[
 8Q_\phi^2+16\sqrt2\,|Q_uQ_\phi|<1.
\]
This is the condition stated in \cref{eq:XY-K-allowed}.
Equivalently, $D>0$ and $8Q_\phi^2<1$: the discriminant
ensures distinct roots, while the additional inequality
places them in the physical interval. At equality, the
two turning values coincide and $v$ remains constant.

To integrate the motion, set $x=\sin^2v$. Using
$\dot x=2\sqrt{x(1-x)}\,\dot v$ converts the first integral
into a quadratic polynomial:
\begin{equation}
\begin{aligned}
 \dot x^2&=16[-\alpha x^2+\beta x-8Q_\phi^2],\\
 \ddot x+16\alpha x&=8\beta.
\end{aligned}
 \label{eq:XY-K-oscillator}
\end{equation}
The second equation follows from the geodesic acceleration
equation and remains valid at the turns. Thus $x$ undergoes
harmonic motion about $x_c=\beta/(2\alpha)$:
\begin{equation}
\begin{aligned}
 x(s)&=x_c+d_x\cos(z_0+\omega s),\\
 \omega&=4\sqrt\alpha,\qquad
 d_x=\sqrt{x_c^2-\frac{8Q_\phi^2}{\alpha}}.
\end{aligned}
 \label{eq:XY-K-orbit}
\end{equation}
The first integral fixes the amplitude $d_x$, while the
initial position and velocity fix the phase $z_0$.
The turning values are $x_\pm=x_c\pm d_x$, with
$0<x_-<x_+<1$ under the strict inequality above.

Alternatively, the solution can be written directly in
terms of $x_0=\sin^2v_0$ and
$\dot x_0=\sin(2v_0)\dot v(0)$:
\begin{equation}
 x(s)=x_c+(x_0-x_c)\cos(\omega s)
             +\frac{\dot x_0}{\omega}\sin(\omega s).
\end{equation}
This form incorporates the initial direction of motion
without having to choose a branch for the phase $z_0$.
Once $x(s)=\sin^2v(s)$ is known, the remaining coordinates
follow by integrating the conserved-momentum equations:
\begin{equation}
\begin{aligned}
 u(s)-u_0
 &=16Q_u\int_0^s\frac{x(s')}{1-x(s')}\,\dd s',\\
 \phi(s)-\phi_0
 &=8Q_\phi\int_0^s\frac{\dd s'}{x(s')}.
\end{aligned}
 \label{eq:XY-K-primitives}
\end{equation}
Together with \cref{eq:XY-K-orbit}, these expressions specify
the full trajectory. Both coordinates evolve monotonically,
with directions set by the signs of their conserved momenta.

\paragraph{Recurrence within the physical phase.}
\label[appendix]{app:XY-K-count-proof}

Successive turns occur when the oscillator phase advances
by $\pi$. Integrating \cref{eq:XY-K-primitives} between
these turns gives
\begin{equation}
\begin{aligned}
 \Delta s&=\frac{\pi}{4\sqrt{1+16Q_u^2}},\\
 \Delta u&=\operatorname{sgn}(Q_u)d_u,\\
 d_u&=\pi\left(1-\frac{4|Q_u|}
                         {\sqrt{1+16Q_u^2}}\right)>0.
\end{aligned}
 \label{eq:XY-K-advance}
\end{equation}
The coordinate integrals use
$\int_{n\pi}^{(n+1)\pi}(c+d\cos z)^{-1}\dd z
=\pi/\sqrt{c^2-d^2}$ for $c>|d|$.

Although $v$ oscillates, the monotonic drift in $u$
eventually reaches the Ising critical line at $|u|=\pi/2$, or $|h|=1$. If $u_1$ is the coordinate
at the first future turn from a regular initial point
inside the phase, the number of interior folds is
\begin{equation}
 N_{\rm fold}=\max\!\left\{0,
 \left\lceil
 \frac{\pi/2-\operatorname{sgn}(Q_u)u_1}{d_u}
 \right\rceil\right\}.
 \label{eq:XY-K-count}
\end{equation}
The strict boundary exclusion is included in this count.
Under \cref{eq:XY-K-allowed}, both turning values satisfy
$0<v_-<v_+<\pi/2$.

\section{XY fold catastrophes and their fidelity signatures}
\label[appendix]{app:XY-caustics}

\subsection{The \texorpdfstring{$A_2$}{A2} fold and congruence deformation}
\label[appendix]{app:XY-K-ray-fold}

\paragraph{A smooth family in geodesic phase space.}
To identify the caustic within catastrophe theory, we express
the geodesic motion in Hamiltonian form.
The geodesic momenta are $p_v=\dot{v}/4$, $p_u=Q_u$,
and $p_\phi=Q_\phi$. Define $W(v)=16Q_u^2\tan^2v+8Q_\phi^2\csc^2v$.
The geodesic Hamiltonian is given by
\begin{equation}
 H_{\rm geo}=\frac12g^{\mu\nu}p_\mu p_\nu =2p_v^2+\frac12W(v).
\label{eq:XY-geodesic-Hamiltonian}
\end{equation}
Our arclength parametrization fixes
$g(U,U)=1$, hence $H_{\rm geo}=1/2$.

Fixing $p_u=Q_u$, $p_\phi=Q_\phi$, and $H_{\rm geo}=1/2$ leaves three independent coordinates in the six-dimensional geodesic phase space. At a simple turn, $p_v=0$, $W(v_*)=1$, and $c_*=-2W'(v_*)\ne0$. Expanding the energy constraint $4p_v^2+W(v)=1$ about this point gives
\begin{equation}
 v(p_v)=v_*+\frac8{c_*}p_v^2+O(p_v^4).
 \label{eq:XY-fold-momentum}
\end{equation}
Although $v$ reverses direction at the turn, its momentum continues to change: $\dot p_v(s_*)=c_*/4\ne0$. We can therefore use $(p_v,u,\phi)$ as smooth coordinates on the phase-space family through the turn.

Indeed, its parametrization by $(s,u_0,\phi_0)$ satisfies
\begin{equation}
 \left.
 \det\frac{\partial(p_v,u,\phi)}
          {\partial(s,u_0,\phi_0)}
 \right|_{s_*}
 =\dot p_v(s_*)=\frac{c_*}{4}\ne0.
\end{equation}
If we retain only the state coordinates $(v,u,\phi)$,
the corresponding determinant is instead $\dot v(s_*)=0$.

Variations within the submanifold parametrised by $(p_v,u,\phi)$ gives $dp_u=dp_\phi=0$
and $dv=v'(p_v)\,dp_v$. Consequently, the canonical
symplectic form
\begin{equation}
 \Omega_{\rm can}=dv\wedge dp_v+du\wedge dp_u
                 +d\phi\wedge dp_\phi
\end{equation}
vanishes when evaluated on any two tangent directions
within the family. A submanifold with this property
and half the dimension of phase space is called
\emph{Lagrangian}. Its projection therefore admits the standard
classification of Lagrangian caustics \cite{Arnold1990}.

\paragraph{The cubic generating function.}
The projection from geodesic phase space onto the parameter
manifold retains $(v,u,\phi)$ and discards the momenta.
On the fixed-charge family, this is the map
$(p_v,u,\phi)\mapsto(v(p_v),u,\phi)$.
We encode it through the generating function
\begin{equation}
\begin{aligned}
 \mathscr S(p_v;v,u,\phi)
 &=Q_u u+Q_\phi\phi+p_v\,v
   -\int_0^{p_v}v(\zeta)\,\dd\zeta\\
 &=Q_u u+Q_\phi\phi+(v-v_*)p_v\\
 &\quad-\frac8{3c_*}p_v^3+O(p_v^5).
\end{aligned}
 \label{eq:XY-K-fold-generating-family}
\end{equation}
Here $v$ is the target coordinate, whereas $v(p_v)$ is
the function determined by the energy constraint.
Treating $p_v$ as an independent auxiliary variable,
the stationarity condition
$\partial_{p_v}\mathscr S=v-v(p_v)=0$ selects the
momenta that reach this target. Derivatives with respect
to $(v,u,\phi)$ recover $(p_v,Q_u,Q_\phi)$.

At $(p_v,v)=(0,v_*)$, the first two derivatives with
respect to $p_v$ vanish, while
\begin{equation}
 \partial_{p_v}^3\mathscr S=-\frac{16}{c_*}\ne0,
 \qquad
 \partial_v\partial_{p_v}\mathscr S=1.
\end{equation}
The first condition makes the leading dependence cubic.
The second one ensures that changing the target coordinate
moves the system through the singularity.
After smooth coordinate changes, the part controlling
stationarity takes the standard form
$t^3/3+\lambda t$, where $t$ is a rescaled auxiliary
variable and $\lambda$ measures displacement from the
caustic \cite{Arnold1990,BerryUpstill1980}.
Its stationary equation, $t^2+\lambda=0$, has two
solutions for $\lambda<0$, one merged solution at
$\lambda=0$, and none for $\lambda>0$.
This is the $A_2$ fold catastrophe, which persists under
small smooth perturbations of the Lagrangian family.

\paragraph{Expansion and shear.}\label[appendix]{app:XY-K-geometry}
The deformation identity has a useful general form. Consider $g=\dd r^2+a(r)^2\dd u^2+b(r)^2\dd\phi^2$, with
$a,b>0$, and a unit-speed geodesic field with common conserved momenta $p_u,p_\phi$.
Write $z=\dot r\ne0$. Then
\begin{equation}
 U=z\partial_r+\frac{p_u}{a^2}\partial_u+\frac{p_\phi}{b^2}\partial_\phi,
 \qquad z^2+\frac{p_u^2}{a^2}+\frac{p_\phi^2}{b^2}=1.
\end{equation}
In XY coordinates the velocity one-form is
\begin{equation}
U^\flat=\frac{\dot v}{4}\dd v+Q_u\dd u+Q_\phi\dd\phi,
 \label{eq:XY-K-closed}
\end{equation}
and in the general metric it is $z(r)\dd r+p_u\dd u+p_\phi\dd\phi$.
It is closed on each regular branch. Thus
$\nabla U$ is symmetric and has no longitudinal component. For
$\eta_u^\perp=\partial_u-p_uU$ and $\eta_\phi^\perp=\partial_\phi-p_\phi U$,
\begin{equation}
\begin{aligned}
 \cG_\perp&=\begin{pmatrix}a^2-p_u^2&-p_up_\phi\\-p_up_\phi&b^2-p_\phi^2\end{pmatrix},\\
 \det \cG_\perp&=a^2b^2z^2,\\
 \dot{\cG}_\perp&=z\operatorname{diag}((a^2)_{,r},(b^2)_{,r}).
\end{aligned}
\end{equation}
If $V$ denotes the Jacobi columns in a parallel orthonormal screen,
$\cG_\perp=V^TV$ and $\dot{V}=\cB V$. Since $\cB$ is symmetric,
$\tfrac12\cG_\perp^{-1}\dot{\cG}_\perp=V^{-1}\cB V$. Taking determinants proves
\begin{equation}
 \det\cB=\frac{(a^2)_{,r}(b^2)_{,r}z^2}{4a^2b^2z^2}
 =\frac{a_{,r}b_{,r}}{ab}=-K_{u\phi}.
\end{equation}
For the XY metric, $r=v/2$ and the Ricci contraction is
\begin{equation}
 \Ric(U,U)=4+64Q_u^2-8\csc^2v+96Q_\phi^2\csc^4v.
\end{equation}
For a symmetric two-dimensional operator,
$\norm\sigma^2=\Tr\cB^2-\theta^2/2
=\theta^2/2-2\det\cB$, proving \eqref{eq:XY-K-geometric}.
Substitution into the twist-free Raychaudhuri equation yields
\begin{equation}
 \dot\theta=-\theta^2-4-64Q_u^2-96Q_\phi^2\csc^4v<0.
 \label{eq:XY-K-Ray}
\end{equation}
The negative Ricci term $-8\csc^2v$ is canceled by the
$8\csc^2v$ in the shear norm. Expansion therefore decreases even
where Ricci curvature alone favors defocusing.
The momenta cancel from the determinant identity for this family. The two principal rates are
\begin{equation}
 \lambda_\pm=\frac12\left(\theta\pm
 \sqrt{\theta^2+16\csc^2v}\right),\qquad
 \lambda_+>0>\lambda_-.
 \label{eq:XY-K-rates}
\end{equation}

To obtain the deformation near a simple turn, set
$\delta=s-s_*$ and $c_*=F_{,v}(v_*)/2\ne0$.
The geodesic equation gives
$v=v_*+c_*\delta^2/2+O(\delta^4)$ and
$\dot v=c_*\delta+O(\delta^3)$. Substitution into
\cref{eq:XY-K-H,eq:XY-K-area,eq:XY-K-geometric} yields
\begin{equation}
\begin{aligned}
 \det\cG_\perp
 &=\frac{\cos^2v_*\,c_*^2}{512}\delta^2+O(\delta^4),\\
 \theta&=\frac1\delta+O(\delta),\\
 \frac{\norm{\sigma}^2}{\theta^2/2}
 &=1+16\csc^2v_*\,\delta^2+O(\delta^4).
\end{aligned}
\label{eq:XY-K-foldlimit}
\end{equation}
The last line follows directly from
$\norm{\sigma}^2=\theta^2/2+8\csc^2v$.
Thus expansion tends to $-\infty$ on approach to the fold
and to $+\infty$ on the outgoing branch. One principal
stretching rate diverges as $1/\delta$, while the other
is $-4\csc^2v_*\,\delta+O(\delta^3)$.
The ratio approaches unity as the fold is approached, since only
one transverse channel collapses. At the fold itself, the ratio
is undefined.
At a regular zero of $\theta$, the same ratio can diverge
without any singularity of the shear.

\subsection{Fidelity and quantum Fisher signatures}
\paragraph{Full and transverse sensitivity.}
\label[appendix]{app:XY-relative-channels}

Write $\boldsymbol Q=(Q_u,Q_\phi)^T$ and
$\cG=\operatorname{diag}(a^2,b^2)$. The transverse
label metric is
$\cG_\perp=\cG-\boldsymbol Q\boldsymbol Q^T$,
so
\begin{equation}
 \cG^{-1/2}\cG_\perp\cG^{-1/2}
 =I-\boldsymbol q\boldsymbol q^T,
 \qquad
 \boldsymbol q=\cG^{-1/2}\boldsymbol Q.
\end{equation}
Its eigenvalues are $1$ and
$1-|\boldsymbol q|^2=\dot v^2/4$, proving
\eqref{eq:XY-K-relativeFisher}. These give the transverse
response relative to the equal-arclength response.

At a turn, the label direction
$n=(Q_u/a_*^2,Q_\phi/b_*^2)$ satisfies
$n^A\eta_A=U_*$ because $\dot v_*=0$. Its transverse
separation therefore vanishes, although its full separation
does not. The equal-arclength label metric remains positive
definite.

\paragraph{Compensated fidelity.}
\label[appendix]{app:XY-K-overlap-proof}

For nearby parameters in the interior of the gapped patch,
expanding the logarithmic fidelity per site gives
$\mathfrak f=g_{\mu\nu}\delta\lambda^\mu\delta\lambda^\nu
+O(|\delta\lambda|^3)$.
We use this expansion to compare the equal-arclength
and compensated displacements at a fold.

For the label direction $n=(\dot u_*,\dot\phi_*)$, hold its
components fixed and define
$\xi=n^u\partial_u+n^\phi\partial_\phi$.
This is a Killing field with $\xi_*=U_*$, but its orbit
is generally not a geodesic. Its squared norm is
\begin{equation}
 |\xi|_g^2(v)
 =\frac{Q_u^2a^2(v)}{a_*^4}
  +\frac{Q_\phi^2b^2(v)}{b_*^4}.
\end{equation}
At the turn,
$\partial_v|\xi|_g^2=F_{,v}(v_*)/4=c_*/2$.
The Killing identity and $g^{vv}=4$ therefore give
\begin{equation}
\begin{aligned}
 \left.\nabla_\xi\xi\right|_*&=-c_*\partial_v,\\
 \norm{\nabla_\xi\xi}_g^2\big|_*&=\frac{c_*^2}{4}.
\end{aligned}
\label{eq:XY-K-orbit-acceleration}
\end{equation}

At equal arclength, a label shift $\epsilon n$ produces
the displacement $\epsilon U_*$ to first order.
Evaluating the shifted path at $s_*-\epsilon$ cancels
this term. Since $\ddot u_*=\ddot\phi_*=0$, the remaining
coordinate differences are
\begin{equation}
 \Delta v=\frac{c_*}{2}\epsilon^2+O(\epsilon^4),
 \qquad
 \Delta u,\Delta\phi=O(\epsilon^3).
\end{equation}
The leading residual displacement is therefore
$-\tfrac12(\nabla_\xi\xi)_*\epsilon^2$.
Using $g(U_*,U_*)=1$ and $g_{vv}=1/4$ in the fidelity
expansion gives
\begin{equation}
\begin{aligned}
 \mathfrak f_{\rm eq}
 &=\epsilon^2+O(\epsilon^3),\\
 \mathfrak f_{\rm comp}
 &=\frac{c_*^2}{16}\epsilon^4+o(\epsilon^4).
\end{aligned}
\end{equation}
This proves \cref{eq:XY-K-overlap}: removing the longitudinal
separation suppresses the quadratic response, while the
symmetry-orbit acceleration fixes the remaining quartic term.

\paragraph{Compensation as comparison on a common wavefront.}
On each regular branch, the closed velocity one-form defines
a local stationary Hamilton-Jacobi function $S$ through
$dS=U^\flat$. Since $U$ has unit norm, $dS/ds=1$.
Pulling this relation back to the family coordinates
$(s,\lambda^A)$, with $\lambda^A=(u_0,\phi_0)$, gives
\begin{equation}
 dS=ds+Q_A\,d\lambda^A,
 \qquad S=s+Q_A\lambda^A+\mathrm{const}.
\end{equation}
Here $g(U,\eta_A)=Q_A$, and the momenta are fixed throughout
the family. A comparison at fixed $S$ therefore requires
$\delta s=-Q_A\delta\lambda^A$. Its first-order state
displacement is
\begin{equation}
 \delta P
 =U\,\delta s+\eta_A\delta\lambda^A
 =(\eta_A-Q_AU)\delta\lambda^A.
\end{equation}
Thus $\cG_\perp$ measures the local fidelity response along
a common geometrical wavefront, whereas $\cG$ compares
states at equal arclength.

At the fold, unit speed and $\dot v_*=0$ imply
\begin{equation}
 Q_An^A=\frac{Q_u^2}{a_*^2}
       +\frac{Q_\phi^2}{b_*^2}=1.
\end{equation}
Consequently, the label shift $\epsilon n$ and arclength
shift $-\epsilon$ preserve $S$ exactly in the family
coordinates. Their first-order displacement vanishes,
leaving the second-order mismatch evaluated above.

The function $S$ is defined locally on each regular
projected branch. Through the fold it remains smooth on
the lifted phase-space family, but need not define a
single-valued smooth function of $(v,u,\phi)$.
Here ``wavefront'' means a surface of constant geodesic
action \cite{Arnold1990,BerryUpstill1980}.

\bibliographystyle{quantum}
{\bibliography{QPC_references}}

\end{document}